\magnification=\magstephalf

%
% NOVA Toponderzoekschool, August 1997
%
% Set of macros to change the point size globally
% (and not just one special font family)
%
% First, the fonts
%
\font\ninerm=cmr9      \font\eightrm=cmr8     \font\sixrm=cmr6
\font\ninei=cmmi9      \font\eighti=cmmi8     \font\sixi=cmmi6
\font\ninesy=cmsy9     \font\eightsy=cmsy8    \font\sixsy=cmsy6
\font\ninebf=cmbx9     \font\eightbf=cmbx8    \font\sixbf=cmbx6
\font\ninett=cmtt9     \font\eighttt=cmtt8
\font\nineit=cmti9     \font\eightit=cmti8
\font\ninesl=cmsl9     \font\eightsl=cmsl8
\font\tenrm=cmr10      \font\twelverm=cmr10 scaled 1200
\font\teni=cmmi10      \font\twelvei=cmmi10 scaled 1200
\font\tensy=cmsy10     \font\twelvesy=cmsy10 scaled 1200
\font\tenbf=cmbx10     \font\twelvebf=cmbx10 scaled 1200
\font\tentt=cmtt10     \font\twelvett=cmtt10 scaled 1200
\font\tenit=cmti10     \font\twelveit=cmti10 scaled 1200
\font\tensl=cmsl10     \font\twelvesl=cmsl10 scaled 1200
\font\tenex=cmex10     \font\twelveex=cmex10 scaled 1200

\font\fourteenrm=cmr10 scaled 1440
\font\fourteeni=cmmi10 scaled 1440
\font\fourteensy=cmsy10 scaled 1440
\font\fourteenbf=cmbx10 scaled 1440
\font\fourteentt=cmtt10 scaled 1440
\font\fourteenit=cmti10 scaled 1440
\font\fourteensl=cmsl10 scaled 1440
\font\fourteenex=cmex10 scaled 1440

%
%      Some special font parameters
%
\skewchar\teni='177    \skewchar\twelvei='177  \skewchar\fourteeni='177
\skewchar\ninei='177   \skewchar\eighti='177   \skewchar\sixi='177
\skewchar\tensy='60    \skewchar\twelvesy='60  \skewchar\fourteensy='60
\skewchar\ninesy='60   \skewchar\eightsy='60   \skewchar\sixsy='60
\hyphenchar\tentt=-1   \hyphenchar\twelvett=-1 \hyphenchar\fourteentt=-1
\hyphenchar\ninett=-1  \hyphenchar\eighttt=-1  \hyphenchar\tentt=-1
%
%      Insert here other special fonts
%
%%%%%%%%%%%%%%%%%%%%%%%%%%%%%%%%%%%%%%%%%%%%%%%%%%%%%%%%%%%%%%%%%%%%%%%%
%
%      Size-swithching macros --- support math in each size, and also provide
%      small caps pseudofont (\sc)
%      Huib: included \ssc for small capitals in superscript
%      This being the TeXbook format, a special glue is set up for \tt;
%      should be switched off if true constant spacing is used
%      (from p.414 of the TeXbook)
%
\catcode`@=11%      necessary to access some internal macros of plain TeX
\newskip\ttglue
\def\tenpoint{\def\rm{\fam0\tenrm}
  \textfont0=\tenrm \scriptfont0=\sevenrm  \scriptscriptfont0 = \fiverm
  \textfont1=\teni  \scriptfont1=\seveni   \scriptscriptfont1 = \fivei
  \textfont2=\tensy \scriptfont2=\sevensy  \scriptscriptfont2 = \fivesy
  \textfont3=\tenex   \scriptfont3=\tenex  \scriptscriptfont3 = \tenex
  \textfont\itfam=\tenit  \def\it{\fam\itfam\tenit}%
  \textfont\slfam=\tensl  \def\sl{\fam\slfam\tensl}%
  \textfont\ttfam=\tentt  \def\tt{\fam\ttfam\tentt}%
  \textfont\bffam=\tenbf  \scriptfont\bffam=\sevenbf%
   \scriptscriptfont\bffam=\fivebf   \def\bf{\fam\bffam\tenbf}%
  \tt \ttglue=0.5em plus 0.25em minus 0.15em
  \normalbaselineskip=12pt
  \setbox\strutbox=\hbox{\vrule height 8.5pt depth 3.5pt width 0pt}%
  \let\sc=\eightrm \let\ssc=\sixrm  \let\big=\tenbig  \normalbaselines\rm}
\def\ninepoint{\def\rm{\fam0\ninerm}
  \textfont0=\ninerm \scriptfont0=\sixrm  \scriptscriptfont0 = \fiverm
  \textfont1=\ninei  \scriptfont1=\sixi   \scriptscriptfont1 = \fivei
  \textfont2=\ninesy \scriptfont2=\sixsy  \scriptscriptfont2 = \fivesy
  \textfont3=\tenex   \scriptfont3=\tenex  \scriptscriptfont3 = \tenex
  \textfont\itfam=\nineit  \def\it{\fam\itfam\nineit}%
  \textfont\slfam=\ninesl  \def\sl{\fam\slfam\ninesl}%
  \textfont\ttfam=\ninett  \def\tt{\fam\ttfam\ninett}%
  \textfont\bffam=\ninebf  \scriptfont\bffam=\sixbf%
   \scriptscriptfont\bffam=\fivebf   \def\bf{\fam\bffam\ninebf}%
  \tt \ttglue=0.5em plus 0.25em minus 0.15em%
  \normalbaselineskip=11pt%
  \setbox\strutbox=\hbox{\vrule height 8pt depth 3pt width 0pt}%
  \let\sc=\sevenrm   \let\ssc=\fiverm \let\big=\ninebig  \normalbaselines\rm}
\def\eightpoint{\def\rm{\fam0\eightrm}
  \textfont0=\eightrm \scriptfont0=\sixrm  \scriptscriptfont0 = \fiverm
  \textfont1=\eighti  \scriptfont1=\sixi   \scriptscriptfont1 = \fivei
  \textfont2=\eightsy \scriptfont2=\sixsy  \scriptscriptfont2 = \fivesy
  \textfont3=\tenex   \scriptfont3=\tenex  \scriptscriptfont3 = \tenex
  \textfont\itfam=\eightit  \def\it{\fam\itfam\eightit}%
  \textfont\slfam=\eightsl  \def\sl{\fam\slfam\eightsl}%
  \textfont\ttfam=\eighttt  \def\tt{\fam\ttfam\eighttt}%
  \textfont\bffam=\eightbf  \scriptfont\bffam=\sixbf%
   \scriptscriptfont\bffam=\fivebf   \def\bf{\fam\bffam\eightbf}%
  \tt \ttglue=0.5em plus 0.25em minus 0.15em%
  \normalbaselineskip=9pt%
  \setbox\strutbox=\hbox{\vrule height 7pt depth 2pt width 0pt}%
  \let\sc=\sixrm  \let\ssc=\fiverm \let\big=\eightbig  \normalbaselines\rm}
\def\tenbig#1{{\hbox{$\left#1\vbox to 8.5pt{}\right.\n@space$}}}
\def\ninebig#1{{\hbox{$\textfont0=\tenrm\textfont2=\tensy
  \left#1\vbox to 7.25pt{}\right.\n@space}}}
\def\eightbig#1{{\hbox{$\textfont0=\ninerm\textfont2=\ninesy
  \left#1\vbox to 6.5pt{}\right.\n@space}}}
\def\twelvebig#1{{\hbox{$\textfont0=\twelverm\textfont2=\twelvesy
  \left#1\vbox to 11pt{}\right.\n@space}}}
\def\fourteenbig#1{{\hbox{$\textfont0=\fourteenrm\textfont2=\fourteensy
  \left#1\vbox to 13pt{}\right.\n@space}}}
% for 10-point math in 9-point territory
%
\def\twelvepoint{\def\rm{\fam0\twelverm}
  \textfont0=\twelverm \scriptfont0=\ninerm  \scriptscriptfont0 = \sevenrm
  \textfont1=\twelvei  \scriptfont1=\ninei   \scriptscriptfont1 = \seveni
  \textfont2=\twelvesy \scriptfont2=\ninesy  \scriptscriptfont2 = \sevensy
  \textfont3=\twelveex \scriptfont3=\twelveex  \scriptscriptfont3 = \twelveex
  \textfont\itfam=\twelveit  \def\it{\fam\itfam\twelveit}%
  \textfont\slfam=\twelvesl  \def\sl{\fam\slfam\twelvesl}%
  \textfont\ttfam=\twelvett  \def\tt{\fam\ttfam\twelvett}%
  \textfont\bffam=\twelvebf  \scriptfont\bffam=\ninebf%
   \scriptscriptfont\bffam=\sevenbf   \def\bf{\fam\bffam\twelvebf}%
  \tt \ttglue=0.5em plus 0.25em minus 0.15em%
  \normalbaselineskip=15pt%
  \setbox\strutbox=\hbox{\vrule height 11pt depth 4pt width 0pt}% these values
                                                           % should be checked
  \let\sc=\ninerm  \let\ssc=\sevenrm  \let\big=\twelvebig  \normalbaselines\rm}
\def\fourteenpoint{\def\rm{\fam0\fourteenrm}
  \textfont0=\fourteenrm \scriptfont0=\tenrm  \scriptscriptfont0 = \sevenrm
  \textfont1=\fourteeni  \scriptfont1=\teni   \scriptscriptfont1 = \seveni
  \textfont2=\fourteensy \scriptfont2=\tensy  \scriptscriptfont2 = \sevensy
  \textfont3=\fourteenex \scriptfont3=\fourteenex
                                              \scriptscriptfont3 = \fourteenex
  \textfont\itfam=\fourteenit  \def\it{\fam\itfam\fourteenit}%
  \textfont\slfam=\fourteensl  \def\sl{\fam\slfam\fourteensl}%
  \textfont\ttfam=\fourteentt  \def\tt{\fam\ttfam\fourteentt}%
  \textfont\bffam=\fourteenbf  \scriptfont\bffam=\tenbf%
   \scriptscriptfont\bffam=\sevenbf   \def\bf{\fam\bffam\fourteenbf}%
  \tt \ttglue=0.5em plus 0.25em minus 0.15em%
  \normalbaselineskip=12pt%
  \setbox\strutbox=\hbox{\vrule height 13pt depth 4.5pt width 0pt}% these values
                                                             % should be checked
  \let\sc=\tenrm   \let\ssc=\eightrm \let\big=\fourteenbig  \normalbaselines\rm}
\catcode`@=12%      make @ sign again special character
%
% This added by MRH to get into 10-point family by default.
%
% In principle this is taken care of by plain-TeX, but somehow not all
% goodies defined in \tenpoint are available (e.g. \sc is undefined)
%
% Futhermore, definition of \sc changed into true 'caps and small caps' font
%
\tenpoint
\let\sc=\tencsc

\input psfig.tex
\footline={\ifnum\pageno=1 \hfil \else {\hss \tenrm \folio \hss}\fi }
\hsize 6.2 true in
\vsize 8.5 true in
\baselineskip 13 pt
\hyphenation{mo-le-cu-les}
\hyphenation{para-me-triz-ed}
\hyphenation{u-bi-qui-tous}
\hyphenation{Goi-co-e-che-a}
\hyphenation{Bausch-li-cher}
\def\lapprox{{_<\atop{^\sim}}}
\def\grapprox{{_>\atop{^\sim}}}
\newdimen\digitwidth
\setbox0=\hbox{\rm0}
\digitwidth=\wd0
\catcode`#=\active
\def#{\kern\digitwidth}
\def\df{{\dotfill}}
\parindent 0pt
\parskip 5pt
\centerline{\bf ISO Spectroscopy of Gas and Dust: From
Molecular Clouds to Protoplanetary Disks}
\bigskip
\centerline{\it Ewine F. van Dishoeck}
\centerline{\it Leiden Observatory, P.O. Box 9513, 2300 RA Leiden, The
                Netherlands}
\centerline{\tt ewine@strw.leidenuniv.nl}
\bigskip
\bigskip
\centerline{\bf CONTENTS}
\bigskip
{\baselineskip 13 pt
\settabs \+ xxxxxx xxxxxx xxxxxx xxxxxx xxxxxx xxxxxxxxxx
   xxxxxxxxxxxxxxxxxxxxxxxxxxxxxxxxxx& xxxx & \cr
\+ 1.\ INTRODUCTION \df & ## &\cr
\+ 2.\ INFRARED SPECTROSCOPIC FEATURES \df &\cr
\+ 3.\ GAS-PHASE MOLECULES AND THEIR CHEMISTRY \df &\cr
\+ #####3.1. New Detections \df &\cr
\+ #####3.2. Other Molecules \df &\cr
\+ #####3.3. H$_2$ and HD \df &\cr
\+ #####3.4. H$_2$O, OH and O \df &\cr
\+ 4.\ INTERSTELLAR ICES \df &\cr
\+ #####4.1. Inventory of Ice Features \df &\cr
\+ #####4.2. Ice Abundances and Chemistry \df &\cr
\+ #####4.3. Heating and Processing of Ices \df &\cr
\+ 5.\ POLYCYCLIC AROMATIC HYDROCARBONS \df &\cr
\+ #####5.1. PAH Spectroscopy \df &\cr
\+ #####5.2. PAH Feature Variations \df &\cr
\+ #####5.3. Hydrogenated Amorphous Carbon \df &\cr
\+ 6.\ SILICATES \df &\cr
\+ #####6.1. Amorphous Silicates \df &\cr
\+ #####6.2. Crystalline Silicates \df &\cr
\+ 7.\ PHOTON-DOMINATED REGIONS\df &\cr
\+ #####7.1. Low-Excitation PDRs and the Gas Heating Efficiency\df &\cr
\+ #####7.2. H$_2$ Pure Rotational Lines \df &\cr
\+ 8.\ SHOCKS \df &\cr
\+ #####8.1. Shock Structure and Physical Conditions \df &\cr
\+ #####8.2. Gas Cooling in Shocks \df &\cr
\+ 9.\ EMBEDDED YOUNG STELLAR OBJECTS \df &\cr
\+ #####9.1. Low-Mass YSOs  \df &\cr
\+ #####9.2. Intermediate and High-Mass YSOs \df &\cr
\+ 10.\ CIRCUMSTELLAR DISKS AROUND YOUNG STARS \df &\cr
\+ #####10.1. Dust Evolution and Disk Structure \df &\cr
\+ #####10.2. Gas in Disks \df &\cr
\+ 11.\ CONCLUDING REMARKS \df &\cr
}
\vskip 0.5 cm
\centerline{\bf To appear in 2004 Annual Reviews of Astronomy and Astrophysics}
\medskip
\centerline{\it Submitted January 2004; subject to minor changes}

\vfill \eject
\centerline{\bf ABSTRACT}
Observations of interstellar gas-phase and solid-state species in the
2.4--200 $\mu$m range obtained with the spectrometers on board the
Infrared Space Observatory are reviewed. Lines and bands due to ices,
polycyclic aromatic hydrocarbons, silicates and gas-phase atoms and
molecules (in particular H$_2$, CO, H$_2$O, OH and CO$_2$) are
summarized and their diagnostic capabilities illustrated. The results
are discussed in the context of the physical and chemical evolution of
star-forming regions, including photon-dominated regions, shocks,
protostellar envelopes and disks around young stars.


\vfill \eject
\centerline{\bf 1. INTRODUCTION}
\medskip
During the formation of stars deep inside molecular clouds, the
surrounding gas and dust become part of the infalling envelope feeding
the central object. The protostars themselves are extinguished by
hundreds of magnitudes, so that the circumstellar gas and dust are the
only tracers of the physical processes happening inside.  A study of
their characteristics is therefore key to understanding the origin of
stars.  Part of this gas and dust ends up in the rotating disks
surrounding young stars, and forms the basic material from which icy
planetesimals, and ultimately rocky and gaseous planets, are
formed. Spectroscopic surveys of star-forming regions at different
evolutionary stages therefore also provide quantitative information on
the building blocks available during planet formation.  The
Infrared Space Observatory (ISO)\footnote{$^1$}{\eightpoint ISO was a
project of the European Space Agency with instruments funded by ESA
member states (especially the PI countries: France, Germany, The
Netherlands, and the United Kingdom) and with the participation of
ISAS and NASA (Kessler et al. 1996, 2003).} has provided the first
opportunity to obtain {\it complete} infrared spectra from 2.4--200
$\mu$m unhindered by the Earth's atmosphere.

At the temperatures characteristic of star-forming regions, most of
the radiation emerges at mid- and far-infrared wavelengths, the
majority of which is blocked from Earth.  Most young stellar objects
(YSOs) have been found through ground-based infrared surveys and the
InfraRed Astronomical Satellite (IRAS).  The spectra of the coldest
protostellar objects (ages of $\sim 10^4$ yr since collapse began)
peak around 100 $\mu$m and such sources are best studied with
ground-based submillimeter telescopes and the longest wavelength
instruments on ISO.  Once the dense envelopes start to dissipate due
to the effects of outflows, the objects become detectable at
mid-infrared wavelengths, around ages of $\sim 10^5$ yr, and are
accessible to the shorter wavelength instruments on ISO.  The outflows
also create shocks in the surrounding molecular cloud, which emit
brightly in mid-infrared lines. If ultraviolet (UV) radiation from the
young stars can escape, it heats the neighboring gas and
photodissociates molecules, creating so-called photon-dominated or
photodissociation regions (PDRs).  The tremendous range in physical
conditions in these different phases, with densities from $10^4$ to
$10^{13}$ cm$^{-3}$ and temperatures from 10 to 10,000 K, are
reflected in the abundances and excitation of the atoms and
molecules. The ISO spectroscopic data therefore contain a wealth of
information with which to unravel the physical and chemical processes
during star- and planet formation.

{\parindent 20pt
{\narrower
\topinsert{
\null
%\vskip -3.5 true cm
\centerline{\hbox{
\psfig{figure=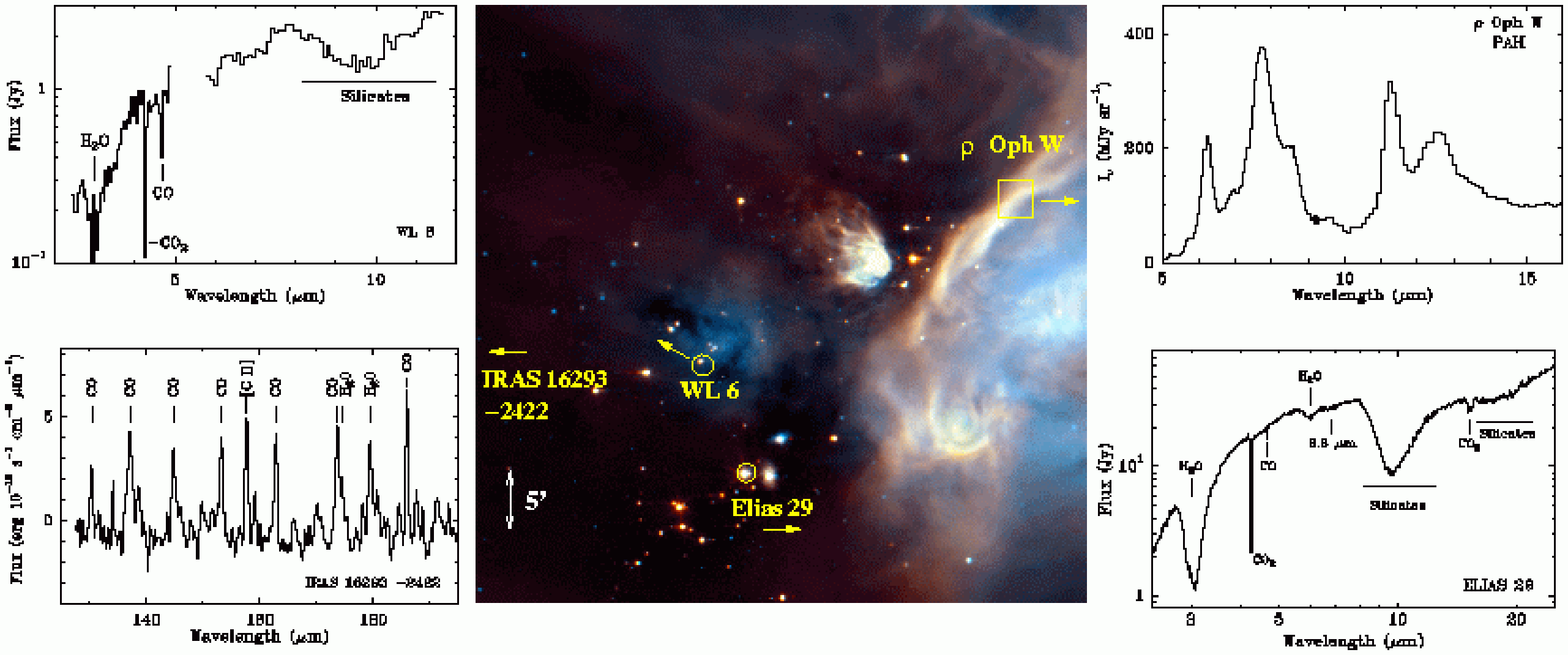,height=7cm,angle=-90}
}} 
%\vskip -3.5 true cm
\noindent
{\bf Figure 1 (color).} ISOCAM composite 7 and 15 $\mu$m image of
the $\rho$ Oph molecular cloud. The bright filaments in the west trace
the hot dust illuminated by the B2V star HD 147889 located just
outside the map.  The dense cloud containing IRAS 16293 -2422 is to
the east. The dark patches are very dense cores which are optically
thick even at 15 $\mu$m, whereas the bright point sources throughout
the image are low-mass YSOs. Most of the extended 7$\mu$m emission is
due to PAHs (Abergel et al.\ 1996).  Spectra of various objects in
$\rho$ Oph are indicated on the side.  Left Top: PHOT-S spectrum of WL
6 (G\"urtler et al., unpublished); Right Top: CAM-CVF spectrum of
$\rho$ Oph W (Boulanger et al. 2000); Left Bottom: LWS spectrum of
IRAS 16293 -2422 (Ceccarelli et al. 1998a); Right Bottom: SWS spectrum
of Elias 29 (Boogert et al. 2000b).}
\endinsert 
}}

ISO's main contributions have been in the following areas.  First, the
complete wavelength coverage has provided an unbiased overview of the
major gas- {\it and} solid-state species in star-forming regions,
including an inventory of the reservoirs of the major elements (C, O,
N, ...). Indeed, the identifications of many species have become much
more secure because of the broad spectral range without gaps due to
the Earth's atmosphere.  Second, ISO was particularly well suited to
study the physical structure of warm gas in the important 100--2000 K
regime, which is difficult to probe with ground-based telescopes:
(sub-)millimeter CO line observations trace cold ($<100$ K) gas,
whereas near-IR H$_2$ and optical atomic lines refer to much hotter
($\geq 2000$ K) gas.  Third, ISO has quantified the energetics of
star-forming regions, by measuring directly the photoelectric heating
efficiency as well as the total cooling power in all relevant
lines. Finally, the sensitivity of the instruments has allowed much
larger samples to be probed, so that systematic trends and
characteristics at each stage of evolution could be determined.  Much
of this progress would not have been possible without dedicated
laboratory astrophysics and theoretical chemistry studies to provide
the basic molecular and solid-state data needed to analyze the ISO
spectra.

{\parindent 20pt
{\narrower
\topinsert{
\settabs \+ xxxxxxxxxxx & xxxxxxxxxxxxxxx & xxxxxxxxxxxxxxxx
  & xxxxxxxxxxxxxxx & xxxxxxxxx \cr
\centerline{\bf Table 1. Spectrometers on the Infrared Space Observatory}
\bigskip
\hrule
\vskip 1 pt
\hrule
{\ninepoint
\smallskip
\+ Instrument & Wavelength  & Resolving  power& Aperture & Ref.\cr
\+         & range ($\mu$m) & ##($\lambda/\Delta \lambda$)& (arcsec) \cr
\smallskip
\hrule
\smallskip
\+ SWS & 2.4--45.2 & #1500$^a$ & $14\times 20$ -- $20\times 33$ & 1\cr
\+    & 11.4--44.5 & 30000 & $10\times 39$ -- $17\times 40$ \cr
\+ LWS & 43--197 & ##200 & ####$\sim 80^b$ & 2\cr
\+    &        & 10000 & ####$\sim 80^b$ \cr
\+ CAM-CVF & 2.3--16.5 & 35--50 & ##$12\times 12$ & 3 \cr
\+ PHOT-S & 2.5--5, 5.8--11.6 & ##90 & ##$24\times 24$ & 4 \cr
\smallskip
\hrule
\bigskip
\noindent
$^a$ Varying with wavelength from 1000---2500; full spectral scans have
usually been obtained with a factor 2 lower spectral resolution.
         
\noindent                                                                      
$^b$ Diameter of circular aperture.

\smallskip \noindent
References. 1. de Graauw et al. 1996a; 2. Clegg et al. 1996;
3. Cesarsky et al. 1996; 4. Lemke et al.\ 1996
}}
\endinsert 
}}

This review summarizes the spectroscopic results from all four ISO
instruments (see Table~1 for characteristics). The bulk of the data
come from the Short Wavelength Spectrometer (SWS) and Long Wavelength
Spectrometer (LWS), but relevant spectra obtained with the
Camera-Circular Variable Filter (CAM-CVF) and Photometer-Spectrometer
(PHOT-S) are included.  It is heavily biased toward ISO results only,
but the science case builds on pioneering observations performed prior
to ISO with ground-based and airborne infrared telescopes, in
particular the Kuiper Airborne Observatory (KAO) (Haas et al.\ 1995),
and with the IRAS and the InfraRed Telescope in Space (IRTS) missions.
A historical review of those data in the context of the ISO results
has been given by van Dishoeck \& Tielens (2001).  Also, more recent
results from complementary ground-based facilities and new
scientifically related space missions such as the Submillimeter Wave
Astronomical Satellite (SWAS) and ODIN are mentioned only in
passing.

Due to space limitations, this review focuses on spectroscopy of
galactic molecular clouds and the immediate surroundings of
YSOs\footnote{$^2$}{\eightpoint The term `protostar' refers strictly
only to objects which derive most of their luminosity from accretion
and have L$_{\rm submm}$/L$_{\rm bol}> 0.5\%$ and $M_{\rm env}/M_*>1$
(Andr\'e et al. 2000). The term `young stellar object' refers to all
embedded stages of star formation and associated phenomena such as
outflows. In this review, the term protostar will be used loosely to
also refer to the more evolved embedded stages, especially in the
context of high-mass objects.}. Figures 1--3 illustrate the striking
variety of spectral features associated with active star-forming
regions.  ISO imaging has also contributed to our understanding of the
interstellar medium, especially through spatial variations in grain
components (e.g., Abergel et al.\ 2002), the discovery of infrared
dark clouds (Bacmann et al.\ 2000, Omont et al.\ 2003) and studies of
the coldest pre-stellar cores (Ward-Thompson et al.\ 2002, Krause et
al.\ 2003), but details are not covered here. This paper also does not
include ionized regions of the interstellar medium, nor the
circumstellar material around evolved (post-) AGB stars and planetary
nebulae (Molster \& Waters 2003).  A general overview of interstellar
dust has been given by Draine (2003) whereas extragalactic ISO spectra
have been summarized by Genzel \& Cesarsky (2000).  Even with these
limitations, it is not possible to refer to all ISO spectroscopy
results. Many of them have been summarized in dedicated ISO
conferences, including Heras et al.\ (1997), Yun \& Liseau (1998),
Waters et al.\ (1998), Cox
\& Kessler (1999), d'Hendecourt et al.\ (1999) and Salama et al.\
(2000).

{\parindent 20pt
{\narrower
\topinsert
\centerline{\hbox{
\psfig{figure=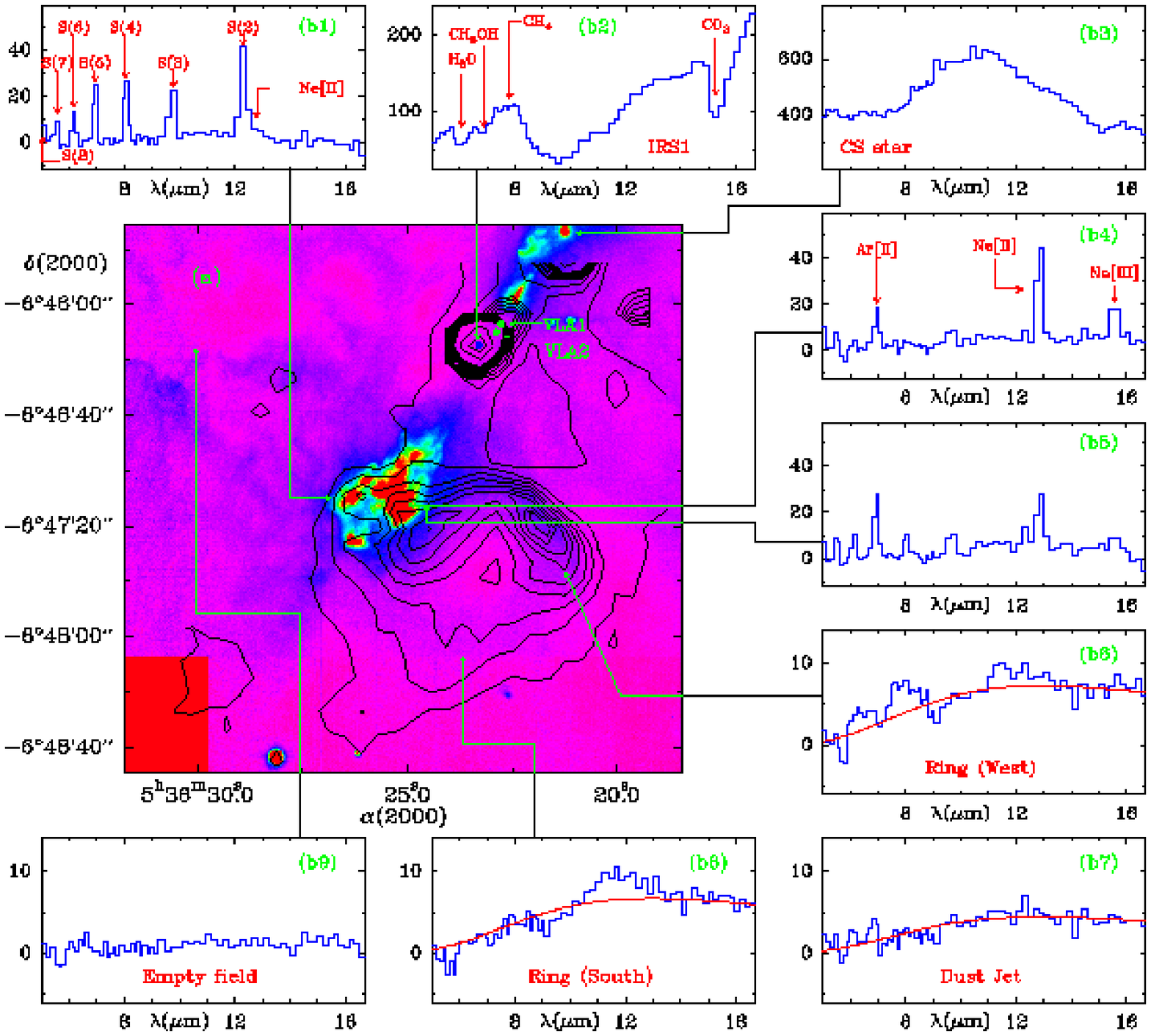,height=10cm,angle=0}
}}
%\vskip -5.0 true cm 
\noindent
{\bf Figure 2 (color).} CAM-CVF image of the HH 1-2 region by
Cernicharo et al.\ (2000b), illustrating a wide variety of types of
spectra, ranging from ice absorptions toward the class 0 protostar
VLA1 (spectrum b2) to silicate emission from a disk around the
Cohen-Schwarz star (spectrum b3) and gas emission from shocks (spectra
b1, b4, b5).  Both low-velocity shocks ($v\approx 10-15$) km s$^{-1}$
as traced by the strong H$_2$ lines in panel b1 and high-velocity
shocks ($v\approx 100-140$ km s$^{-1}$, as traced by the [Ne II] and
[Ne III] lines in panel b4 are seen.  Note the complete absence of any
continuum emission due to hot dust and PAH emission in both panels.
The color image is a [Si II] 6717 \AA \ optical image, whereas the
black contours indicate the 7 $\mu$m CAM image. The intensity scale is
in mJy.
\endinsert
}}

\bigskip
\medskip
\vfill \eject
\centerline{\bf 2. INFRARED SPECTROSCOPIC FEATURES}
\medskip
The mid- and far-infrared wavelength ranges are spectacularly rich in
spectral features, many of which probe unique species which cannot be
observed at any other wavelength. The wealth of sharp gas-phase lines
and broad solid-state bands is illustrated in the SWS spectrum of one
of the brightest mid-infrared sources in the sky, the Orion-KL region
(Figure 3). Each of these features, summarized in Table~2, has its
own diagnostic capability. 

\pageinsert
{
\settabs \+ xxxxxxxxxxx & xxxxxxxxxxxxx & xxxxxxxxxxxxxx
  & xxxxxxxxxxxxxxxxxxx & xxxxxx & xxxxxx & xxxxxxxxx & xxxxxx & xxxxxxxxx \cr
\centerline{\bf Table 2. Selected mid- and far-infrared
spectral features observed by ISO}
\medskip
\hrule
\vskip 1 pt
\hrule
{\eightpoint
\smallskip
\+ Category & \ \ $\lambda$  & Species/ & Diagnostic$^a$ &
     PDR & Shock & Embedded & Disk &  \cr
\+         & ($\mu$m) & Line & & & & ##YSO & & \cr
\hrule
\smallskip
\+ Atoms &  25.2 & [S I] & shock- vs. & - & + & - & - & \cr
\+ & 34.8 & [Si II] & photon heating  & + & + & - & - & \cr
\+ & 63.2 & [O I] &                   & + & + & +  & - &  \cr
\+ & 145.5 & [O I] &                  & + & + & + & - &\cr
\+ & 157.7 & [C II] &                 & + & + & + & ? &\cr
\smallskip
\+ H$_2$  & 6.9 & 7$\to$5 & mass and temperature & + & + & - & - \cr
\+       & 8.0 & 6$\to$4 & of warm gas,        & + & + & - & -\cr
\+     & 9.7 & 5$\to$3 & shock- vs.            & + & + & + & - \cr
\+     & 12.2 & 4$\to$2 & photon heating       & + & + & + & - \cr
\+     & 17.0 & 3$\to$1 &                      & + & + & + & ? \cr
\+     & 28.2 & 2$\to$0 &                      & + & + & + & ? \cr
\smallskip
\+ HD & 19.4 & 6$\to$5 & [D]/[H]               & - & + & - & - \cr
\+   & 112.0 & 1$\to$0 &                       & + & - & - & - \cr
\smallskip
\+ Gas-phase & 6.0 & H$_2$O & temperature + density, & - & + & + & - \cr
\+ Molecules &7.7 & CH$_4$ &  ice evaporation,       & - & - & + & - \cr
\+        & 13.7 & C$_2$H$_2$ & organic chemistry,    & - & - & + & - \cr
\+        & 14.0 & HCN &  depletion                  & - & - & + & - \cr
\+        & 15.0 & CO$_2$   &                        & - & + & + & - \cr
%\+        & 16.0, 16.5 & CH$_3$                      & + & - & - & - \cr
%\+        & 121.7 & HF                               & - & + & - & - \cr
\+        & 104.4 & CO 25$\to$24   &                 & - & + & + & - \cr
\+        & 108.1 & $o$-H$_2$O $2_{21}-1_{10}$  &    & - & + & + & - \cr
\+        & 119.3 & OH $^2\Pi_{3/2}$ ${5\over 2} \to {3\over 2}$
     && - & + & + & - \cr
%\+        & 113.5 & $o$-H$_2$O $4_{14}-3_{03}$       & - & + & + & - \cr
\+        & 130.4 & CO 20$\to$19                    & & - & + & + & - \cr
\+        & 138.5 & $p$-H$_2$O $3_{13}-2_{02}$      && - & + & + & - \cr
\+        & 162.8 & CO 16$\to$15                    && - & + & + & - \cr
%\+        & 163.3 & OH $²\Pi_{1/2}$ ${3\over 2} \to {1\over 2}$ && - & + & + &- \cr
\+        & 174.6 & $o$-H$_2$O $3_{03}-2_{12}$      & & - & + & + & - \cr
\+        & 179.5 & $o$-H$_2$O $2_{12}-1_{01}$       && - & + & + & - \cr
\+        & 186.0 & CO 14$\to$13                    && - & + & + & - \cr
 
\smallskip
\+ PAHs &  3.3, 6.2, 7.7,  & C-H and &carbonaceous material, & + & - & - & +\cr
\+     &   8.6, 11.3, 12.7, & C-C modes & UV radiation       & + & - & - & +\cr
\+     &  14.2, 16.2        &     &                          & + & - & - & +\cr
\smallskip
\+ Silicates    & 9.7   & &               bulk of dust   & - & - & + & + \cr
\+ (Amorphous)  & 18.0  & &                                 & - & - & + & + \cr
\smallskip
\+ Silicates & 11.3, 16.4, 23.9, & Forsterite & mineralogy, & - & - & - & + \cr
\+ (Crystalline)$^b$  &  27.7, 33.8, 69 &  &grain growth and & - & - & - & +\cr
\+    &  18.5, 21.5, 24.5 & Enstatite  & processing/heating, & - & - & - & +\cr
\+    & 8.6            & Silica    & solar system & - & - & - & + \cr
\+    & 65             & Diopside & connection & - & - & ? & ? \cr
\smallskip
\+ Oxides  & 11.6    & Al$_2$O$_3$ & solar system   & - & - & - & + \cr
\+         &    23 & FeO   & connection             & - & - & - & +\cr
\+ Sulfides  & 23.5 & FeS &                          & - & - & - & +\cr
\+ Carbonates & 92.6 & Calcite &                     & - & - & ? & - \cr
\smallskip
\+ Ices & 4.27 & CO$_2$ & volatile solids,            & - & - & + & - \cr
\+      & 4.38 & $^{13}$CO$_2$ & organic chemistry,   & - & - & + & - \cr
\+      & 4.67 & CO & thermal history,                & - & - & + & - \cr
\+      & 6.0, 13 & H$_2$O & solar system             & - & - & + & - \cr
\+      & 6.85 & CH$_3$OH + NH$_4^+$?& connection     & - & - & + & - \cr
\+      & 7.7 & CH$_4$     &                          & - & - & + & - \cr
%\+      & 9.3 & NH$_3$ &                              & - & - & + & - \cr
%\+      & 9.7 & CH$_3$OH &                            & - & - & + & - \cr
\+      & 15.2 & CO$_2$ &                             & - & - & + & - \cr
\+      & 44, 63 & H$_2$O (cryst.) &                  & - & - & + & + \cr
\smallskip
\hrule
\smallskip
$^{a}$ Diagnostic properties of category of species; individual
species or lines probe a subset of these properties.
\smallskip
$^b$ Position may vary depending on composition
 
}}

\endinsert
 
{\parindent 20pt
{\narrower
\topinsert
\centerline{\hbox{
\psfig{figure=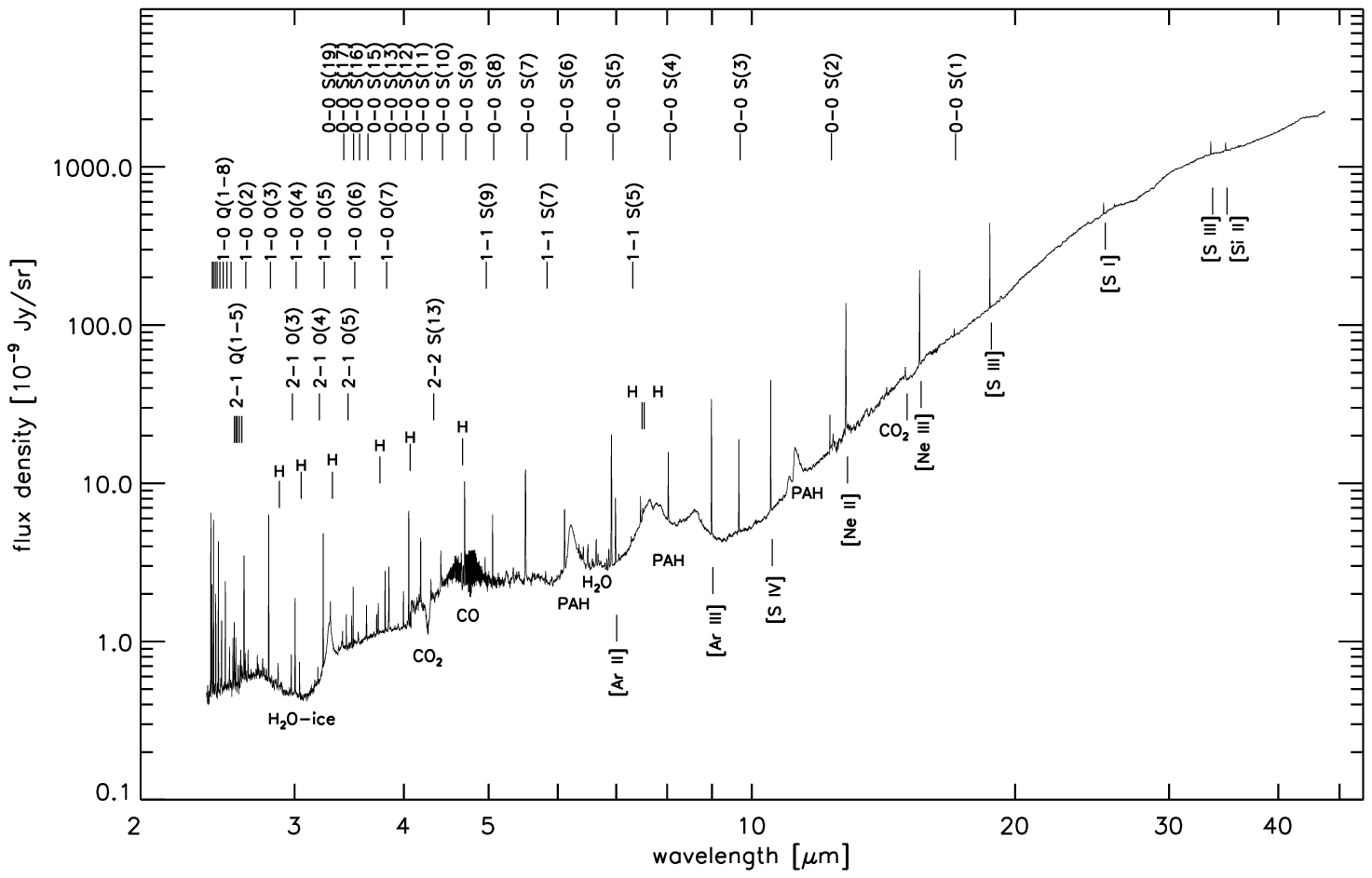,width=14cm,angle=0}
}}
\noindent
{\bf Figure 3.} SWS grating scan of the Orion `Peak 1' shock, showing a rich
forest of H$_2$ lines and other features (Rosenthal et al.\ 2000).
\endinsert
}}

{\bf Atomic lines:} Fine-structure transitions within the lowest
electronic term of most astrophysically relevant atoms and ions occur
at mid-infrared wavelengths. Many of them are found in H II regions
where they are used to probe the hardness of the radiation field, but
some lines are observed in the neutral clouds discussed here, e.g., [S
I] 25.2, [Si II] 34.8, [O I] 63.2 and 145.5 and [C II] 157.7
$\mu$m\footnote{$^3$}{\eightpoint The following nomenclature is used:
X refers to the atom or molecule as a chemical species, [X] indicates
the abundance of the element in all forms with respect to total
hydrogen, and [X I] 100 $\mu$m denotes its forbidden mid- or
far-infrared line.}. These are key diagnostics of photon- versus
shock-heating. H I recombination lines are often seen in mid-infrared
spectra of ionized gas, but not in those of neutral clouds.

{\bf H$_2$ and HD pure rotational lines:} The lowest transitions of
the most abundant molecule in the universe, H$_2$, occur at
mid-infrared wavelengths, and probe directly the bulk of the warm gas
and its temperature.  Most of the deuterium in dense clouds is in HD,
whose fundamental transition lies at 112 $\mu$m.

{\bf Gas-phase bands:} Fundamental vibrational transitions of
important molecules such as H$_2$O, CH$_4$, C$_2$H$_2$, HCN and CO$_2$
occur at mid-infrared wavelengths.  This is the only way to observe
symmetric molecules like CH$_4$ and C$_2$H$_2$ which have no dipole
moment and thus cannot be observed through rotational transitions at
millimeter wavelengths.  Also, CO$_2$ and H$_2$O are so abundant in
the Earth's atmosphere that their interstellar lines can only be
detected from space. The pure rotational transitions of H$_2$O, OH and
CO lie at far-infrared wavelengths. Together, the vibrational and
rotational lines are good probes of the physical conditions of the
gas, its cooling and its chemistry.

{\bf PAHs:} The C-C and C-H stretching and bending modes of polycyclic
aromatic hydrocarbons (PAHs) at 6.2, 7.7, 8.6, 11.3, ...  $\mu$m
dominate the mid-infrared spectra of many objects and are indicators
of the presence of complex carbonaceous material excited by UV
radiation.

{\bf Solid-state vibrational bands:} The characteristic vibrational
bands of ices, silicates, oxides, carbides, carbonates and sulfides
occur uniquely at infrared wavelengths. They can be used as
diagnostics of heating as well as grain growth. Solid-state species
can be distinguished from gas-phase molecules because their bands lack
the characteristic ro-vibrational structure and are broadened (see
Figure 4).

{\parindent 20pt
{\narrower
\topinsert
\null
%\vskip -2.5 true cm
\centerline{\hbox{
\psfig{figure=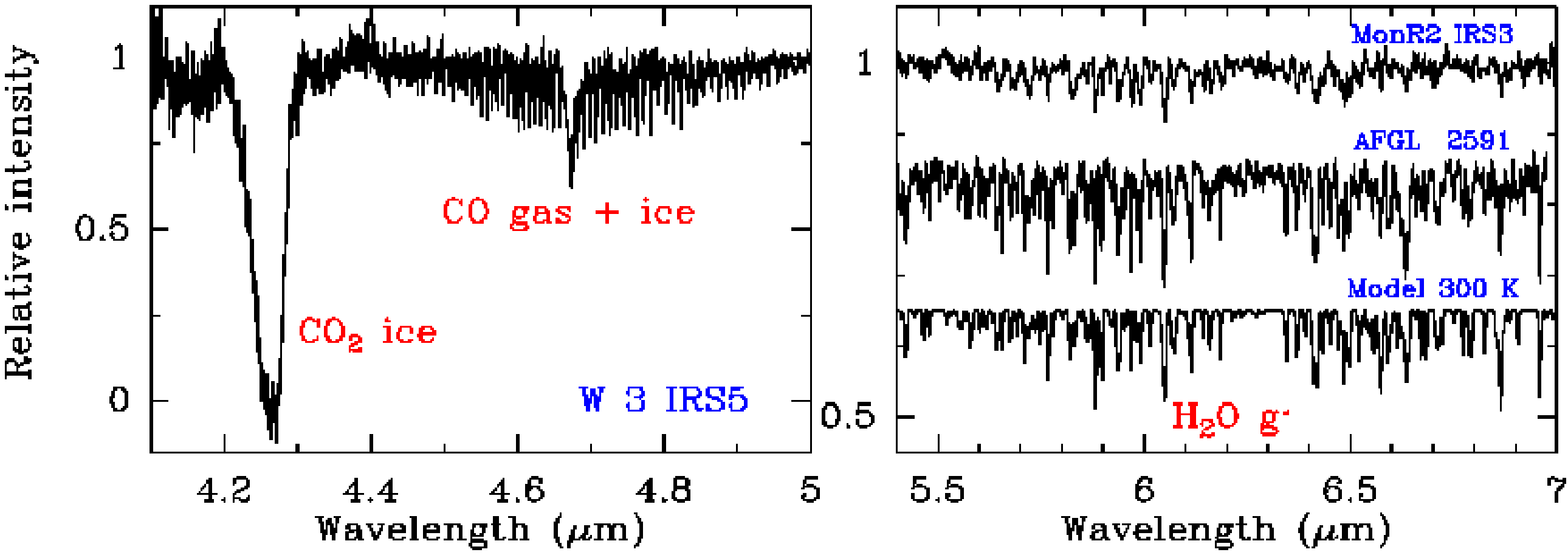,width=14cm,angle=-90}
}}
%\vskip -3.0 true cm
\noindent
{\bf Figure 4.}  Normalized SWS spectra toward massive protostars
showing absorption by various species. Left: CO$_2$ $\nu_3$ and CO
$\nu_1$ bands; Right: H$_2$O $\nu_2$ band. The spectra in the
right panel have been shifted vertically for clarity. Gas-phase molecules show
sharp unresolved ro-vibrational lines in a characteristic pattern,
whereas ices have a single broad absorption (Based on van Dishoeck
1998, Helmich et al. 1996, and Boonman \& van Dishoeck 2003).
\endinsert
}} 

Not all features listed above have been observed with all ISO
instruments.  In general, the detection of intrinsically narrow
gas-phase lines in emission or absorption against a strong continuum
such as Orion-KL required the highest spectral resolution provided by
the Fabry-Perots of the SWS and LWS.  For most other sources,
sensitivity considerations dictated the use of the grating.  The SWS
spectral resolving power of $1000-2500$ was particularly well matched
to that needed to resolve the solid-state features and PAH bands, and
analyse their substructure.  At the low spectral resolving power
provided by CAM-CVF and PHOT-S, the PAHs and broad solid state bands
are readily detected in much weaker sources than accessible with the
SWS, but the analysis is limited because the profiles are unresolved.
Strong gas-phase lines are only detected with CAM-CVF if the continuum
emission is nearly absent, such as in shocks or PDRs.

The main limitations of the ISO data are their relatively low spatial
and spectral resolution compared with modern ground-based
telescopes. ISO's telescope was only 60 cm, with spectrometer
apertures ranging from 20$''$ to $>1'$. This means that the different
physical components associated with star formation (envelopes,
outflows, disks) are often unresolved and blurred together. For
extended sources, aperture changes cause jumps in the fluxes with
wavelength. The spectral resolution was generally insufficient to
obtain kinematic information.

In Sections 3--7, each of the species from Table~2 is discussed in
detail, focussing on their spectroscopy.  Subsequently, the features
are put in the context of the different physical regions associated
with star formation ---PDRs, shocks, protostellar envelopes and
circumstellar disks--- illustrating their diagnostic capability and
ISO's contribution to our understanding of physical and chemical
processes.

\bigskip
\medskip
\centerline{\bf 3. GAS-PHASE MOLECULES AND THEIR CHEMISTRY}
\medskip
Although hampered by limited spectral resolution, the SWS and LWS have
observed many gas-phase species.  ISO's main contributions have been
of three types: (i) Detections of new interstellar molecules not seen
prior to ISO (e.g., CO$_2$, HF, CH$_3$; see Figure 5); (ii) Infrared
detections of molecules seen previously at other wavelengths (e.g.,
H$_2$O, SO$_2$, HD); (iii) Infrared bands of molecules observed
previously from the ground or KAO (e.g., H$_2$, CH$_4$, C$_3$,
C$_2$H$_2$, OH).

{\parindent 20pt
{\narrower
\topinsert
\null
\vskip -0.6 true cm
\centerline{\hbox{
\psfig{figure=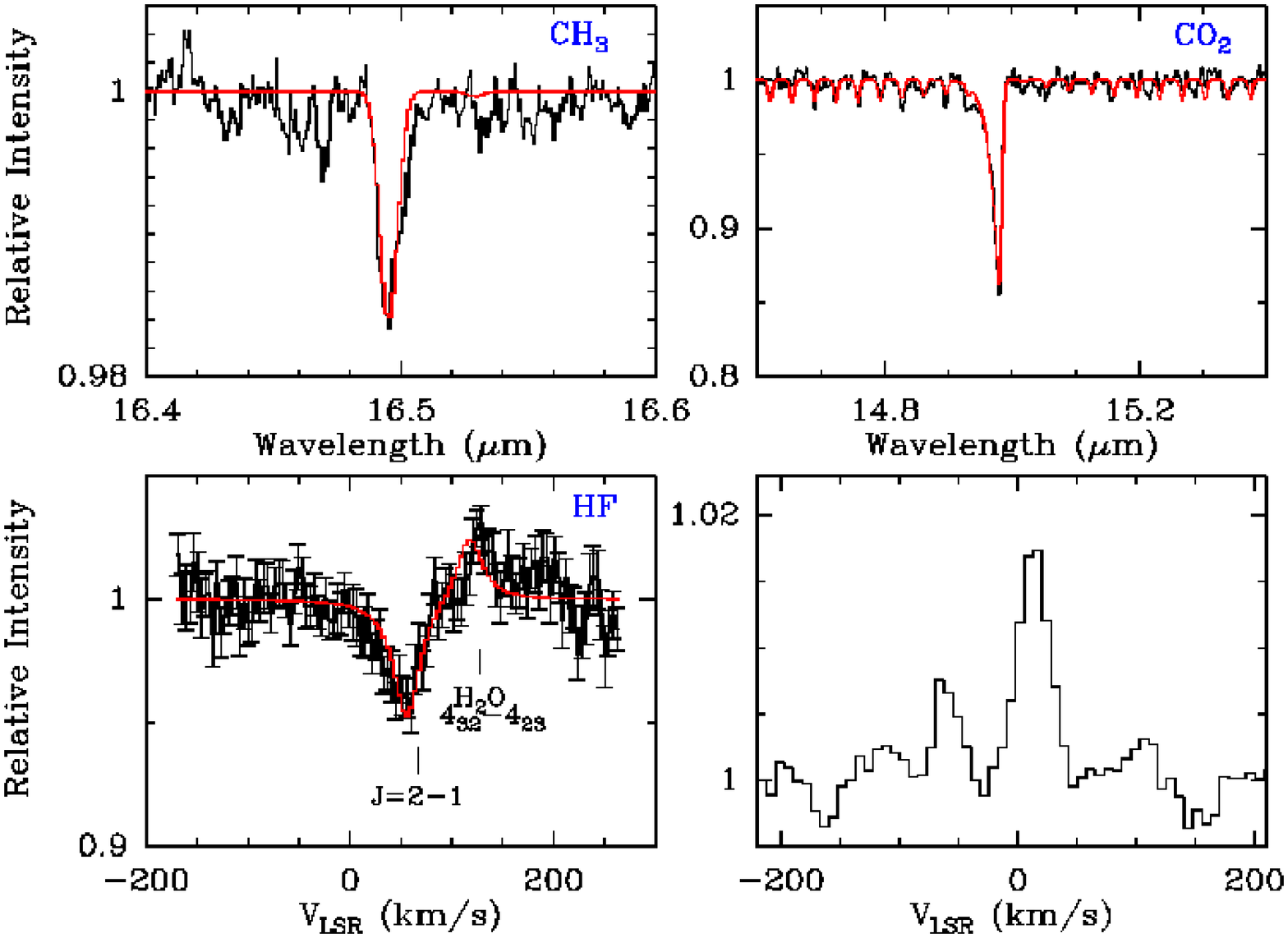,width=14cm,angle=-90}
}} 
\noindent
{\bf Figure 5.} New gas-phase molecules detected by ISO. The light
line in each panel indicates the model spectrum.  Top Left: CH$_3$
$\nu_2$ $Q$-branch toward Sgr A* (Feuchtgruber et al. 2000); Top
Right: CO$_2$ $\nu_2$ band toward AGFL 2136 (van Dishoeck et al. 1996,
Boonman et al. 2003b); Bottom Left. HF $J=2-1$ toward Sgr B2 (Neufeld et al.\
1997); Bottom Right: HD $J=1-0$ line toward the Orion Bar (Wright et
al.\ 1999).
\endinsert
}}

\bigskip
\centerline{\it 3.1. New Detections}
\medskip
Searches for gaseous CO$_2$ formed one of the main astrochemical goals
of the SWS. This molecule is potentially one of the more abundant
carbon- and oxygen-containing species, but ground-based searches were
limited to the chemically related species HOCO$^+$. Detection of the
CO$_2$ $\nu_2$ band at 14.98 $\mu$m was reported by van Dishoeck et
al.\ (1996) and confirmed by van Dishoeck (1998), Dartois et al.\
(1998b) and Boonman et al.\ (2003b,c) in a variety of star-forming
regions.  The inferred abundances with respect to H$_2$ are typically
$\sim 2\times 10^{-7}$, up to two orders of magnitude lower than those
of solid CO$_2$ in the same regions (see \S 4).  The line profiles
indicate that the abundances can be a factor of 10 higher in warm gas
at $T>300$~K, whereas it is at least an order of magnitude lower in
shocks. Its chemistry is not yet fully understood.

Gas-phase CH$_3$ has been detected with the SWS through its $\nu_2$
bending mode lines at 16.0 and 16.5 $\mu$m along the line of sight
toward Sgr~A* with an abundance of $\sim 10^{-8}$ (Feuchtgruber et
al.\ 2000). This molecule is an important building block in the
formation of small hydrocarbon molecules. Together with observations
of CH, CH$_4$ and C$_2$H$_2$ along the same line of sight, it forms a
good test of basic gas-phase astrochemistry networks in diffuse and
translucent clouds.

HF is interesting because it is the main reservoir of fluorine in
dense molecular clouds and thus a direct measure of the fluorine
depletion, supposedly characteristic of other species. The detection
of the HF $J$=2--1 line at 121.7 $\mu$m in absorption toward Sgr B2 by
Neufeld et al.\ (1997) implies an abundance of $\sim 3\times
10^{-10}$, indicating that only $\sim 2\%$ of the fluorine is in the
gas-phase.

A new band at 57.5 $\mu$m has been reported by Cernicharo et al.\
(2002) toward Sgr B2 and may be due to either C$_4$ or C$_4$H. If
ascribed to C$_4$, this would be the first detection of this molecule
in outer space, at an abundance comparable to that of C$_3$. The
latter molecule has been seen toward Sgr B2 through many LWS lines by
Cernicharo et al.\ (2000a), with an abundance of $\sim 3 \times
10^{-8}$.

Several other new molecules have been detected in the envelopes around
late-type stars and proto-planetary nebulae. Although not formally
part of this review, the exciting SWS discoveries of C$_6$H$_6$
(benzene), C$_6$H$_2$ and C$_4$H$_2$ (Cernicharo et al.\ 2001a) and
C$_2$H$_4$ (Cernicharo et al.\ 2001b) toward CRL 618 should be
mentioned.

\bigskip
\centerline{\it 3.2. Other Molecules} 
\medskip
Most interstellar molecules with dipole moments are detected
through their pure rotational emission lines at millimeter
wavelengths. Complementary infrared absorption data provide
valuable constraints on the physical structure and geometry of the
region. Using ISO, this has been possible for a few species,
most notably HCN (Lahuis \& van Dishoeck 2000; see also ground-based
observations by Evans et al.\ 1991) and SO$_2$ (Keane et al.\
2001a). In both cases, the inferred abundances from the infrared data
are two orders of magnitude higher than those obtained from the
millimeter lines. Since the pencil-beam absorption data are more
sensitive to the warm gas close to the protostar, the natural
conclusion is that the abundances `jump' by a factor of $\sim 100$ in
the inner region of the envelope due to a combination of ice
evaporation and high-temperature gas-phase chemistry (Boonman et al.\
2001, see Figure 6).

{\parindent 20pt
{\narrower
\topinsert
\null
\vskip -1.5 true cm
\centerline{\hbox{
\psfig{figure=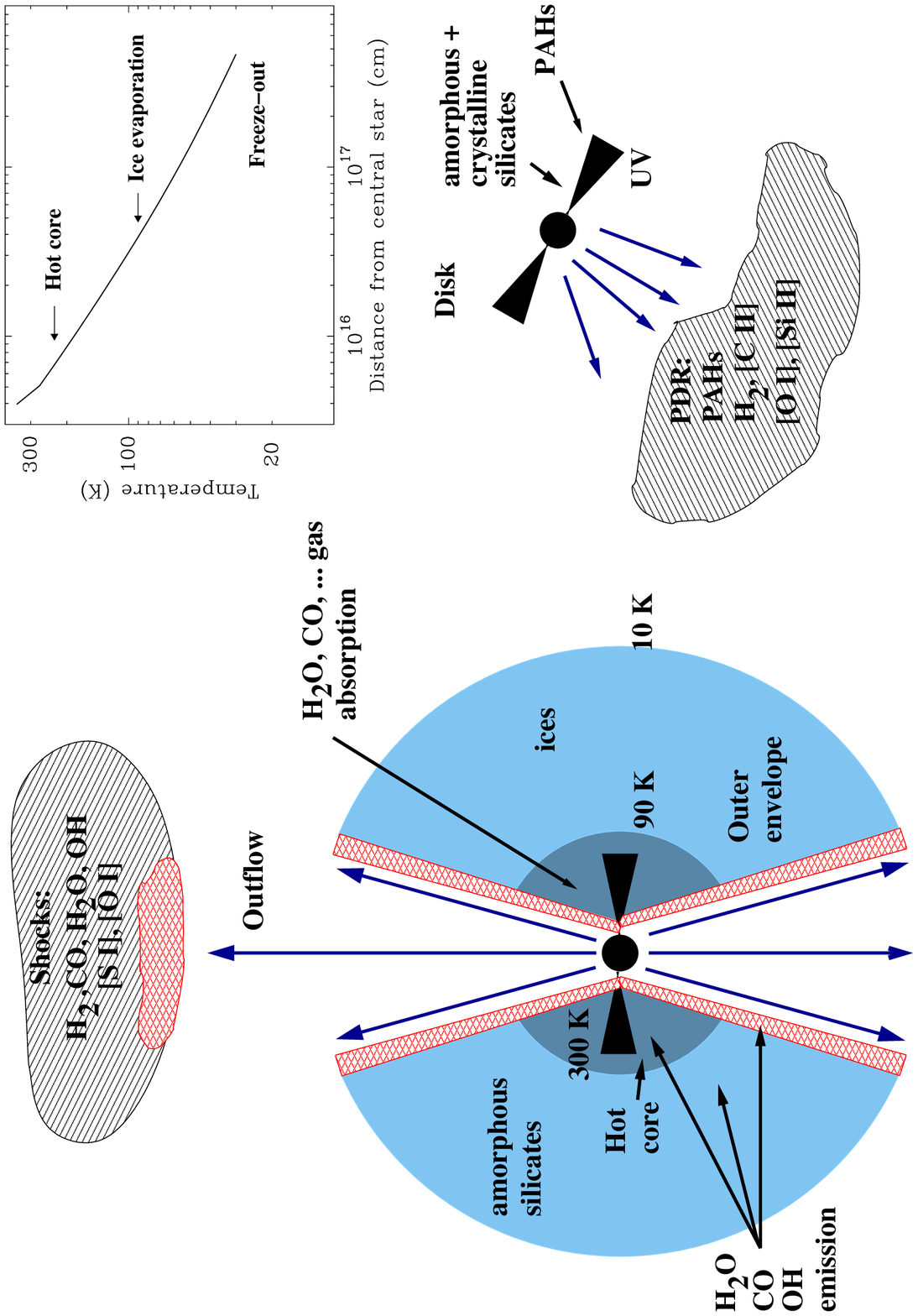,width=14cm,angle=-90}
}}
\vskip -0.7 true cm
\noindent
{\bf Figure 6.} Cartoon of the environment of young stellar objects
and the origin of the various spectral features observed by ISO. The
top diagram indicates the typical temperature structure of a
protostellar envelope, with the critical temperatures for freeze-out,
ice evaporation, and high-temperature chemistry indicated.
\endinsert
}}

Molecules without a permanent dipole moment observed with ISO include
CH$_4$ (e.g., Boogert et al.\ 1998) and C$_2$H$_2$ (e.g., Lahuis \&
van Dishoeck 2000, Boonman et al.\ 2003c). Both species have been
detected from the ground, but ISO allowed surveys in a larger number
of sources. Analysis of the strong C$_2$H$_2$ $Q-$branch profiles
gives rotational temperatures up to 1000~K. These temperatures
correlate well with those of CO and other temperature tracers,
indicating that C$_2$H$_2$ is a good probe of the inner envelope.

Higher rotational lines of NH, NH$_2$ and H$_3$O$^+$ have been seen in
absorption with the LWS toward Sgr B2 (Cernicharo et al.\ 2000a,
Goicoechea \& Cernicharo 2001a,b).  Together with the detection of a
wealth of far-infrared lines of hot NH$_3$ (Ceccarelli et al.\ 2002a),
this allows tests of the basic nitrogen gas-phase chemistry. The
interpretation is complicated, however, by the presence of many
absorption components along the line of sight, including diffuse
clouds, a warm dense envelope and a newly-revealed layer of hot
shocked gas. Several other molecules including SH, H$_2$O$^+$ and
CH$_2$ have been searched for but not detected (Cernicharo et al.\
2000a). The analysis of the full Sgr B2 Fabry-Perot spectrum is still
in progress and may lead to further discoveries in this spectral
`goldmine´.  Higher-lying pure rotational transitions of CO$^+$ have
been seen with the LWS toward the low-mass protostar IRAS 16293 -2422
by Ceccarelli et al.\ (1998b), where its high abundance indicates
either the presence of a dissociative shock or the importance of
UV or X-rays in the chemistry.

\bigskip
\centerline{\it 3.3. H$_2$ and HD} 
\medskip
Interstellar H$_2$ is commonly observed through UV absorption lines in
diffuse gas, and through near-infrared emission lines in warm clouds,
but only few mid-infrared pure rotational lines had been seen prior to
ISO (e.g., Parmar et al.\ 1991). As discussed in detail in \S 7 and 8,
the SWS has routinely observed many H$_2$ pure rotational lines in a
variety of regions, providing an excellent new probe of their physical
conditions.  The detection of the fundamental $J$=2$\to 0$ line at
28.22 $\mu$m was first reported by Valentijn et al.\ (1996) in the
galaxy NGC 6946, and was subsequently seen in several galactic
sources.  This line probes the bulk of the warm gas at a temperature
that is often lower than that inferred from the higher-lying H$_2$
lines.

Like H$_2$, interstellar HD has been observed through ultraviolet
absorption lines prior to ISO, but only in diffuse clouds where the
bulk of the deuterium is still in atomic form. The LWS has allowed the
first detection of the fundamental HD $J$=1--0 line at 112 $\mu$m in
emission in a dense cloud in Orion where most of the deuterium is in
molecular form (Wright et al.\ 1999, see Figure 5). In addition, the
higher-lying HD $J$=6--5 line at 19.4 $\mu$m has been seen with the
SWS in the Orion shock (Bertoldi et al.\ 1999). Together with
observations of the pure-rotational H$_2$ lines, an accurate HD/H$_2$
abundance ---and thus [D]/[H] abundance--- can be derived.  For both
sources, a [D]/[H] ratio of $(0.5-1.0) \times 10^{-5}$ has been found,
up to a factor of two lower than the [D]/[H] ratio of $(1.5\pm
0.1)\times 10^{-5}$ measured in local diffuse clouds (Moos et al.\
2002). The lower [D]/[H] ratio in an active star-forming region like
Orion may be evidence of deuterium destruction by nuclear burning
since its creation in the Big Bang.  The HD 112 $\mu$m line has also
been reported in emission in Sgr B2 (Polehampton et al.\ 2002) and in
absorption toward W 49 (Caux et al.\ 2002). In these latter cases, H$_2$
is not observed directly and the determination of [D]/[H] is
complicated by the fact that alternative tracers of H$_2$, in
particular CO, may be significantly frozen out onto grains.

\bigskip
\centerline{\it 3.4. H$_2$O, OH and O} 
\medskip
Various H$_2$O lines have been seen prior to ISO from the ground and
KAO at radio and millimeter wavelengths, but in virtually all cases
these observations concern maser transitions. ISO has opened up a
flood of data on thermal H$_2$O lines, either in emission or
absorption, allowing much more accurate determinations of the H$_2$O
abundance and excitation. Pure rotational emission and absorption
lines at 25--200 $\mu$m have been seen with the LWS (e.g., Liseau et
al.\ 1996, Cernicharo et al.\ 1997, Harwit et al.\ 1998) and SWS
(Wright et al.\ 2000), and vibration-rotation lines at 6 $\mu$m with
the SWS (e.g., Helmich et al.\ 1996, Dartois et al.\ 1998b,
Gonz\'alez-Alfonso et al.\ 1998, 2002, Moneti \& Cernicharo 2000, see
Figure 4).  Large variations in the H$_2$O abundances are found,
ranging from $<10^{-8}$ in the coldest clouds to $>10^{-4}$ in warm
gas and shocks, illustrating the extreme sensitivity of this molecule
to the physical conditions.  These variations can be explained by
freeze-out of water at the lowest temperatures, ice evaporation at
warmer temperatures ($\grapprox 90$~K) and gas-phase reactions driving
most of the oxygen into water at high temperatures ($\grapprox 230$~K)
(see Figure~6, \S 9.2).

The chemically related OH molecule has been detected through its pure
rotational far-infrared transitions in various objects (see \S 8),
most prominently in absorption toward Sgr B2 (Goicoechea \& Cernicharo
2002). Along this line of sight, [O I] 63 $\mu$m absorption has been
seen as well, both with the LWS grating (Baluteau et al.\ 1997) and
the LWS Fabry-Perot (Lis et al.\ 2001, Vastel et al.\ 2002).  The
conclusion from both sets of data is that the bulk of the gas-phase
oxygen is in atomic form with O/CO$\approx 3-9$, consistent with
previous KAO results for DR 21 by Poglitsch et al.\ (1996).  Even
higher O/CO ratios have been inferred for W~49N by Vastel et al.\
(2000) and for the cold cloud L1689N by Caux et al.\ (1999). Although
care should be taken in analyzing highly optically thick, potentially
self-absorbed [O I] emission lines, the inferred high atomic oxygen
abundances are consistent with non-detection of O$_2$ by the SWAS
satellite (Goldsmith et al.\ 2000).

\bigskip
\medskip
\centerline{\bf 4. INTERSTELLAR ICES}
\medskip
Interstellar H$_2$O ice was detected at 3 $\mu$m by Gillett \& Forrest
(1973), and the study of ices has subsequently been pursued from the
ground and with the KAO (e.g., Willner et al.\ 1982).  ISO has doubled
the number of detected ice bands, bringing the current total to nearly
40 features. Most of the bands are observed in absorption arising in
the cold outer envelopes of high-mass protostars (see Figures 6 and
7), but ISO has also allowed the first glimpse of ices toward low-mass
objects (Cernicharo et al.\ 2000b, see Figure~2).  The absorptions are
deep: the total integrated optical depth is often larger than that of
the 9.7 and 18 $\mu$m silicate absorptions and can amount to more than
50\% of the total integrated flux. Only a few `windows' in between the
ice bands remain to probe close to the protostar itself.  Ice bands
are also seen with the LWS, in absorption toward protostars at 44
$\mu$m (Dartois et al.\ 1998a) and in emission in disks around young
stars at 44 and 63 $\mu$m (Malfait et al.\ 1998, 1999, Creech-Eakman
et al.\ 2002) and one shock (Molinari et al.\ 1999). In the latter
cases, water ice is in its crystalline form rather than in the
amorphous phase which is thought to dominate most of interstellar
ice. In terms of abundances, the amount of ice (mostly H$_2$O) can be
comparable to that of the most abundant gas-phase molecule containing
heavy atoms (CO), making it the second most abundant species after
H$_2$ in cold clouds. Thus, a good understanding of the processes
involving ices is highly relevant to understanding the physical and
chemical characteristics of star-forming regions. Conversely, because
ices are such a major component and show large variations in
abundances and profiles, they are particularly powerful diagnostics of
changes in environment. Recent reviews on interstellar ices and
summaries of ISO data are given by Schutte (1999), Ehrenfreund \&
Schutte (2000), Boogert \& Ehrenfreund (2004), and Gibb et al.\
(2004).

{\parindent 20pt
{\narrower
\topinsert
\null
\vskip -3.0 true cm
\centerline{\hbox{
\psfig{figure=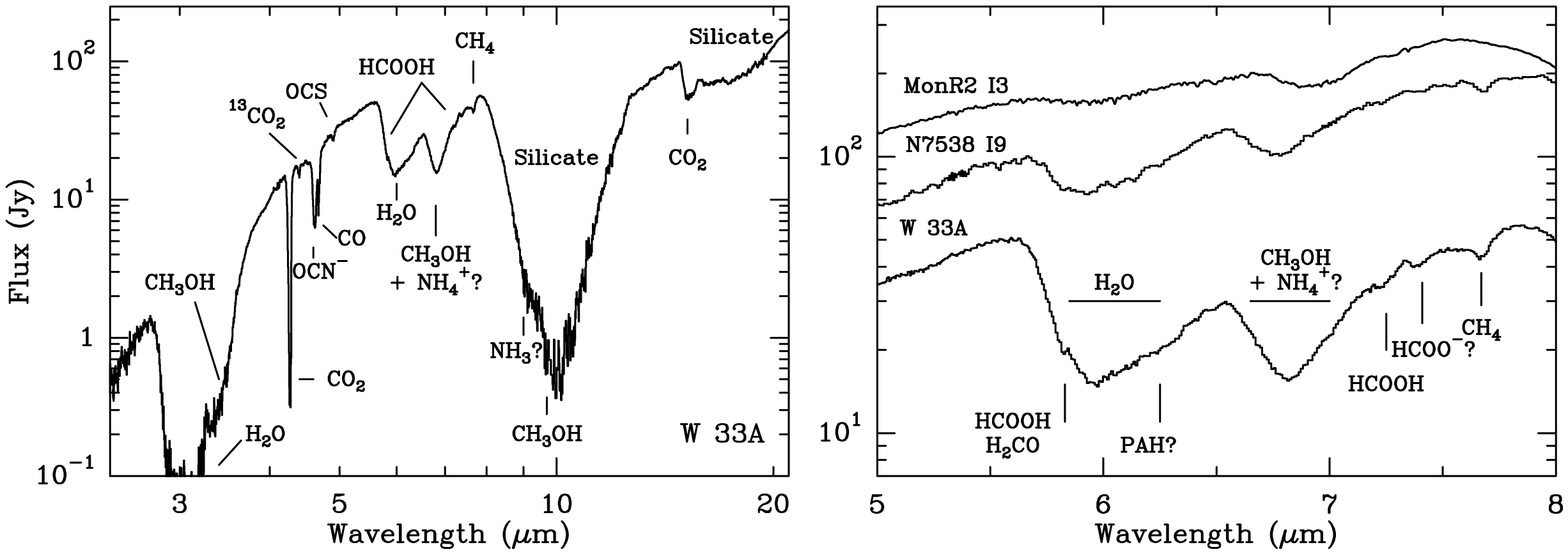,width=14cm,angle=-90}
}} 
\vskip -3.0 true cm
\noindent
{\bf Figure 7.} SWS spectrum of the embedded massive protostar W
33A, illustrating the wealth of ice features observed by ISO (Gibb et
al.\ 2000).  Blow-ups of the 5--8 $\mu$m region are shown for W 33A,
NGC 7538 IRS9 (shifted in flux for clarity) and Mon R2 IRS3. Note the
shift in the 6.85 $\mu$m band toward longer wavelengths for warmer
sources like Mon R2 IRS3 (Keane et al.\ 2001b).
\endinsert
}}

ISO's main contribution to the study of ices has been in two areas:
(i) The first complete inventory of the main ice features; and (ii)
Clear evidence that changes in profiles and abundances can trace the
gradual heating of protostellar envelopes.

\bigskip
\centerline{\it 4.1. Inventory of Ice Features}
\medskip
Table~2 and Figure~7 show a number of features with their
identifications (see Boogert \& Ehrenfreund 2004 for a complete list).
Several of them are securely identified on the basis of multiple
vibrational modes and isotopic bands, such as H$_2$O, CO, CO$_2$
and CH$_3$OH. For example, Dartois et al.\ (1998a) are able to fit five
H$_2$O bands simultaneously.  Cleanly isolated
single bands such as that due to solid CH$_4$ at 7.67 $\mu$m and OCS
at 4.92 $\mu$m can also be confidently assigned if they are well
matched by laboratory spectra. The identification of other species
like H$_2$CO and NH$_3$ is more problematic because all their
strong vibrational modes are blended with those of other species.

High quality SWS profiles of the 6.0 and 6.8 $\mu$m bands --- prominent
toward all ice sources --- show that they are clearly composites of
several features (Schutte et al.\ 1996, Keane et al.\ 2001b). A good
fit to the 6.0 $\mu$m band requires not only H$_2$O (known to be
present from the 3 $\mu$m band), but also H$_2$CO at 5.85 $\mu$m
and/or HCOOH at 5.83 $\mu$m to fit the blue wing and NH$_3$ at 6.15
$\mu$m or some carbonaceous material to fit the red wing. The 6.85
$\mu$m feature is still not firmly identified, although NH$_4^+$ is a
plausible candidate and CH$_3$OH a minor contributor (Demyk et al.\
1998, Keane et al.\ 2001b, Schutte \& Khanna 2003). An empirical
decomposition of the profile shows two components with fixed position
and widths, whose relative intensities change with temperature of the
source (see Figure 7).

The shape, exact position and intrinsic strength of each band depend
on the chemical environment of the molecule. For example, the position
of the CO$_2$ stretching mode shifts by 20 cm$^{-1}$ and its width
broadens by a factor 2 when CO$_2$ is mixed with H$_2$O ice rather
than in its pure form.  Analysis of the band profiles has led to a
picture of layered ices, in which some molecules (e.g., CO) are not
mixed with H$_2$O (Tielens et al.\ 1991).  Ice profiles of crystalline
material can be distinguished from their amorphous counterparts by
being sharper and red-shifted.  Laboratory simulations of
astrophysically relevant ice mixtures have been essential to provide
the basic data to quantitatively analyze the ice bands (e.g., Hudgins
et al.\ 1993, Gerakines et al.\ 1995, Ehrenfreund et al.\ 1996, 1997a).

A number of weak minor features have been found in the 7--8 $\mu$m
region. Schutte et al.\ (1999) and Keane et al.\ (2001b) suggest HCOOH
and HCOO$^-$ or CH$_3$HCO as plausible identifications. The presence
of ions such as NH$_4^+$, OCN$^-$ and HCOO$^-$ has been subject to
considerable debate over the last 20 years, with the major controversy
centering around the charge balance in the ice (not all counter ions
have been found) and the constancy of the position of the observed
ionic features. Schutte \& Khanna (2003) show that heavier counterions
have only weak spectral features and likely remain undetected, whereas
the observed 6.85 $\mu$m shift is well reproduced in the laboratory by
heating NH$_4^+$-containing ices.

Upper limits can be just as interesting as detections.  The region
between 5.0 and 5.8 $\mu$m (Figure 7) is remarkably void of features
and places limits on minor ice constituents. For example, the absence
of the 5.62 $\mu$m band of the simplest amino acid glycine puts a
limit on its abundance of less than 0.3\% with respect to H$_2$O ice
(or $<3\times 10^{-7}$ with respect to H$_2$).  Boudin et al.\ (1998)
have put significant upper limits on C$_2$H$_6$, C$_2$H$_5$OH and
H$_2$O$_2$.  Solid HNCO, the likely precursor of OCN$^-$ and widely
observed in the gas in hot cores, absorbs at 4.42 $\mu$m and is
undetected down to 0.5\%. In these regions, the `confusion' limit has
certainly not yet been reached and higher $S/N$ data may reveal new
weak bands. Similarly, very few bands have been seen superposed on the
strong continuum longward of the 9.7 $\mu$m silicate band, with the
strong CO$_2$ bending mode at 15 $\mu$m being the notable exception.

Searches for absorption by more complex solid organic material are of
obvious interest as a link with the large organic gas-phase molecules
observed toward high-mass protostars.  PAHs have been proposed to
contribute to the red wing of the 6.0 $\mu$m absorption.  Refractory
organic residues, put forward by Greenberg et al.\ (1995) as resulting
from UV processing of ices, have been suggested as an alternative
explanation for the 6.0 $\mu$m excess absorption (Gibb \& Whittet
2002).

Two very important molecules, O$_2$ and N$_2$, are largely
``invisible'' because their vibrational modes are dipole forbidden.
When surrounded by other molecules, however, the transitions become
weakly allowed.  Searches for the N$_2$ band at 4.295 $\mu$m are
hampered by overlap with the strong solid CO$_2$ stretching mode
(Sandford et al.\ 2001). Upper limits on solid O$_2$ feature at 6.45
$\mu$m, combined with an analysis of the effects of O$_2$ on the solid
CO profiles, indicate that solid O$_2$ contains less than 6\% of the
total interstellar oxygen budget (Vandenbussche et al.\ 1999).

The isotopic species $^{13}$CO$_2$ and $^{13}$CO have been detected
and provide an independent determination of the [$^{12}$C]/[$^{13}$C]
abundance gradient as a function of galactocentric radius as well as
constraints on grain shape effects which complicate the analysis of
the main isotopes (Boogert et al.\ 2000a). The potential detection of
solid HDO by Teixeira et al.\ (1999) has not been confirmed by a
re-analysis of the ISO data nor by independent ground-based data
(Dartois et al.\ 2003).

\bigskip
%\vfill \eject
\centerline{\it 4.2. Ice Abundances and Chemistry}
\medskip
H$_2$O ice is clearly dominant, with abundances of $10^{-5}-10^{-4}$
with respect to H$_2$, thus containing a significant fraction of the
oxygen (van Dishoeck 1998).  The combined contribution of CO and
CO$_2$ is typically 10--30\%, but may be as high as 60\%.  For some
sources, especially the most massive protostars, other carbon-bearing
species (CH$_3$OH, HCOOH, CH$_4$ and H$_2$CO) may add an additional
20--30\%, bringing the total abundance up to $\sim$50\% of that of
H$_2$O (Gibb et al.\ 2004). Nitrogen appears much less abundant in the
ice.  Although the amount of NH$_3$ ice is under debate for some
sources, it is certainly less than 10\% and often less than 5\% of
that of H$_2$O ice (e.g., Taban et al.\ 2003).  The NH$_4^+$ abundance
may be of order 10\% (Schutte \& Khanna 2003), but this still makes
the identified nitrogen content at least a factor of 4 less than that
of carbon.

A striking aspect of the ice composition is its simplicity. Although there
is some observational bias against detection of low-abundance complex
molecules in ices, all of the identified species are those expected
from hydrogenation and oxidation of the main species to be accreted
from the gas, i.e., O, C, N and CO (Tielens \& Hagen 1982).  Various
groups (e.g., Charnley \& Tielens 1987, Stantcheva et al.\ 2002) have
attempted to model this grain surface chemistry quantitatively with
varying degrees of success.

The SWS has provided detailed inventories of ices in about 30 sources.
The majority of them are high-mass protostars, with only a limited
number of intermediate mass YSOs (LkH$\alpha$ 225: van den Ancker et
al.\ 2000c; Elias 18, HH 100, R CrA: Nummelin et al.2001; AFGL 490:
Schreyer et al.\ 2002) and low-mass sources (Elias 29: Boogert et al.\
2000b; L1551 IRS5: White et al.\ 2000; T Tauri: van den Ancker et al.\
1999; see also van den Ancker 2000 for unpublished SWS results). Using
the higher sensitivity provided by CAM-CVF, Alexander et al.\ (2003)
surveyed 42 low luminosity sources ($L<$100 L$_{\odot}$), whereas
PHOT-S observed a handful of cases (e.g., Feldt et al.\ 1998,
G\"urtler et al.\ 1996, 1999, 2002).  Although the lower spectral
resolution does not allow a decomposition of the intrinsically blended
6.0 and 6.8 $\mu$m profiles, their overall spectra look remarkably
similar to those of the higher luminosity counterparts. The 6.8 $\mu$m
feature clearly correlates with H$_2$O, confirming that it is largely
due to an ice.  Significant variations are found between different
clouds. For example, YSOs in the Chamaeleon I cloud show very little
ice absorption overall, whereas those in $\rho$ Oph appear to have a
lower H$_2$O ice abundance and a larger scatter in the CO$_2$/H$_2$O
abundance ratio compared with, for example, Serpens and CrA. $\rho$
Oph may be peculiar because it has a large foreground layer of
molecular gas with $A_V \grapprox 10$~mag which may be relatively poor
in ices due to its lower density and larger exposure to UV radiation
(Boogert et al.\ 2002).

\bigskip
\centerline{\it 4.3. Heating and Processing of Ices}
\medskip
The 15 $\mu$m bending mode of solid CO$_2$ toward high-mass protostars
shows considerable variation from source to source, with a pronounced
double-peak structure and a red wing seen in some sources (de Graauw
et al.\ 1996b, Gerakines et al.\ 1999). Extensive laboratory
simulations show that these variations can be well reproduced by
gradual heating of a mixture of H$_2$O, CO$_2$ and CH$_3$OH ices in
roughly equal proportions (Ehrenfreund et al.\ 1998, Dartois et al.\
1999). At higher temperatures, the laboratory ices appear to
segregate, leading to near-pure H$_2$O and CO$_2$ structures. Although
this explanation seemed far-fetched at first, it is corroborated by a
wealth of other evidence.  In particular, the profiles of solid
$^{13}$CO$_2$ at 4.38 $\mu$m (Boogert et al. 2000a), H$_2$O at 3
$\mu$m (Smith et al.\ 1989), and the 6.8 $\mu$m feature (Keane et al.\
2001b) show similar changes due to heating.  Moreover, these changes
are well correlated with increased gas/solid ratios derived from ISO
(van Dishoeck \& Helmich 1996, Boonman \& van Dishoeck 2003, see \S
9.2) and submillimeter data (van der Tak et al.\ 2000a) as well as
increased dust color temperatures derived from the 45/100 $\mu$m ISO
flux ratios (Boogert et al.\ 2000a). Indeed, the predictive power of
these heating indicators is large: if one of them is found, so are the
others. The emission data refer to the entire source rather than
absorption along the line of sight, indicating that the entire
envelope undergoes `global warming' rather than locally.  While
geometry (e.g., where does the mid-infrared continuum become optically
thick?) and line-of-sight effects (e.g., absorption looking down the
outflow cone) may play a role in individual cases, they cannot be
responsible for these general trends.  As discussed further in \S 9.2,
the most likely scenario is that the higher temperatures are due to a
decreased envelope mass relative to the source's luminosity, perhaps
caused by the dispersion of the envelope with time.

The presence of some features, e.g., that due to OCN$^-$ at 4.62
$\mu$m, has often been cited as an indicator of `energetic'
processing, i.e., changes induced in the ice matrix composition or
profiles due to UV or particle irradiation rather than thermal heating
(e.g., Pendleton et al. 1999). There have been extensive laboratory
studies of these processes (see Gerakines et al. 1996 for UV; Moore \&
Hudson 1998 for energetic particle bombardment), but no `smoking gun'
has yet been found observationally.  In particular, van Broekhuizen et
al.\ (2004) show that the OCN$^-$ data can be quantitatively
reproduced by thermal acid-base reactions of HNCO with NH$_3$ starting
at temperatures as low as 10~K.

\bigskip
\medskip
\centerline{\bf 5. POLYCYCLIC AROMATIC HYDROCARBONS}
\bigskip
Broad, strong infrared emission features at 3.3, 6.2, 7.7, 8.6, 11.3
and 12.7 $\mu$m from a variety of astronomical sources have been known
since the 1970's, and are commonly known as the Unidentified InfraRed
(UIR) bands. These features coincide with the vibrational modes of
aromatic materials (Duley \& Williams 1981), with the preferred
identification that of free-flying Polycyclic Aromatic Hydrocarbons
(PAHs) (L\'eger \& Puget 1984, Allamandola et al.\ 1985).  The fact
that the bands are detected far from any illuminating source with
roughly the same intensity ratios as close to bright stars proves that
the carriers are excited non-thermally by single UV photons
(Sellgren 1984). Indeed, both CAM-CVF, PHOT-S and the IRTS have shown
that the mid-infrared spectra (and thus also the IRAS 12 $\mu$m
emission) of even the most diffuse interstellar medium are dominated
by PAH features (Mattila et al.\ 1996, Onaka et al.\ 1996, Boulanger
et al.\ 2000, Kahanp\"a\"a et al.\ 2003). Any large ($>$100
\AA) grain does not become hot enough under these conditions to emit
at mid-infrared wavelengths, so the carriers must be small, of order
10 \AA, corresponding to PAHs with 30 to a few hundred carbon atoms
containing up to 10--15\% of the available interstellar carbon. Larger
PAH clusters are seen through the plateaus underlying the discrete
features.  PAHs have also been seen in absorption at 6.2 $\mu$m in the
diffuse medium toward objects in the Galactic Center and several dusty
late-type Wolf-Rayet stars (Schutte et al.\ 1998).

{\parindent 20pt
{\narrower
\topinsert
\centerline{\hbox{
\psfig{figure=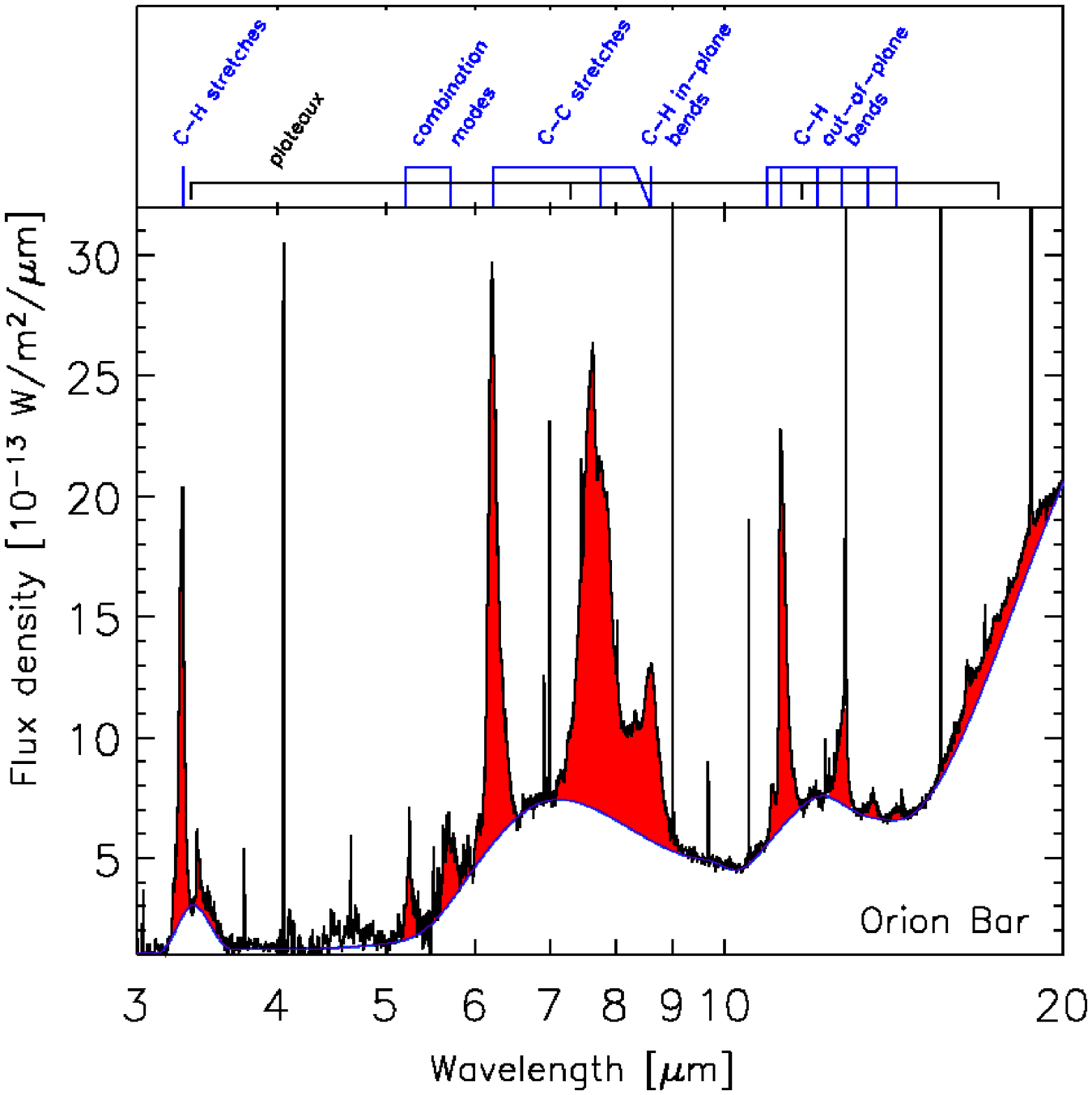,width=10cm,angle=0}
}}
\noindent
{\bf Figure 8 (color).} SWS spectrum of the Orion Bar PDR,
illustrating the richness of PAH features. Various atomic and
molecular lines are seen as well, originating in the PDR and in the
Orion H II region along the line of sight. The spectrum is taken at the
position of the peak of the H$_2$ $v=1-0$ S(1) emission (Peeters et al. 2004b).
\endinsert
}}

The main contributions of ISO in this area have been to (i) Allow a
systematic analysis of the spectral characteristics of PAHs, revealing
many new weak features, subbands and shoulders; (ii) Survey the
similarities and differences between the features for a wide variety
of sources, relating them to physical and chemical changes; and (iii)
Extend PAH studies to more diffuse regions exposed to much lower
intensity radiation fields (see above). PAHs have been detected in
many interstellar ISO spectra, indicating that they are ubiquitous,
abundant and able to survive in a wide range of conditions.
Deeply-embedded YSOs and shocks are the only types of regions where
PAH emission is not seen.  Detailed reviews of ISO PAH results have
been given by Tielens et al.\ (2000) and Peeters et al.\ (2004b).

\bigskip
\centerline{\it 5.1. PAH Spectroscopy} 
\medskip
Figure 8 shows the rich PAH spectrum of the Orion Bar PDR.  In
addition to the main features, weaker bands or substructures can be
found at 5.2, 5.7, 6.0, 7.2--7.4, 7.6, 7.8, 8.2, 10.8, 11.0, 11.2,
12.0, 12.7, 13.2, 14.5 and/or 16.4 $\mu$m, with a weak plateau between
15 and 20 $\mu$m (see also Moutou et al.\ 2000, van Kerckhoven et al.\
2000). Assignments in terms of C-H and C-C stretches are indicated.
The C-H out-of-plane bending modes at 11--14 $\mu$m are particularly
useful to characterize the PAH edge structure since their position
depends on the number of adjacent H atoms bonded to neighboring C
atoms on the ring. Solo-CH groups at 11.3 $\mu$m are indicative of
straight edges on compact `condensed' PAHs like circumcoronene,
whereas trio-CH groups at 12.7 $\mu$m arise at corners and thus imply
more irregular PAHs.

Heroic efforts in laboratory spectroscopy (e.g., Hudgins \&
Allamandola 1999, see review by Tielens \& Peeters 2004) have led to a
large database of PAH spectra of various sizes and charges.  It is
clear that no single species fits all the features, but Allamandola et
al.\ (1999) have succeeded in reproducing ISO spectra with a
reasonable mixture of a dozen neutral and positively ionized PAHs,
where each UIR feature is due to a superposition of many vibrational
bands. The fact that the relative band strengths do not appear to vary
much between PDRs exposed to orders of magnitude different radiation
fields has been used to argue that the UIR features cannot be due to
such an extended family of PAHs whose composition must change with
physical conditions (e.g., Boulanger et al.\ 2000). However, as long
as the ratio of UV intensity over electron density
$G_0$/$n_e$ (see \S 7) is similar and $\leq 10^3$ cm$^3$, the spectra
of a PAH family do not change much (Bakes et al.\ 2001, Li \& Draine
2002).  

The high-quality spectra allow strict limits to be placed on
functional groups attached to the PAHs, e.g., -CH$_3$, -OH, -NH$_2$
and -C=O groups are at most 1\% relative to aromatic C. The absence of
strong 3.4, 6.85 and 7.3 $\mu$m bands indicates that they do not
contain large aliphatic moieties.  The fact that the C-C stretch
occurs at wavelengths as short as 6.2 $\mu$m may be indicative of
nitrogen substitution in the ring: pure PAHs start absorbing at 6.3
$\mu$m.  Deuterated PAHs, in which an aromatic H is replaced by a D,
have possibly been detected through bands at 4.4 and 4.65 $\mu$m at
high abundances, PAD/PAH$>0.1$, in two sources (Peeters et al.\
2004a).  Other forms of large carbon molecules, e.g. C$_{60}$ or
C$_{60}^+$, are not seen down to 0.3\% of total carbon (Moutou et al.\
1999a).

{\parindent 20pt
{\narrower
\topinsert
\null
\vskip -3.0 true cm
\centerline{\hbox{
\psfig{figure=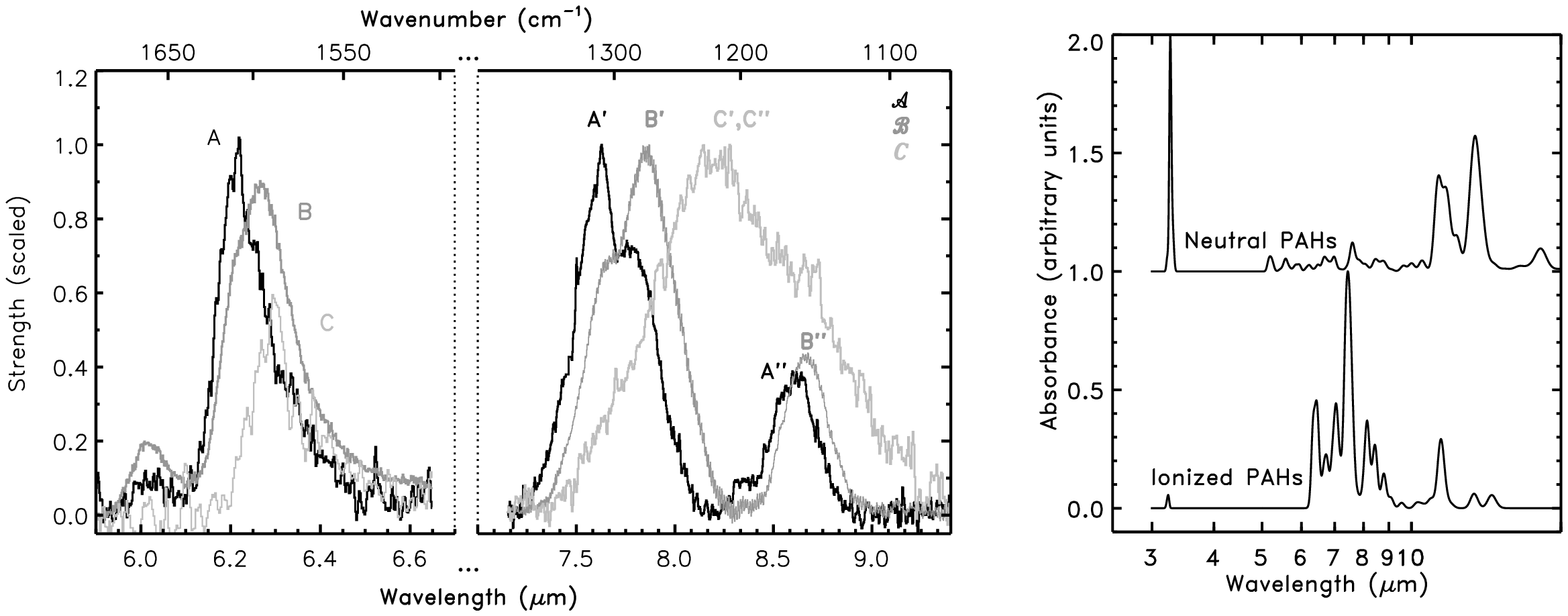,width=14cm,angle=-90}
}}
\vskip -3.0 true cm
\noindent
{\bf Figure 9.} Left: Source to source variations observed for the
6--9 $\mu$m PAH features with the SWS. Profile A is representative of
all PDRs and reflection nebulae, including the diffuse interstellar
medium. Profile B refers to the isolated Herbig Ae stars with disks,
as well as most planetary nebulae. Profile C is observed for some
post-AGB objects (Peeters et al. 2002a). Right: Laboratory spectra of
a mixture of neutral PAHs compared to that of positively ionized
PAHs. Note the stronger 3.3 $\mu$m feature for neutral PAHs and the
stronger 6--9 $\mu$m complex for ionized PAHs (figure adapted from
Allamandola et al.\ 1999).
\endinsert
}}

\bigskip
\centerline{\it 5.2. PAH Feature Variations} 
\medskip
Whereas the CH 3.3 and 11.2 $\mu$m bands correlate well with each
other over a wide range of conditions, significant variations in the
relative strengths of the C-H and C-C features are found, both between
different types of sources and at different positions within a source
(e.g., Hony et al.\ 2001).  An excellent example of the latter case is
provided by the M17 PDR, where the ratio of the 11.2/6.2 $\mu$m ratio
changes by more than a factor of 2 across the ionization front
(Verstraete et al.\ 1996).  Since the C-H modes have much lower
intrinsic band strengths for ionized PAHs compared with neutral PAHs,
whereas the opposite holds for the C-C-modes (see Figure 9), this
variation reflects a change in the degree of ionization: at the H II
region interface, the PAHs are more ionized than deeper in the
molecular cloud where they are also re-hydrogenated. The actual charge
state depends on the ratio $G_0/n_e$ and may include doubly- or even
triply-ionized PAHs. The smaller PAHs may be destroyed by intense
UV radiation inside the H II region (Verstraete et al.\
2001), although observational evidence is still controversial (Abergel
et al.\ 2003).

The positions of individual C-H modes are relatively stable with peak
wavelengths, changing by at most 4--6.5 cm$^{-1}$. In contrast, the
C-C modes at 6--9 $\mu$m vary by 25--50 cm$^{-1}$. Moreover, the
variations in the different C-C profiles are linked to each other (see
Figure 9, Peeters et al.\ 2002a). Specifically, reflection nebulae and
PDRs show a different 7.7 $\mu$m complex than isolated young stars and
post-AGB stars. Some H II regions contain a 8.6 $\mu$m band which
dwarfs the commonly much brighter 7.7 $\mu$m band. While these changes
clearly indicate variations in the composition of the PAH family in
response to different physical conditions, the details remain to be
understood, especially since few PAHs reveal strong spectral features
longward of 7.7 $\mu$m in the laboratory.

\bigskip
\centerline{\it 5.3. Hydrogenated Amorphous Carbon} 
\medskip
The C-H stretching mode of aliphatic (chain-like) hydrocarbons at 3.4
$\mu$m has been detected from the ground along many diffuse lines of
sight with extinctions $\grapprox 10$ mag (e.g., Pendleton \&
Allamandola 2002).  ISO has provided further contraints on the
composition of this material by detecting new features at 6.85 and
7.25 $\mu$m toward the Galactic Center (Chiar et al.\ 2000). A good
fit is obtained with Hydrogenated Amorphous Carbon (HAC) with H/C=0.5,
with little evidence for strong C=O modes which would be the relic of
first-generation photoprocessed ices.  The total amount of carbon
locked up in this material is uncertain, but may be as much as 25\%.
The relation with the aromatic material in the diffuse medium and the
transformation between the two forms of carbon remains to be clarified
(e.g., Mennella et al.\ 2001).

\bigskip
\medskip
%\vfill \eject
\centerline{\bf 6. SILICATES}
\bigskip
The detection of strong features at 9.7 and 18 $\mu$m bands due to the
Si-O stretch and O-Si-O bending modes (Gillett \& Forrest 1973),
coupled with the almost complete absence of silicon, magnesium and
iron from the interstellar gas, convincingly demonstrates that a large
fraction of the interstellar refractory material is in the form of
silicates. ISO has allowed to (i) Obtain high quality silicate
profiles of lines of sight through the diffuse interstellar medium;
and (ii) Reveal a surprisingly rich mineralogy of crystalline
silicates in material around young and old stars.
\bigskip
\centerline{\it 6.1. Amorphous Silicates}
\medskip
The 9.7 and 18 $\mu$m silicate bands are the deepest broad absorption
features seen in ISO spectra toward embedded YSOs and background stars
and are found to be very smooth.  Demyk et al.\ (1999) fit the 9.7 and
18 $\mu$m bands for dense cloud material toward protostars with
amorphous, porous pyroxenes (Mg$_x$Fe$_{1-x}$SiO$_3$) with some
admixture of aluminum in the silicates. Kemper et al.\ (2004) obtain a
good fit for the diffuse cloud 9.7 $\mu$m band toward Sgr A$^*$ with
$\sim 0.1$ $\mu$m-sized amorphous silicates consisting mostly of
amorphous olivine ($\sim$85\%; MgFeSiO$_4$) with a small admixture of
amorphous pyroxene ($\sim$15\%).  The limits on the degree of
crystallinity are very strict, $<$1--2\% in the case of dense clouds,
and $<$0.4\% for the more diffuse medium. Alternative fits by Bowey \&
Adamson (2002) with a complex mixture of crystalline material,
constructed such that the many individual peaks blend into a smooth
broad profile, seem less likely.

This low degree of crystallinity is in stark contrast with the
significant amounts of crystalline material of typically 10--20\%
(with extremes up to 75\%) found in the envelopes around evolved stars
(e.g., Molster et al.\ 2002).  Since these stars
are thought to provide a significant fraction of the interstellar
dust, the observed lack of crystallinity has important
implications. One explanation is that amorphization due to energetic
particles occurs in the interstellar medium on a timescale of $<$10
Myr, significantly shorter than the dust destruction time
scale. Another possibility is that the production rates of amorphous
silicates from other sources (e.g., supernovae) are much larger than
thought before, diluting the crystalline silicate fraction originating
from (post-)AGB stars.

{\parindent 20pt
{\narrower
\topinsert
\null
\vskip -3.0 true cm
\centerline{\hbox{
\psfig{figure=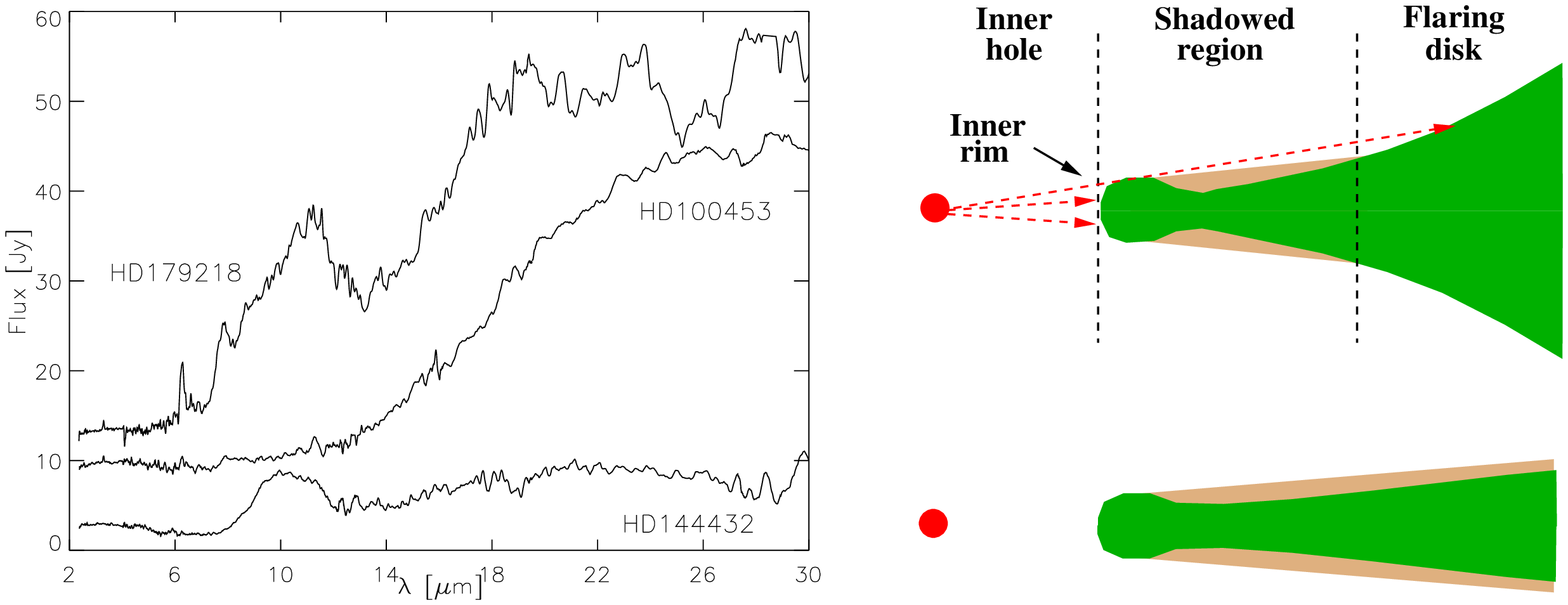,width=14cm,angle=-90}
}} 
\vskip -3.0 true cm
\noindent
{\bf Figure 10.} Left: SWS spectra of disks around isolated Herbig
Ae stars, illustrating their different spectral characteristics.  HD
179218 is an example of a source with crystalline silicates, whereas
HD 100453 does not show any 10 $\mu$m emission.  Both sources have
rising SEDs (Group I). HD 144432 is an example of a source with a flat
SED (Group II). The HD 144432 and HD 100453 spectra have been shifted
by $-5$ and $-10$ Jy for clarity. Right: Sketch of the different
geometries for Group I and II sources. The silicate emission is
thought to arise from the warm upper layers in the inner disk, whereas
the PAH emission can also come from the flaring outer part (Meeus et
al. 2001, Dullemond \& Dominik 2004).
\endinsert
}}

\bigskip
%\vfill \eject
\centerline{\it 6.2. Crystalline Silicates}
\medskip
In dense clouds close to hot stars, the dust particles become
sufficiently warm at a few hundred K to show silicate bands in
emission. ISO examples are provided by Jones et al.\ (1999) for M 17,
Cesarsky et al.\ (2000a) in Orion, Lefloch et al.\ (2002) in the
Trifid nebula and Onaka \& Okada (2003) for Carina and S 171.
Although the bulk of the emission is thought to be due to small
amorphous grains, some evidence for discrete features has been
claimed. For example, Onaka \& Okada find a feature at 65 $\mu$m which
may be due to diopside, a calcium-bearing crystalline
silicate. Ceccarelli et al.\ (2002c) show that the spectral energy
distribution of one protostar, NGC 1333 IRAS4, has excess emission at
95 $\mu$m, which may be ascribed to calcite ---a calcium-containing
carbonate.  Calcite has also been found in two planetary nebulae by
Kemper et al.\ (2002) and is common in meteorites, with abundances of
$\sim 0.3-1$\% of the (warm) dust mass in all types of objects.  If
confirmed, the detection of calcite in a cold protostellar envelope
would be additional evidence that its formation does not require
liquid water.

{\parindent 20pt
{\narrower
\topinsert
\null
\vskip -1.0 true cm
\centerline{\hbox{
\psfig{figure=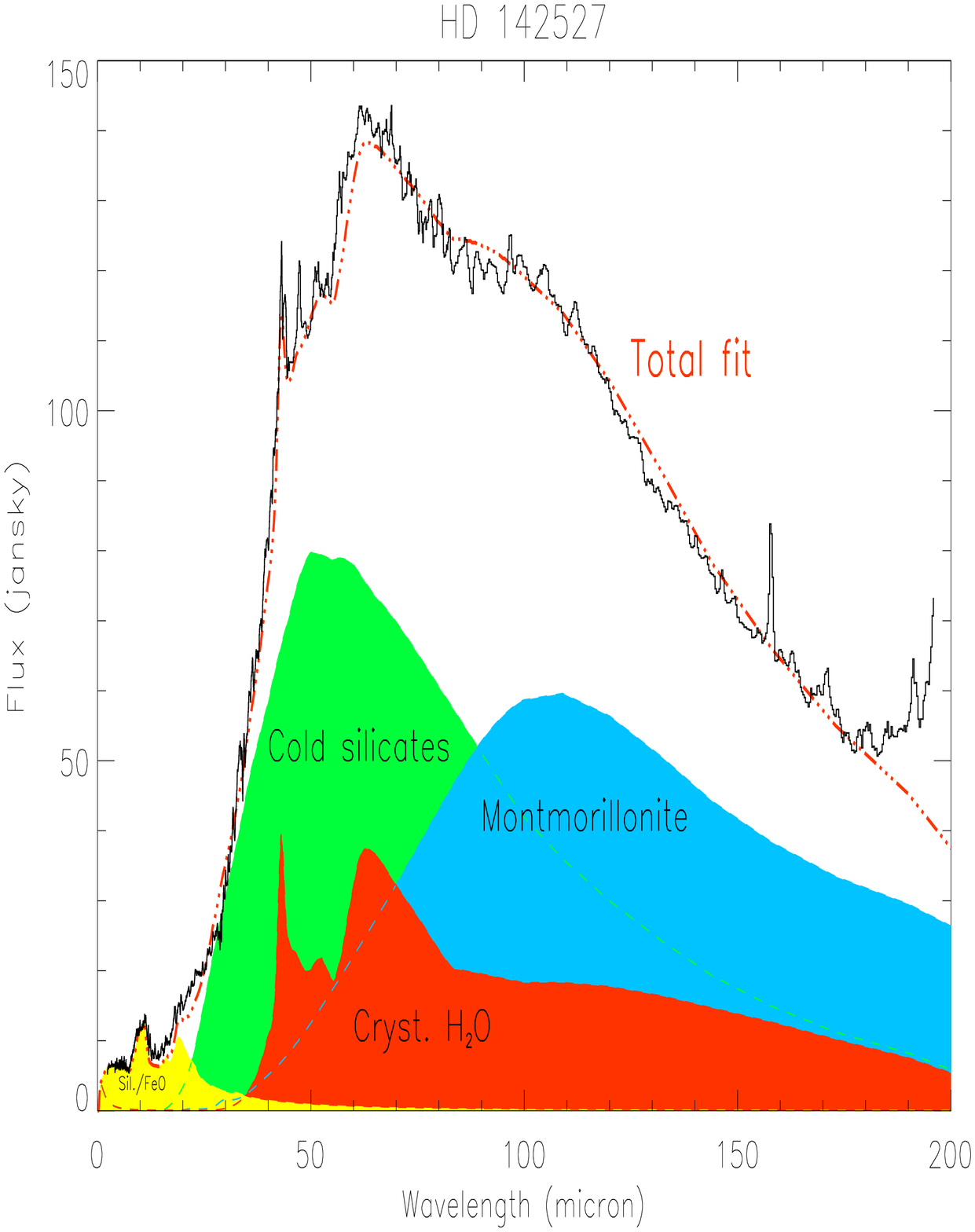,width=9cm,angle=0}
}} 
\vskip 1.0 true cm
\noindent
{\bf Figure 11 (color).} Complete SWS-LWS spectrum of the isolated
Herbig Ae star HD 142527, together with the decomposition into the
different components, including hot and cold amorphous silicates,
crystalline H$_2$O ice, and tentatively the crystalline hydrous
silicate montmorillonite (Malfait et al.\ 1999).
\endinsert
}}

The richest silicate spectra are found in disks around young and old
stars. A survey by Meeus et al.\ (2001) of a sample of isolated young
intermediate mass stars away from molecular clouds ---the so-called
Herbig Ae stars (Waters \& Waelkens 1998)--- reveals silicate emission
for at least 70\% of the sources.  In some objects, only the broad 10
and 18 $\mu$m bands due to amorphous silicates are seen, whereas
others reveal a large variety of narrower solid-state emission bands
(see Figure 10). The most spectacular example is provided by HD 100546,
which shows at least 8 bands which can be ascribed to forsterite,
Mg$_2$SiO$_4$ (Waelkens et al.\ 1996, Malfait et al.\ 1998).  Other
minerals identified in disks around Herbig Ae stars include enstatite
(MgSiO$_3$; Bouwman et al.\ 2001), hydrous silicates such as
montmorillonite (Malfait et al.\ 1999), silica (SiO$_2$; Bouwman et
al.\ 2001), metallic iron and iron oxide (van den Ancker et al.\
2000a, Bouwman et al.\ 2000), although the FeO 23 $\mu$m feature can
also be ascribed to FeS (Keller et al.\ 2002). The necessary
laboratory spectroscopy is summarized in Henning (2003). Long
wavelength data are often essential to make firm identifications
(Figure 11), but LWS spectra exist for only a few sources.  Typical
mass fractions of the crystalline material are 5--10\% compared with
amorphous olivine, with a few \% of silica. Several spectra also show
features due to PAHs (see \S 5) and crystalline H$_2$O ice (see \S 4),
some of which may be blended with the crystalline silicate
bands. Remarkably, no trend has been found between the age of the star
and the presence of crystalline silicate or PAH material over the
0.1--10 Myr range.

{\parindent 20pt
{\narrower
\topinsert
\null
\vskip -1.0 true cm
\centerline{\hbox{
\psfig{figure=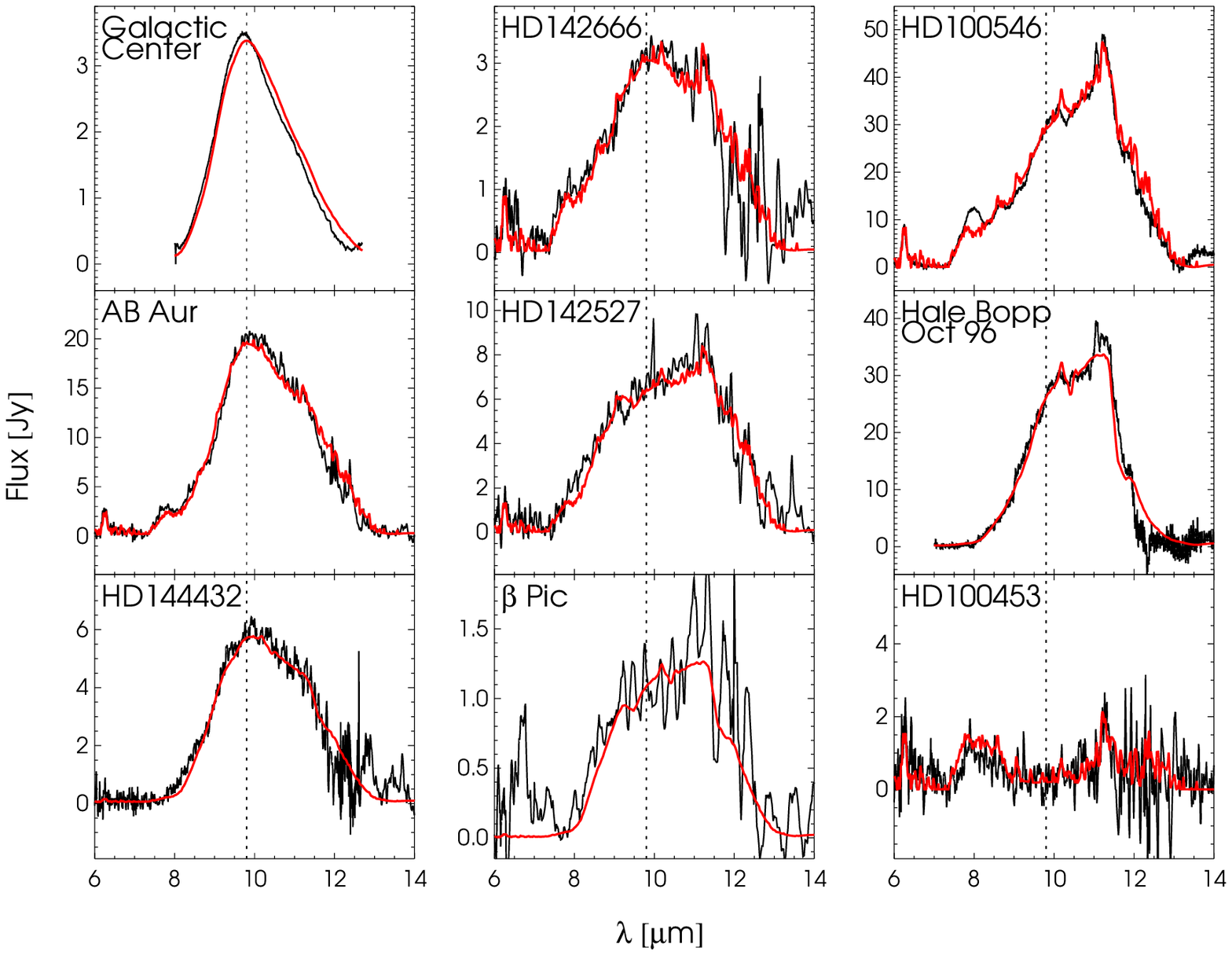,width=14cm,angle=0}
}} 
\vskip -0.4 true cm
\noindent
{\bf Figure 12.} Continuum-subtracted 10 $\mu$m silicate emission
bands observed toward a selected sample of isolated Herbig Ae stars
with the SWS. For reference, the silicate bands of interstellar medium
dust and comet C/1995 O1 Hale-Bopp are included. The dashed line
indicates the position of the 9.8 $\mu$m amorphous silicate band
observed in the interstellar medium.  The full lines indicate the
best-fit models using a mixture of amorphous and crystalline material
with different grain sizes. The sharp 11.3 $\mu$m feature is evidence
for crystallization; the shift toward longer wavelengths for grain
growth (Bouwman et al.\ 2001).
\endinsert
}}

Although not related to age, the 10 $\mu$m silicate profiles do show
variations from source to source and are significantly different from
those found in the diffuse medium (\S 6.1). In Figure 12, a collection
of silicate profiles is presented, arranged such that the peak
emission occurs at progressively longer wavelengths.  These shifts can
be caused by two effects, collectively called `dust processing':
coagulation of grains resulting in larger average grain sizes, and
crystallization resulting in discrete peaks, for example by forsterite
at 11.3 $\mu$m. Bouwman et al.\ (2001) argue on the basis of
correlations between features that grain growth from sub-micron to
micron-sized grains dominates the shift.  The timescales for
coagulation and annealing are not coupled, with crystallization having
a longer timescale. The SiO$_2$ band at 8.6 $\mu$m may be a by-product
of the annealing process leading to crystalline forsterite, as has
been observed in the laboratory (Rietmeijer et al.\ 2002).

The absence of the 10 $\mu$m band in some sources is intriguing and
may be due either to the absence of any small ($\lapprox$ few $\mu$m)
silicate grains or to low temperatures caused, for example, by
shadowing (see \S 10).  A recent re-analysis of the SWS spectrum of HD
100453 shows that both effects likely play a role (Meeus et al.\ 2002,
Vandenbussche et al.\ 2004): features due to large ($\sim 2$ $\mu$m)
crystalline forsterite grains are detected at longer ($>$30 $\mu$m)
wavelengths, but no emission is seen at 11 $\mu$m.

Silicate emission from disks around low-mass T Tauri stars has been
observed with PHOT-S (Natta et al.\ 2000, G\"urtler et al.\ 1999).
Although the quality of the spectra is limited, broad amorphous
silicate features from (sub)-micron sized grains consisting of a mix
of olivines and pyroxenes are found in all cases. There are hints of
11.3 $\mu$m crystalline forsterite and 8.6 $\mu$m silica features in
some sources, but confirmation requires higher $S/N$ and higher
spectral resolution data. Indeed, crystalline silicates in disks
around solar-mass stars have recently been found from ground-based
data by Honda et al.\ (2003). Mid-infrared spectroscopy of such
objects is an obvious target for future space missions.

\bigskip
\medskip
\centerline{\bf 7. PHOTON-DOMINATED REGIONS}
\bigskip
In Photon-Dominated or Photo-Dissociation Regions (PDRs),
far-ultraviolet photons (6--13.6 eV) control both the thermal and
chemical structure of the neutral gas. Except for the densest shielded
regions, the bulk of the molecular gas in our Galaxy and external
galaxies has only a moderate opacity to UV radiation.  Indeed, PDRs
dominate the diffuse infrared emission from molecular clouds, through
conversion of the ambient UV photons to infrared radiation from
dust and gas. The study of PDRs was triggered by early airborne observations
of strong far-infrared atomic fine-structure lines (e.g., Melnick et
al.\ 1979). Prominent [C II] 158 $\mu$m and [O I] 63 and 145 $\mu$m
lines from regions like Orion stimulated the development of detailed
PDR models (Tielens \& Hollenbach 1985). Together with subsequent
observations of UIR/PAH emission (see \S 5) and ground-based [C I] and
high-excitation CO lines, a flurry of activity on PDRs resulted in the
1980's and 1990's.  Comprehensive reviews have been given
by Hollenbach \& Tielens (1997, 1999).

PDRs have been widely observed with ISO using all four instruments.
CAM images at 7 $\mu$m (dominated by PAHs) and 15 $\mu$m (dominated by
small grains) beautifully reveal the cloud surfaces set aglow by the
surrounding UV radiation (see Fig.\ 1).  Both CAM-CVF and
PHOT-S have obtained spectra of the diffuse medium (Boulanger et al.\
2000, Mattila et al.\ 1996). LWS and SWS spectra show a wealth of
atomic fine-structure lines of neutral or singly ionized species, as
well as the mid-infrared H$_2$ lines.  Bright PDRs studied with ISO
spectroscopically include S 140 (Timmermann et al. 1996), NGC 2023
(Moutou et al.\ 1999b, Draine \& Bertoldi 1999), NGC 7023 (Fuente et
al.\ 1999, 2000), NGC 2024 (Giannini et al.\ 2000), W 49N (Vastel et
al.\ 2001), S 106 (Schneider et al.\ 2003), S 125 (Aannestad \& Emery
2001, 2003), S 171 (Okada et al.\ 2003), IC 63 (Thi et al.\ 1999), the
Trifid nebula (Lefloch et al.\ 2002), and Trumpler 14 (Brooks et al.\
2004).

Models have shown that the strength of the emission lines depends
mainly on two parameters: the gas density $n$ and the intensity of the
incident radiation field, characterized by an enhancement factor $G_0$
compared with the standard interstellar radiation field. Here $G_0$=1
refers to an integrated ultraviolet (6--13.6 eV) intensity of
$1.6\times 10^{-3}$ erg s$^{-1}$ cm$^{-2}$ (Habing 1968), a factor of
1.7 lower than the reference field by Draine (1978). Most of the PDRs
mentioned above have densities of at least $10^4$ cm$^{-3}$ and
radiation fields $G_0>10^3$.  Since many of these bright PDRs have
also been observed with the KAO, the ISO data have helped mostly to
refine the physical structure of the models (e.g., more accurate
density determinations, constraints on clumpy or layered structures)
and geometry (e.g., inclination, face-on versus edge-on, 3D
geometry). Care should be taken not to overinterpret the data,
however.  Comparison of results from different PDR codes shows
differences in predicted line intensities up to a factor of 2--3 owing
to different assumptions in the input data and chemistry.  Also, some
lines, in particular the [O I] 63 $\mu$m line, are optically thick so
that model results are sensitive to the details of the radiative
transfer and possible self-absorption by cold foreground gas.

The higher spatial resolution ISO data have confirmed earlier
suggestions that most of the observed [Si II] 34.8 $\mu$m emission
originates from the outer PDR layer with $A_V \leq$2 mag rather than
from any ionized gas in the region (e.g., Fuente et al.\ 2000, Lefloch
et al.\ 2002), with the enhanced gas-phase silicon most likely due to
photodesorption of Si-containing grain-mantle material (Walmsley et
al.\ 1999). Not all PDRs show [Si II] emission, however (Rosenthal et
al.\ 2000).

The main new insights due to ISO have been (i) Extension of PDR
studies to the lower $n$ and $G_0$ regime; (ii) Direct measurement of
the efficiency of heating the gas through the photoelectric effect;
and (iii) Detailed studies of the H$_2$ mid-infrared pure rotational
lines leading to new insights into the temperature structure and H$_2$
formation rate at high temperatures.  PAHs are discussed mostly in \S
5.

\bigskip
\centerline{\it 7.1. Low-Excitation PDRs and the Gas Heating Efficiency}
\medskip
Low excitation PDRs with $n\leq 10^4$ cm$^{-3}$ and/or $G_0<100$
studied by ISO include L1457 (MBM12) (Timmermann et al.\ 1998), the
S~140 extended region (Li et al.\ 2002), L1721 (Habart et al.\ 2001),
$\rho$ Oph main (Liseau et al.\ 1999), $\rho$ Oph W (Habart et al.\
2003), and a set of translucent clouds also studied by optical
absorption lines (Thi et al.\ 1999).  Cederblad 201 (Kemper et al.\
1999, Cesarsky et al. 2000b) is an example of a PDR exposed to a
cooler B9.5 star where the radiation field has fewer far-UV
photons. The physical structure of these clouds is often well
characterized from complementary data, so that they form good tests of
the basic PDR models. While some ISO lines (e.g., [C II]) are well
explained, others (e.g., H$_2$) cannot be reproduced
with standard model parameters (see \S 7.2.3).

Comparison of the total intensity of the main cooling lines (principly
[C II] 158 $\mu$m for the low-excitation PDRs) with the far-infrared
continuum gives directly the heating efficiency $\epsilon$ of the gas
due to the photoelectric effect.  Assuming that the lines are
optically thin, values ranging from 1--3\% are obtained.  These
efficiencies are somewhat higher than the range of 0.1--1\% found for
a sample of higher $n, G_0$ PDRs by Stacey et al.\ (1991) with the KAO
and that of 0.7\% for S 106 studied with ISO.  For very high values of
$G_0$/$n$ appropriate for clouds like W 49N, the photoelectric heating
efficiency is clearly suppressed by an order of magnitude to
(0.01--0.04)\%.  This is consistent with models by Bakes \& Tielens
(1994), who show that intense radiation results in more highly charged
grains which have a larger barrier for the photoelectrons to
escape. So far, W 49N is the only PDR for which such low efficiencies
have been found.  Even Trumpler 14, a PDR powered by at least 13 O
stars, and 30 Doradus in the LMC, powered by at least 30 O stars, have
efficiencies of 0.2--0.5\% (Brooks et al.\ 2004). One possible
explanation is that the ionization front in these cases is at such a
large distance from the stellar cluster that the PDRs at the outer
edges of the cloud do not have extreme parameters in terms of $G_0$ or
$n$.

Habart et al. (2001) went one step further by spatially correlating
the cooling lines with infrared emission due to PAHs (4--12 \AA; IRAS
12 $\mu$m), very small grains (VSGs, 12--150 \AA; IRAS 25 and 60
$\mu$m) and big grains ($>150$ \AA; IRAS 100 $\mu$m). They
conclusively show that the PAHs are the most efficient heating agents,
with $\epsilon_{\rm PAH}$=3\%, $\epsilon_{\rm VSG}$=1\% and
$\epsilon_{\rm BG}$=0.1\%. Habart et al.\ (2003) find $\epsilon_{\rm
PAH}$=4\% for the $\rho$ Oph West cloud. Both studies also find
evidence for an increase in the PAH abundance near edge of cloud,
consistent with other studies based on IRAS and/or ISO imaging (e.g.,
Boulanger et al.\ 1990, Abergel et al.\ 2002).

{\parindent 20pt
{\narrower
\topinsert
\null
\vskip -2.8 true cm
\centerline{\hbox{
\psfig{figure=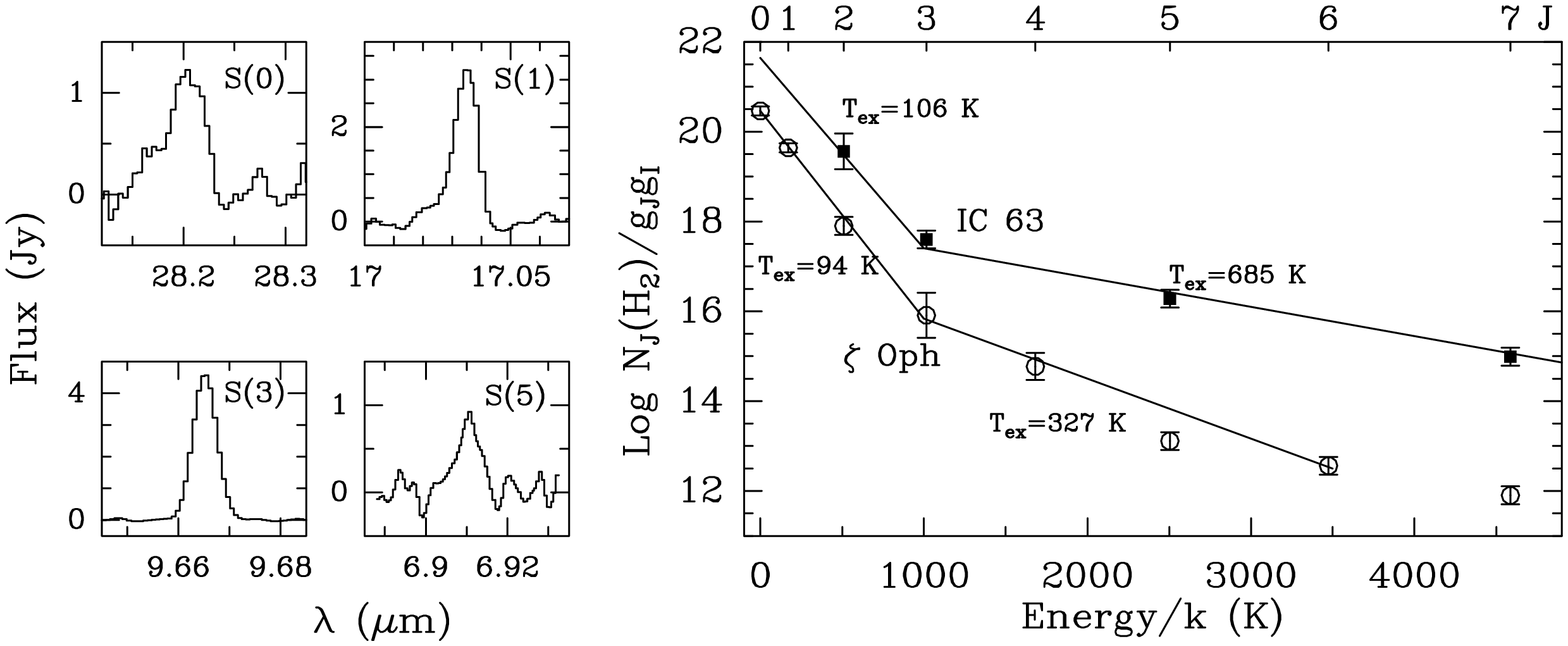,width=14cm,angle=-90}
}}
\vskip -3.0 true cm
\noindent
{\bf Figure 13.} Left: Pure rotational H$_2$ lines observed toward
IC 63 with the SWS (Thi et al.\ 1999). Right: H$_2$ excitation diagram
for IC 63 compared with that derived from UV observations of the
diffuse cloud $\zeta$ Oph (Spitzer \& Cochran 1973). The statistical
weights are $g_J=2J+1$ with $g_I$=1 for para-H$_2$ and $g_I$=3 for
ortho-H$_2$.
\endinsert
}}

\bigskip
\centerline{\it 7.2. H$_2$ Pure Rotational Lines}
\medskip
{\bf 7.2.1. Rotational temperatures.} The H$_2$ pure rotational
quadrupole lines are readily detected from bright PDRs, in spite of
their small intrinsic Einstein $A-$values (see Figure 13). In contrast
with the vibration-rotation lines at 2 $\mu$m seen from the ground,
the lowest rotational levels are primarily excited by collisions in
warm gas rather than by UV pumping (e.g., Black \& van
Dishoeck 1987, Sternberg \& Dalgarno 1989, Draine \& Bertoldi
1996). This is directly reflected in their excitation diagrams (see
Figure 13, right), which show that the $v$=0 rotational distribution
is characterized by a temperature $T_{\rm rot}$=400--700~K (e.g.,
Timmermann et al.\ 1996, Fuente et al.\ 1999, Wright 2000), much lower
than the vibrational temperatures of $T_{\rm vib}\approx$2000~K. The
inferred $T_{\rm rot}$ for various PDRs are remarkably similar in
spite of their different characteristics.  It should be noted,
however, that only few sources have been studied in detail and that in
several cases the integrations were not deep enough to reveal the
lowest $J=2-0$ line at 28.22 $\mu$m. Where detected, this line shows
evidence for a lower temperature component with $T_{\rm rot}
\approx$100 K. It is interesting to note that
the rotation diagrams derived from H$_2$ UV absorption line data of
diffuse (e.g., Spitzer \& Cochran 1973, see Figure 13) and translucent
(e.g., Snow et al.\ 2000) clouds are also similar.  These data, which
include the populations of the lowest two $J$=0 and $J$=1 levels,
clearly reveal the multi-temperature structure.  Similar diagrams have
been found for extragalactic regions in normal (e.g., Valentijn
\& van der Werf 1999) and starburst galaxies (e.g., Rigopoulou et
al. 2002).

In spite of their simple appearance, the interpretation of these
diagrams has to take account of several factors. First, the gas
kinetic temperature varies rapidly through the PDR layer from a value
of several hundred K at the edge to less than 30~K when $A_V>1$ mag.
The pure rotational H$_2$ lines arise primarily from the outer warm
layer, but its temperature depends strongly on $n$ and $G_0$, as well
as the assumed PAH heating parameters (e.g., Kaufman et al.\ 1999).
Second, only the lowest rotational levels are collisionally
excited. For diffuse clouds, this holds only up to $J$=2 whereas for
denser PDRs like S~140 levels up to $J$=5 may be populated by
collisions.  Finally, the rotational populations are affected by UV
pumping in at least two ways: in an absolute sense by providing more
population in the higher-$J$ levels, and in a relative sense by the
fact that the excitation rates out of the various $J$ levels differ
with depth into the cloud.  No PDR model has yet been published which
fits all the ISO data and takes all these effects into account in a
fully self-consistent manner.  For example, the excellent fits to the
S 140 H$_2$ data by Timmermann et al.\ (1996) use a fitted gas
temperature profile rather than a computed one.

One general conclusion of all PDR models is that the gas temperature
needs to be higher in the zone where H$_2$ is formed than can be
explained by the standard gas heating and cooling processes. Draine \&
Bertoldi (1999) discuss possible remedies. In addition to changes in
geometry and density structure, possible uncertainties in the
microphysics include the H$_2$ inelastic collisional rate
coefficients, the grain photoelectric heating efficiency, and an
increased dust-to-gas ratio in the PDR zone due to anisotropic
illumination of grains which causes them to drift through the gas. An
additional option, namely an increased H$_2$ formation rate, is
discussed below.  Yet another explanation, namely that of weak shocks
in diffuse gas, remains to be quantified.

{\bf 7.2.2. Ortho/para ratio.} There has been significant confusion in
the literature on the ortho/para ratio of H$_2$, with several claims
of ortho/para ratios that are lower than the equilibrium value
corresponding to the temperature derived from excitation diagrams
(e.g., Fuente et al.\ 1999).  As pointed out by Sternberg \& Neufeld
(1999), care has to be taken with such statements since a low
ortho/para ratio in excited levels naturally results from the fact
that the optical depth in the UV pumping lines is higher for the
ortho- than for the para-H$_2$ lines (see their Figure 5).  None of
the ISO data measure the bulk of ortho- and para-H$_2$, which resides
in $J$=0 and 1. Additional concerns are mentioned by Wright (2000).
Nevertheless, some of these claims of `non-equilibrium' ortho/para
ratios in high temperature gas may be valid (see also \S 8), in which
case they are interesting indicators of dynamical processes, such as
the advection of colder gas into the PDR.

{\bf 7.2.3. Increased H$_2$ formation rate.} Detailed studies of several
well-characterized, low-excitation PDRs have shown convincingly that
the intensities of the lowest H$_2$  lines cannot be
reproduced with standard PDR parameters (e.g., Bertoldi 1997, Kemper et
al.\ 1999, Thi et al.\ 1999, Habart et al.\ 2001, 2003, Li et al.\
2002).  The discrepancies are significant, at least an order of
magnitude. Since for these regions the gas-grain drifts discussed
above are not significant, the discussion has focussed on two
solutions. First, an increase in the photoelectric heating rate, for
example through a larger PAH or very small grain abundance.  Second,
an increase in the H$_2$ formation rate at high temperatures.  A
combination of these two solutions has the effect of moving the
H--H$_2$ transition zone closer to the edge of the PDR where the
temperatures are higher. The most quantitative study is by Habart et
al.\ (2003a,b, 2004), who show that for most (but not all) sources an
increase in the H$_2$ formation rate by a factor of $\sim$ 5 at
temperatures of a few hundred K is needed, compared with the standard
rate of (3--4)$\times 10^{-17}$ cm$^3$ s$^{-1}$ derived from
UV absorption lines by Jura (1975) and Gry et al.\ (2002) for
diffuse clouds.

Since the residence times of physisorbed H on grains at such high
temperatures are much too short, this conclusion has fundamental
implications for our understanding of basic molecule formation
processes. One possibility is that direct reactions of atomic H
arriving from the gas on top of chemisorbed species dominate in an
Eley-Rideal type process (see e.g., Duley 1996). Another option is
that physisorbed H diffuses over the surface to find a chemisorbed H
in a Langmuir-Hinshelwood-type process (Hollenbach \& Salpeter 1971).
Quantitative estimates of the rates of both processes have been made
(Cazaux \& Tielens 2002, Habart et al.\ 2004). The efficiencies depend
sensitively on the number of chemisorbed sites and the activation
energy barrier to molecule formation, both of which are
uncertain.

\bigskip
\medskip
\centerline{\bf 8. SHOCKS}
\medskip
Shocks in molecular clouds can be caused by a large variety of
processes including outflows and jets from young stars, supernovae and
expanding H II regions. Recent reviews of the theory of shocks have
been given by Draine \& McKee (1993) and Hollenbach (1997). In this
section, ISO observations of well-characterized shocks away from the
driving sources are discussed; those in the inner parts of
protostellar envelopes are included in \S 9.  The continuum emission
due to warm dust offset from the YSOs is generally very low so that
all lines are in emission (see Figure 2).  Early models by Hollenbach
et al. (1989) show that slow, non-dissociative shocks have many strong
diagnostic lines at infrared wavelengths. In these so-called $C-$type
shocks, the hydrodynamical variables change continuously and the
maximum temperature is only 2000--3000~K, too low to dissociate
molecules. In contrast, the temperatures in high-velocity $J-$type
shocks jump to more than $10^5$ K, and most of the emission is from
atomic lines at UV wavelengths. Also, these shocks can cause
shattering or non-thermal sputtering of grain cores, resulting in
enhanced abundances og Si, Mg, Fe and other grain constituents.

ISO's main contributions to the study of shocks have been to (i)
Better characterize the shock structure and physical conditions,
especially that found in the slower shocks; (ii) Measure the total
cooling power directly from observations, in particular the individual
contributions from H$_2$O and H$_2$, and thus understand the overall
energy budget; and (iii) Study global and systematic trends by
observing much larger samples.  A good overview has been given by
Noriega-Crespo (2002).

{\parindent 16pt
{\narrower
\topinsert
\null
\vskip -1.0 true cm
\centerline{\hbox{
\psfig{figure=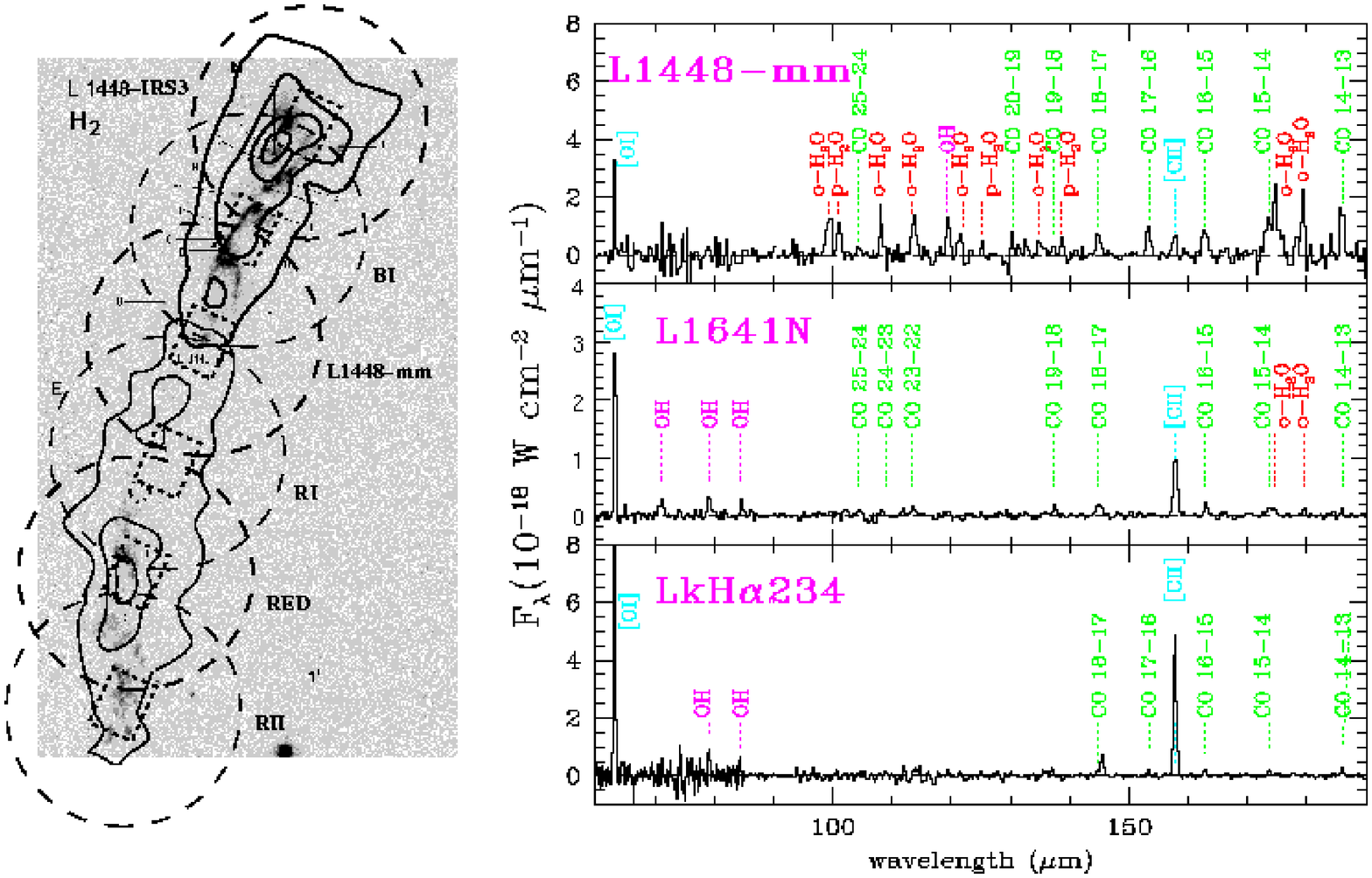,width=14cm,angle=-90}
}} 
\vskip -0.4 true cm
\noindent
{\bf Figure 14.} Left: narrow-band image of the L1448 outflow at
2.12 $\mu$m (H$_2$+continuum), with superposed the LWS (long-dashed
circles) and SWS (short-dished rectangles) apertures. The thick and
thin solid lines delineate the contours of the blue- and red-shifted
CO outflow from the L1448-mm source (Nisini et al.\ 2000). Right: LWS
continuum-subtracted spectrum toward the Class 0 source L1448-mm
compared with the more evolved L1641N and LkH$\alpha$ 234 sources.
Note the decrease of molecular emission, in particular H$_2$O, toward
the latter sources (Benedettini et al.\ 2003).
\endinsert
}}

The ISO data have also convincingly shown that all broad band emission
from shocks is due to lines (e.g., Cesarsky et al.\ 1999).  For
example, the ISOCAM 7 and 15 $\mu$m bands are dominated by the H$_2$
S(4) and S(1) lines, whereas the IRAS 12, 25, and 60 $\mu$m emission
is due to H$_2$ S(2), S(0), and [O I] 63 $\mu$m, respectively.

\bigskip
\centerline{\it 8.1. Shock Structure and Physical Conditions}
\medskip
Shocks associated with outflows studied by ISO include Orion (Harwit
et al.\ 1998, Rosenthal et al.\ 2000, Sempere et al. 2000,
Gonz\'alez-Alfonso et al.\ 2002), HH 1-2 (Cernicharo et al. 2000b,
Lefloch et al.\ 2003, Cabrit et al.\ 2004), HH 7--11 (Molinari et al.\
2000), HH 46 (Nisini et al.\ 2002), HH 52--54 (Liseau et al.\ 1996,
Nisini et al.\ 1996), HH 80-81 (Molinari et al.\ 2001), Cep A West
(Wright et al.\ 1996, Froebrich et al.\ 2002), Cep E (Moro-Mart{\'\i}n
et al.\ 2001, Smith et al.\ 2003), DR 21 (Wright et al.\ 1997), L1448
(Nisini et al.\ 1999a, 2000; Froebrich et al.\ 2002), B 335 (Nisini et
al.\ 1999b), L1551 (White et al.\ 2000), NGC 1333 (Bergin et al.\
2003) and a variety of other low-mass YSOs including L1157 (Giannini
et al.\ 2001).  Shocks associated with supernova remnants investigated
by ISO include IC 443 (Cesarsky et al.\ 1999) and 3C 391, W 44 and W
28 (Rho \& Reach 2003). In all types of shocks, molecular (H$_2$, CO,
H$_2$O, OH) and atomic ([O I], [C II], [S I]) lines are often detected
(see Figure 14 and 15), although there are sources in which no
molecular lines except H$_2$ are seen (e.g., L1551 IRS5, HH 80-81).

{\parindent 20pt
{\narrower
\topinsert
\null
\vskip -5.0 true cm
\centerline{\hbox{
\psfig{figure=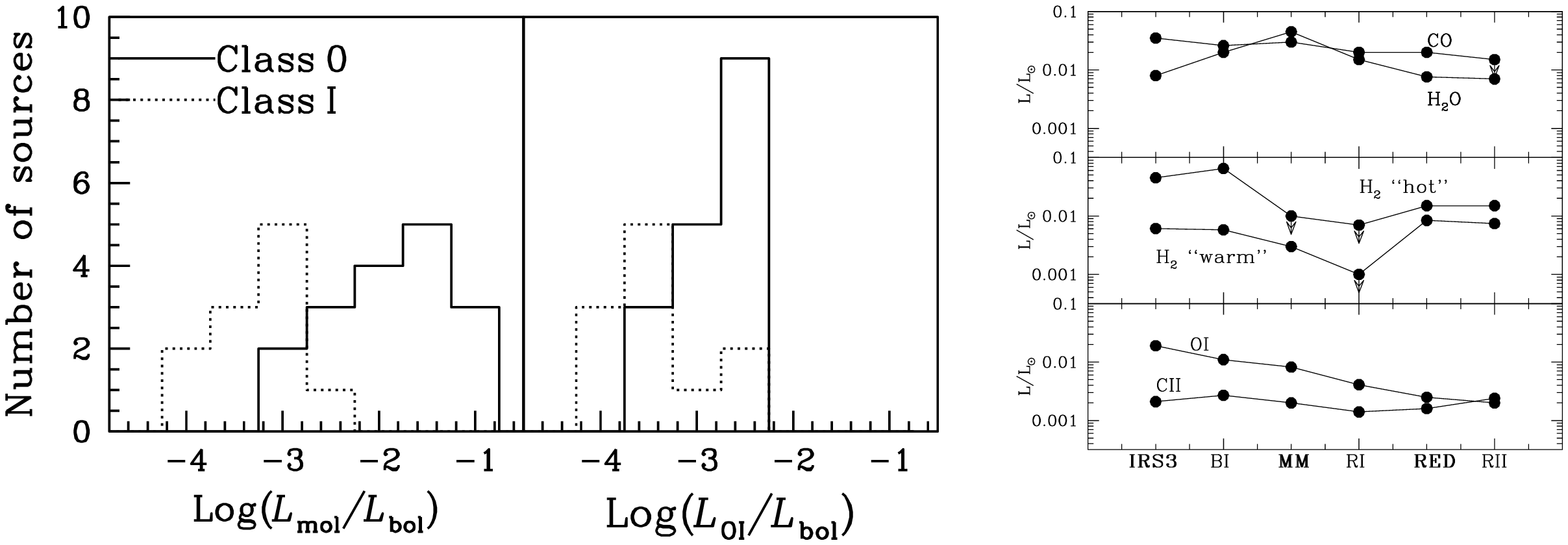,width=14cm,angle=-90}
}}
\vskip -1.7 true cm
\noindent
{\bf Figure 15.} Left: histogram of the total molecular cooling,
defined as $L_{\rm mol}=L_{\rm H_2O} + L_{\rm CO} + L_{\rm OH}$, and
of the cooling due to [O I], compared with the source
bolometric luminosity for a set of embedded YSOs (Nisini et al.\
2002). Right: Luminosity emitted by different molecular and atomic
species along the L1448 outflow (see Figure 14 for
positions). ``Warm'' and ``hot'' H$_2$ refer to the pure rotational
and vibration-rotation lines, respectively (Nisini et al.\ 2000)
\endinsert
}}

In spite of the fact that the available diagnostics often differ
depending on whether SWS, LWS and/or CAM-CVF data are obtained, the
conclusion is invariably that no single planar static shock can
explain all the data. Specifically, the H$_2$ excitation and the
presence of other molecules require a combination of slow ($<$20 km
s$^{-1}$) and fast (30--50 km s$^{-1}$) $C-$shocks with different
covering factors, whereas the atomic lines can only be explained with
a $J-$type shock ($70-140$ km s$^{-1}$).  The estimated filling
factors of the shocks are small, typically a few \% of the LWS
beam. It is still unclear whether these different shocks can be
combined in a single unified model. Attempts to do so include bow
shocks with long flanks, where a broad distribution of shock speeds
arises from the curvature (e.g., Froebrich et al.\ 2002),
time-dependent $C-$shocks with embedded transient $J-$ shocks (e.g.,
Chi\`eze et al.\ 1998), or multiple shocks linked to multiple episodes
of outflow activity (e.g., Nisini et al.\ 2000).

{\bf 8.1.1. Atomic lines, J-shock diagnostics.} A clear diagnostic of
a fast dissociative $J$-shock is the [Ne II] 12.8 $\mu$m line, whose
intensity is directly proportional to $v_s$.  This line is commonly
associated with high-mass YSOs, but has been detected in only one
low-mass YSO (HH 1--2: see Figure 2). The lack of [Ne II] or other
highly ionized fine-structure lines toward low-mass YSOs, coupled with
the detection of optical high-excitation lines, indicates that the
area occupied by high velocity $\sim 200$ km s$^{-1}$ shocks is small,
typically $< 1''$. Instead, van den Ancker et al.\ (2000b,c) show that
[S I] 25.2 $\mu$m may be a better shock diagnostic.  Another
often-used probe is the [C II]/[O I] fine-structure line ratio, which
is expected to be low in fast $J$-shocks.  [C II] can also arise in
PDRs, however, either from the surrounding molecular cloud emission or
produced locally by the shock. Indeed, Molinari et al.\ (2001) and
Molinari \& Noriega-Crespo (2002) have argued that much of the
observed [C II] emission in Herbig-Haro sources arises from PDRs at
the walls of the outflow cavity. This radiation can also create a PDR
in the quiescent gas ahead of the shock.

\medskip
{\bf 8.1.2. H$_2$ lines, C-shock diagnostics.} As for PDRs, the pure
rotational H$_2$ lines dominate the mid-infrared spectra and are
characterized by excitation temperatures of $700-1000$~K.  In
contrast, the well-studied H$_2$ vibration-rotation lines usually give
$T_{\rm ex}$=2000--3000~K. Many papers treat these two data sets
separately, but the combined pure-rotation and vibration-rotation data
observed with ISO clearly indicate that two excitation temperatures
are common, with evidence for even higher excitation conditions
($>$10000~K) in some sources (Cep A: Wright et al.\ 1996, Orion:
Rosenthal et al.\ 2000). The most impressive collection of H$_2$ lines
is seen for Orion at the so-called Peak 1 (Rosenthal et al.\ 2000, see
Figure~3) and Peak 2 (CM Wright, private communication) positions,
where 56 H$_2$ lines have been seen including the $v$=0 S(1) to S(25)
lines with upper energy levels up to 42500 K. The detection of the
S(25) line is particularly intriguing and may be indicative of a
non-thermal excitation processes, either by direct population of
high-$J$ levels during the formation of H$_2$ (so-called `formation
pumping') or by non-thermal collisions between ions and molecules in
the shock. However, Le Bourlot et al. (2002) have recently
re-evaluated the chemistry and excitation of H$_2$ in shocks using
much improved H$_2$-H and H$_2$-H$_2$ collisional rate coefficients,
and concluded that the critical velocity at which H$_2$ is dissociated
is significantly higher than previously thought. As a result, higher
excitation of H$_2$ can result from thermal collisions only, and the
Orion H$_2$ data can be reproduced with a two-component $C-$ shock
model without the need for non-thermal excitation processes. CAM-CVF
images show that H$_2$ avoids the high-excitation optical knots and
has a distribution more consistent with shocked gas swept-up by the
bow-shock (e.g., Cabrit et al.\ 1999, 2004; Larsson et al.\ 2002)

Theoretical calculations of the H$_2$ rotational excitation in various
shock models give $T_{\rm ex}$ values of 200--540~K for $J-$shocks,
but a much larger range of 100--1500~K for $C-$shocks (see Figure 8 in
van den Ancker et al.\ 2000b). $T_{\rm ex}$ does not depend much on
density, but increases strongly with shock velocity. Thus, the
observed range of $T_{\rm ex}$ discussed above naturally leads to at
least two $C-$type shocks. Interestingly, the rather narrow range of
$T_{\rm ex}$ predicted for $J-$shocks is similar to that found for
PDRs (see \S 7.2.1). For PDRs, $T_{\rm ex}$ does not depend much on
density and is a good diagnostic of the UV radiation field $G_0$.
Except for the shock associated with T Tau (van den Ancker et al.\
1999), however, no case of a fluorescent contribution to the H$_2$
shock excitation has been found. More recently, Cabrit et al.\ (2004)
have stressed the use of the H$_2$ pure rotational lines to also
constrain the transverse magnetic field strength and abundance of
small grains, both of which affect the shock details.

Since the collisional rate coefficients of H$_2$ with H are
significantly larger than those with H$_2$ or He, the H/H$_2$ ratio in
the pre-shocked gas is critical in the models. In general, much better
fits to the shock data are obtained if this ratio is $\sim$0.1, at
least two orders of magnitude larger than that expected in dark
clouds. For most shocks, the inferred ortho/para ratio is 3,
consistent with the warm conditions. The major exceptions are HH 54
(Neufeld et al.\ 1998) and HH 2 (Lefloch et al.\ 2003), where
significantly lower ortho/para ratios of 1.2--1.6 are found. This low
ratio is thought to be a relic of the earlier cold cloud stage, where
the timescale of the shock passage has been too short to establish an
equilibrium ortho/para ratio.  Thus, the ortho/para ratio can be used
as a chronometer, indicating for HH 54 that the gas has been warm for
at most 5000 yr.  Timmermann (1998) shows that the ortho/para ratio
only attains the high-temperature equilibrium value of 3 for shock
speeds greater than 20--25 km s$^{-1}$.

{\parindent 20pt
{\narrower
\topinsert
\vskip -4.3 true cm
\centerline{\hbox{
\psfig{figure=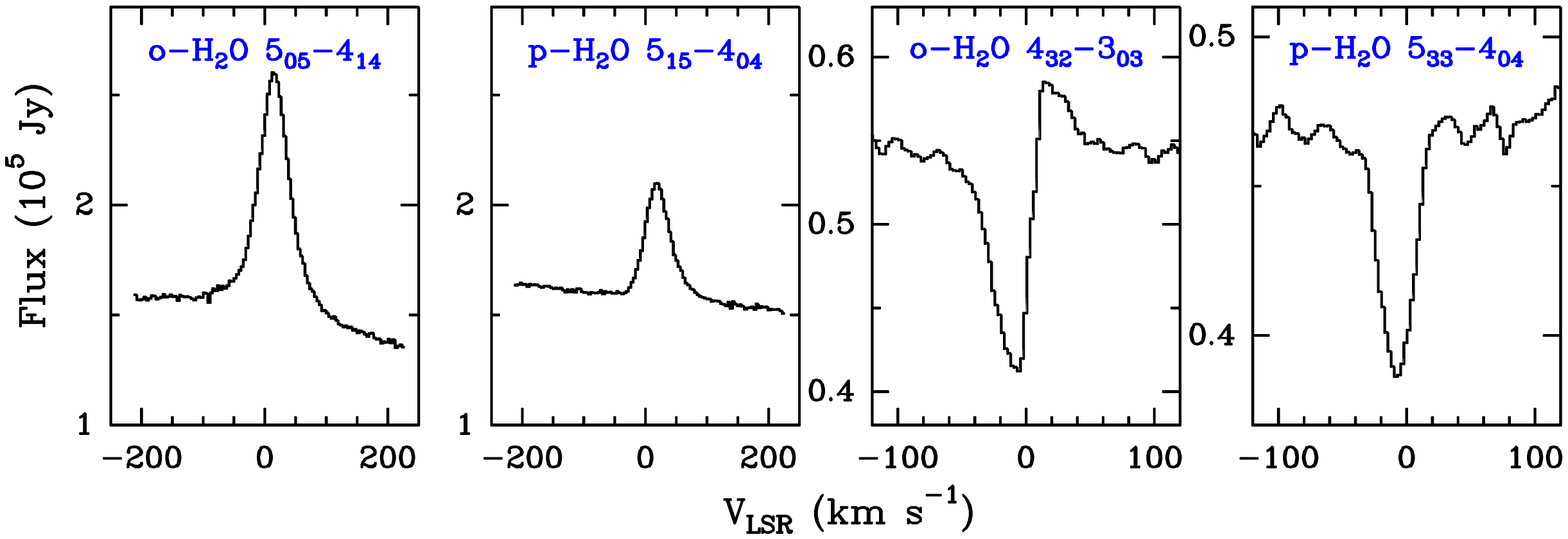,width=14cm,angle=-90}
}} 
\vskip -1.2 true cm
\noindent
{\bf Figure 16.} Examples of high-excitation emission and
absorption lines of H$_2$O for the Orion shock observed with the LWS
($o$-H$_2$O $5_{05}-4_{14}$ at 99.4931 $\mu$m, $p$-H$_2$O
$5_{15}-4_{14}$ at 95.6273 $\mu$m) and SWS ($o$-H$_2$O $4_{32}-3_{03}$
at 40.6909 $\mu$m, $p$-H$_2$O $5_{33}-4_{04}$ at 35.4716 $\mu$m)
Fabry-Perot spectrometers (Harwit et al.\ 1998, Wright et al.\ 2000).
\endinsert
}}

{\bf 8.1.3. CO, H$_2$O, and OH lines, C-shock diagnostics:} The LWS
spectra of shocks associated with nearby low-mass YSOs often contain
several CO lines of comparable strengths, ranging from $J$=14--13 up
to 25--24 (see Figure 1 and 14). The excitation temperature of CO is
generally somewhat lower than that of H$_2$.
%Statistical equilibrium excitation calculations show that the
%relative line strengths can be reproduced in gas with kinetic
%temperatures of a few hundred K and densities of $10^5-10^6$
%cm$^{-3}$. Using these physical conditions, the inferred abundances of
%H$_2$O range from $10^{-5}$ to $>10^{-4}$ and those of OH ---where
%detected --- typically an order of magnitude lower.  
Comparison of the H$_2$O/CO luminosity ratio with shock models
requires slow (10--20 km s$^{-1}$) $C-$type shocks. Since
submillimeter CO line profiles often indicate the presence of much
higher shock velocities, one interpretation is that the slow shocks
are proceeding into a medium which has been put into motion by
a previous generation of shocks. Indeed, Nisini et al.\ (1999, 2000)
associate the strong molecular emission found for L1448-mm directly
with the extremely-high velocity bullets seen in CO which move at
speeds up to 200 km s$^{-1}$.

The most spectacular high-mass example is again provided by Orion,
where Harwit et al.\ (1998) detected many H$_2$O emission lines
resolved with the Fabry-Perot (see Figure 16).  The observed line
strengths are well reproduced by the $C-$shock model of Kaufman \&
Neufeld (1996) in which most of the gas-phase oxygen is driven into
H$_2$O leading to an abundance of $\sim 5\times 10^{-4}$.  At shorter
wavelengths, the pure rotational H$_2$O lines go from emission into
absorption, at least in the direction of Orion-IRc2 (Wright et al.\
2000).  Not all shocks associated with high-mass YSOs resemble Orion,
however, even when scaled for the larger distances: H$_2$O emission
lines are usually not detected with ISO, indicating much smaller
filling factors (Boonman et al.\ 2003a, see also \S 9.2).

\bigskip
\vfill \eject
\centerline{\it 8.2. Gas Cooling in Shocks}
\medskip
The complete infrared spectral coverage provided by the ISO
spectrometers allows a direct determination of the contribution of
each species to the cooling. Figure 15 summarizes the results for a
survey of embedded low-mass YSOs by Nisini et al.\ (2000, 2002).  In
many cases, H$_2$, CO and H$_2$O have comparable contributions
on-source with no single species clearly dominating.  Off source, the
H$_2$O contribution may be less. Together, the molecular lines can
contain up to an order of magnitude more flux than the [O I] 63 $\mu$m
line in the densest regions.  The total observed cooling is often more
than 50\% of the mechanical power input of the shock, illustrating
that the bulk of the kinetic energy is transformed into infrared
radiation.

In $J-$type shocks such as found in Herbig-Haro objects, the [O I] 63
$\mu$m line is usually the dominant coolant.  In that case, its
intensity is a direct measure of the mechanical luminosity, which in
turn is related to the mass-loss rate.  Values of $10^{-7}-
10^{-4}$ M$_{\odot}$ yr$^{-1}$ have been inferred this way, which
correlate well with those derived from CO outflow maps indicating that
the two phenomena are physically associated (Saraceno et
al.\ 1999, Liseau et al.\ 1997).

\bigskip
\medskip
\centerline{\bf 9. EMBEDDED YOUNG STELLAR OBJECTS}
\medskip
ISO spectra taken at the position of the protostars themselves often
show most of the spectral characteristics of cold and warm gas
and dust discussed above, arising from different parts of the
circumstellar environment (see Figure 6, van Dishoeck et al.\
1999). The cold outer envelope is seen prominently in ice absorption,
whereas the PDRs and shocks are revealed through atomic, molecular or
PAH emission lines. As the protostellar evolution progresses, the
relative contribution of each of these components changes.  The main
advances made possible by ISO have been to (i) Quantify the relative
contributions from different physical components; and (ii) Establish a
protostellar classification scheme based on mid- and far-infrared
spectroscopic features.  For low-mass YSOs, this scheme complements
and extends the Class 0, I and II classification based on the spectral
energy distribution (SED) (e.g., Lada 1999). For higher mass objects where
the temporal ordering is much less clear (Churchwell 2002), infrared
spectroscopy is a powerful tool to characterize the earliest stages.

For sensitivity reasons, low-mass YSOs have been studied primarily
with the LWS and CAM-CVF, whereas high-mass YSOs have been intensively
observed with the SWS. These two sets of sources are therefore
discussed separately, but within a common physical and chemical
framework.
\bigskip
\centerline{\it 9.1. Low-Mass YSOs}
\medskip
{\bf 9.1.1. LWS surveys.} Giannini et al.\ (2001), Nisini et al.\
(2002) and Benedettini et al.\ (2003) summarize the LWS observations
of 28 low luminosity ($<$100 L$_{\odot}$) YSOs, consisting of 17 Class
0 and 11 Class I objects. Several of these objects have been studied
individually in earlier papers, e.g., IRAS 16293 -2422 (Ceccarelli et
al.\ 1998a), B 335 (Nisini et al.\ 1999b), L1448-mm (Nisini et al.\
2000), L1551 IRS5 (White et al.\ 2000), Serpens SMM1 (Larsson et al.\
2002) and NGC 1333 IRAS 4 (Maret et al.\ 2002). The atomic [O I] 63
$\mu$m and [C II] 158 $\mu$m lines are detected in all sources and
molecular lines in many cases, most prominently in the Class 0
objects. Specifically, CO, H$_2$O and OH are detected in 94\%, 70\%
and 41\% of the Class 0 sources, respectively, whereas CO and OH are
seen in 64\% and 27\% Class I objects. No H$_2$O has been detected in
any Class I source, except for T Tau (see below). The spectral
evolutionary sequence is illustrated in Figure 14.

The total molecular cooling in Class 0 objects is significantly larger
than that in Class I sources, whereas the atomic cooling through [O I]
63 $\mu$m is similar (see Figure 15). The total atomic + molecular
line cooling $L_{\rm line}/L_{\rm bol} \approx 10^{-2}$
for Class 0 sources, whereas it is an order of magnitude lower for
Class I objects. This decline is consistent with that in the outflow
momentum flux, which in turn is thought to be related to the
decay of the mass accretion activity with time (Andr\'e et al.\ 2000).

Temperatures, densities, and column densities have been obtained by
fitting molecular excitation models to the CO line ratios, giving
values of a few hundred to $>$1000~K and densities of $10^5-10^6$
cm$^{-3}$. The filling factor of this warm and dense gas is small,
$\sim 10^{-9}$ sr or a few hundred AU radius.  Assuming that the
physical parameters derived from CO are also valid for H$_2$O and OH,
the inferred H$_2$O abundances range from $10^{-5}$ to a few $\times
10^{-4}$ and increase with the gas temperature.  Where detected, the
OH abundance is at least an order of magnitide lower than that of
H$_2$O, of order $10^{-6}-10^{-5}$.  For Class I objects, the upper
limit on the H$_2$O abundance is $<10^{-5}$, indicating that the
absence of H$_2$O is not simply due to a lower envelope mass, but is
likely caused by photodissociation of H$_2$O to OH in the more
diffuse, lower density Class I envelopes. Also, the longer timescale
can result in H$_2$O freeze-out onto grain mantles after $\sim 10^5$
yr (Bergin et al.\ 1998).

Two scenarios have been put forward for the origin of the molecular
emission from Class 0 objects. Ceccarelli et al.\ (1999, 2000) and
Maret et al.\ (2002) favor a model of a collapsing envelope in which
the lower-lying lines originate from the outer envelope and the
higher-lying lines from the inner `hot core' where ices evaporate (see
Figure 6). In contrast, Giannini et al.\ (2001), Benedettini et al.\
(2002) and Nisini et al.\ (2000, 2002) argue that most of the emission
arises in the $J-$ and $C-$ shocks associated with the outflow
impacting the envelope, much like discussed in \S 8. Arguments in
favor of the latter explanation include (i) the fact that in some
cases the lines have similar strengths at the source and off-source
outflow positions; (ii) the broad $\sim$20 km s$^{-1}$ H$_2$O line
profiles seen by SWAS; and (iii) the linear relation between total
line luminosity and the kinetic luminosity of the outflow derived from
mapping of CO millimeter lines. The correlation with other molecular
shock tracers such as SiO is less clear.

It should be kept in mind, however, that the observed amount of warm
gas is typically only 0.01 M$_{\odot}$ (a few \% of the total envelope
mass) and that evidence for warm inner envelopes and ice evaporation
is clearly established from other data (see van Dishoeck 2003 for
review). Larsson et al.\ (2002) argue that such models cannot
reproduce the CO emission for one source, but the case for H$_2$O is
less clear. For example, Maret et al.\ (2002) show that for NGC 1333
IRAS4 ---a source where the molecular emission peaks at the YSO---,
the H$_2$O lines can be well explained with an abundance of $5\times
10^{-7}$ in the outer region and $5\times 10^{-6}$ in the inner
envelope.  For high-mass YSOs, ice evaporation is also known to
contribute to the SWAS and ISO H$_2$O observations (see \S 3.4 and
9.2).  Finally, Ceccarelli et al.\ (2002c) suggest that some of the
high-excitation CO may originate from a super-heated disk in the
embedded phase.  Future high spectral and spatial data with the
Herschel Space Observatory are needed to determine the relative
contributions from each of these scenarios for a large sample of
sources.

{\bf 9.1.2. CAM-CVF and SWS surveys.} As noted in \S 4.2, CAM-CVF
observations, together with a few SWS spectra, reveal the presence of
abundant ices in the cold outer envelope of low-mass YSOs. However,
the lack of spectral resolution prevents the establishment of any
trends based on changing line profiles or gas/solid ratios, such as
observed for high-mass YSOs.  Medium resolution mid-infrared
spectroscopy of low-mass YSOs is an area that is wide open for future
studies.

{\bf 9.1.3. T Tau.} A special case is formed by T Tau, the
prototypical Class II object but whose infrared line spectra are more
characteristic of an embedded Class I object. Indeed, T Tau has one of
the richest SWS (van den Ancker et al.\ 1999) and LWS (Spinoglio et
al.\ 2000) spectra, full of atomic and molecular (H$_2$, CO, H$_2$O
and OH) lines. In addition to shocks, the influence of
UV radiation is clearly seen in the fluorescent H$_2$
component and in the H$_2$O--OH chemistry. T Tau may be peculiar
because it is a close binary system with a dusty envelope seen
face-on, filling a large fraction of the beam.

{\bf 9.1.4. FU Orionis objects:} Another interesting sample is formed
by the FU Orionis objects, studied with the LWS by Lorenzetti et al.\
(2000). These are low luminosity young stars which show a sudden
increase in brightness, thought to be caused by a large increase in
the disk accretion rate. In addition to the usual atomic [O I] and [C
II] fine-structure lines, these objects surprisingly show [N II] 122
$\mu$m emission, not seen toward any other type of YSO. [N II] must
arise from ionized gas, and since no higher ionization stages of other
elements are seen, this gas must have a low ionization fraction and
low density ($n(e) \lapprox 100$ cm$^{-3}$). One possibility is that
it is produced by enhanced UV radiation from the disk-star
boundary layer.

{\parindent 20pt
{\narrower
\topinsert
\centerline{\hbox{
\psfig{figure=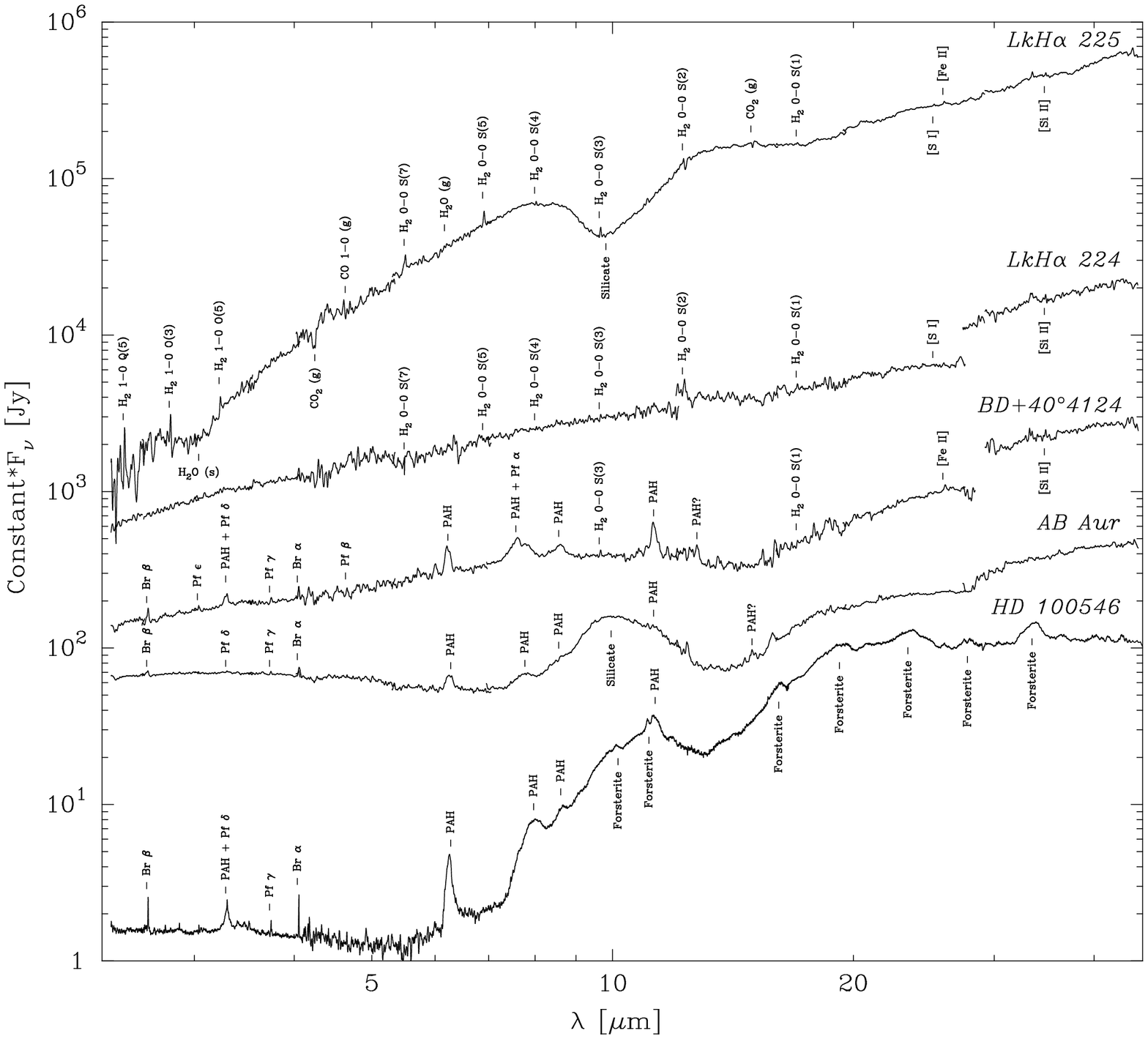,width=14cm,angle=-90}
}}
\noindent
{\bf Figure 17.} SWS spectra of young stars at different stages of
evolution (van den Ancker et al. 2000a,c; Malfait et al. 1998). From
top to bottom --- in a rough evolutionary sequence --- the spectra
change from dominated by solid-state absorption features and gas
(shock) emission lines, to PAH features and PDR lines, to amorphous
and crystalline silicates.
\endinsert
}}

\bigskip
\centerline{\it 9.2. Intermediate and High-Mass YSOs}
\medskip
Several samples of intermediate and high-mass sources have been
studied with ISO, for different purposes. Lorenzetti et al.\ (2002)
summarize the LWS observations on Herbig Ae/Be stars, covering
luminosities from 5 to $10^5$ L$_{\odot}$ (see also Giannini et al.\
1999, Lorenzetti et al.\ 1999). Most of these objects are associated
with diffuse nebulosity and show [C II] and [O I] fine-structure
emission characteristic of PDRs. Some of the extended PAH emission has
been imaged by Siebenmorgen et al.\ (2000).
High-luminosity ($>10^4$ L$_\odot$) deeply embedded sources studied
with the SWS are summarized by Gibb et al.\ (2004), whereas a
catalogue of SWS-LWS spectra of H II regions is presented by Peeters
et al.\ (2002b). LWS spectra of other high-luminosity objects are
summarized by Benedettini et al.\ (2003). A common characteristic of
all these luminous sources is that they show almost no molecular
emission lines, and often just a handful of atomic emission lines,
usually only [O I] 63 $\mu$m and [C II] 158 $\mu$m arising in a PDR
component. If an H II region has started to develop, higher ionization
stages are seen as well and can be used to determine the effective
temperature and spectral type of the embedded source.  Molecular
material, both in the gas and solid state, is however present and is
prominently seen in SWS absorption spectra.

{\bf 9.2.1. From shocks to PDRs.} Two particularly illustrative
examples of the combined SWS and LWS spectral evolution have been
provided by van den Ancker et al.\ (2000b,c). The first concerns the
BD +40$^o$4124 group containing three YSOs with infrared luminosities
of 270--1600 L$_\odot$ (see Figure 17). The LkH$\alpha$ 225 spectrum
is characteristic of a deeply embedded source, with deep silicate and
ice absorptions. LkH$\alpha$ 224 is an example of a new class of
nearly featureless sources, showing no silicate or ice absorption or
emission. The spectrum of BD +40$^o$4124 itself, a B2Ve star, has
strong PAHs and many atomic lines. Van den Ancker et al.\ (2000c) show
that the atomic and H$_2$ lines for the embedded sources originate
from shocks, whereas those for BD +40$^o$4124 are characteristic of a
PDR, with a contribution from the H II region. Thus, as the envelope
disperses, the PDR component starts to dominate compared with the
shock emission.

A similar picture is found from comparing two high luminosity ($2-4
\times 10^4$ L$_{\odot}$) objects, Cep A and S 106 (van den Ancker et
al.\ 2000b). The Cep A spectrum shows deep ice and silicate
absorptions, but no PAH emission nor ionized atomic lines, indicating
that it has not yet started to develop its H II region. In contrast,
the S 106 spectrum is dominated by PAHs and by atomic lines from a
wide range of ionization stages characteristic of a young O8 star. The
atomic and molecular H$_2$ and CO emission lines indicate shock
excitation for Cep A and PDR excitation for S 106. The situation for
other high-mass YSOs is often less clear-cut. For example, the Orion SWS
spectra reveal a mixture of many physical components in the beam,
including the foreground H II region, the PDR, the powerful outflow and
the embedded YSO (van Dishoeck et al.\ 1998).

{\bf 9.2.2. Evolution within the deeply embedded stage.} While the
above characteristics of the deeply embedded versus the more evolved
PDR/H II region stages are readily recognized, more subtle
evolutionary effects {\it within} the deeply embedded stage can be
derived from careful analysis of the ice and gas-phase absorption data
toward high-mass protostars.  These are very young objects, before the
onset of the ultracompact H II region or `hot core' stage. As
discussed in \S 4.3, the profiles of several ice bands show strong
evidence of progressive heating of the envelope. Similarly, the
gas/solid ratios of species like H$_2$O increase systematically for
the same sources, as does the color temperature of the cold dust.  The
most likely scenario, put forward by van der Tak et al.\ (2000b), is
that the warmer sources have a lower ratio of envelope mass $M_{\rm
env}$ compared with the bolometric luminosity of the source $L_{\rm
bol}$, so that the overall temperatures are lower. If the lower
$M_{\rm env}$ are due to the gradual dispersion of the envelope rather
than to initial conditions, the sources can be put in a time sequence with
the warmer sources being more evolved, thus providing a powerful
diagnostic of the first few $\times 10^4$ yr of protostellar
evolution.

Why do high-mass sources not show more prominent molecular emission
lines in the SWS and LWS spectra, like their low-mass counterparts?
The main reason must be beam dilution, since the high-mass sources
often have ten times larger distances.  Detailed modeling shows that
the `hot core' region where the temperature is above that for ice
evaporation of 90~K is typically 1000 AU, less than 1$''$ at $>$1
kpc. Apparently, shocks have an equally low filling factor, as
evidenced from the absence of H$_2$ or [S I] emission lines. Moreover,
the far-infrared continuum of high-mass YSOs is so strong that weak
emission lines are more difficult to detect at low spectral
resolution.

The case of H$_2$O is instructive of the abundance and excitation
variations within high-mass envelopes. Boonman et al.\ (2003a)
analyzed all available H$_2$O SWS, LWS and SWAS data in terms of
possible abundance profiles. The SWS data indicate ice
evaporation in the inner `hot core' with abundances of $>10^{-4}$,
whereas the absence of LWS lines and the weak SWAS emission require
significant H$_2$O freeze-out in the outer cold envelope with
abundances $<10^{-8}$ at $T<90$~K (see also Snell et al.\ 2000). The
SWAS profiles consist of narrow and broad components, with the
outflow contribution at most 50\%. In some cases, only a narrow
($<$few km s$^{-1}$) profile is seen with a line strength consistent
with optically thick H$_2$O emission at $\sim 100$~K arising from a
$\sim 1''$ source. This emphasizes the need for future high spatial
and spectral resolution H$_2$O observations to properly disentangle
the shock, envelope and `hot core' components.

\bigskip
\medskip
\centerline{\bf 10. CIRCUMSTELLAR DISKS AROUND YOUNG STARS}
\bigskip
\centerline{\it 10.1. Dust Evolution and Disk Structure}
\medskip
As discussed in \S 5 and 6, the mid- and far-infrared spectra of disks
around young stars exhibit a wide variety of features due to
crystalline silicates, PAHs and ices (Meeus et al.\ 2001, Bouwman et
al.\ 2001). Although no trend with age within the pre-main sequence
stage has been found, there are clear indications that the dust
evolves from the diffuse cloud to the disk phase. Analysis of the 10
$\mu$m silicate band shows two effects. The first is coagulation of
0.1 $\mu$m to micron-sized grains. Interestingly, small PAHs remain
present during this process, although their spectral characteristics
are clearly different from those of molecular clouds (Peeters et al.\
2002a). The second effect is crystallization, resulting in an
increasing fraction of emission from forsterite and other minerals
when going from the Herbig Ae stars to a debris disk like $\beta$
Pictoris and comets like Halley and Hale-Bopp (Figure 12).  The
near-complete absence of crystalline silicates in interstellar clouds
indicates that the crystallization must occur in the disks
themselves. The correlation of forsterite with silica suggests that
thermal annealing is the dominant process, but this
has to occur at temperatures greater than $\sim$1100 K for the
timescale to be shorter than the disk lifetime. Since the crystalline
material is observed at lower temperatures further out in the disk,
this implies that some radial mixing process must occur (see also
Bockel\'ee-Morvan et al. 2002), although the mechanisms are not yet
well understood.  The data indicate that the timescale for annealing
is longer than that of coagulation. The origin of crystalline water in
cold disks like that around HD 142527 (Figure 11) is also unclear.

The mid infrared spectra, combined with their SEDs, also provide
insight into the physical structure of disks. Natta et al.\ (2000)
noted that the silicate emission seen in the PHOT-S spectra of T Tauri
stars requires the presence of a disk `atmosphere', in which the disk
surface is heated from the top by radiation from the star (Calvet et
al.\ 1992, Chiang \& Goldreich 1997, Chiang et al.\ 2001).  Meeus et
al.\ (2001) divided the isolated Herbig Ae star ISO spectra into two
groups: those with a rising SED (Group I)
and those with a flat continuum (Group II). The SEDs of the latter
group can be fitted with a power-law provided by an optically thick
but geometrically thin disk, whereas the former requires addition of a
black-body component with 100--200~K provided by a disk atmosphere in
the inner $\sim 10$ AU (see Figure 10). 

Natta et al.\ (2001) and Dullemond et al.\ (2001) (see also Dominik et
al.\ 2003) propose a slight modification, in which the inner disk is
truncated at the dust sublimation temperature of $\sim$1400~K at
$\sim$0.5 AU, resulting in a sharp rim exposed to intense radiation
and producing most of the near-infrared radiation. Due to the higher
temperatures, the rim will have an increased scale height (`puffed
up'), shielding the outer parts of the disk and resulting in lower
temperatures in shadowed regions.  Group I sources have a flaring
outer disk directly exposed to the stellar radiation, where the
silicate and PAH emission is thought to arise. In Group II sources,
the disk may be so small that there is no flaring
region. Alternatively, grain growth and/or dust settling may cause the
disk to become self-shadowing (Dullemond \& Dominik 2004).  The
precise origin of the silicate emission in Group II sources is not yet
fully understood. The presence of an additional spherical (remnant)
envelope cannot be excluded for some cases (e.g., Miroshnichenko et
al.\ 1999), and spatially resolved data will be essential to
disentangle the different components.

The disk around HD 100546 is clearly different from that around other
stars: nearly 70\% of the absorbed and re-emitted starlight of HD
100546 emerges at mid- rather than near-infrared wavelengths,
indicating that the disk must intercept the radiation at larger
distances, around 10 AU. Another peculiarity of this object is that
the crystalline silicates have a lower temperature than the amorphous
material and an increasing abundance with radial distance.  One
possibility is that a proto-Jovian planet has cleared out a gap around
10 AU, leading to a `puffed-up' rim on the far-side of the gap
(Bouwman et al.\ 2003).  The wealth of crystalline silicates could
have been produced in a collisional cascade of asteroidal-sized
objects stirred up by the proto-Jupiter, or by shocks caused by tidal
interactions of the planet with the disk producing crystalline
material through flash heating.  Some of this material could have been
lifted and scattered to the disk surface and driven out by radiation
pressure to larger distances High-spatial resolution data are needed
to confirm this scenario, but it illustrates the potential of
mid-infrared spectroscopy to trace planet-formation processes.

The similarity of the HD 100546 and comet Hale-Bopp spectra (Crovisier
et al.\ 1997) has often been cited as evidence of a strong
interstellar medium-solar system connection, although the scenario
described above is quite specific for one object. Additional evidence
for such a connection has been derived from the fact that the ice
abundances found in protostellar regions (see \S 4.2) are comparable
to those observed in cometary comae (e.g., Ehrenfreund et al.\ 1997b,
Bockel\'ee-Morvan et al.\ 2000). This ice similarity is strong
evidence that some fraction of the interstellar material must have
been included largely unaltered, but the origin of the crystalline
silicates in comets is still a mystery in this context.

\bigskip
\centerline{\it 10.2. Gas in Disks}
\medskip
An equally important issue is the timescale over which the gas
disappears from the disk, since this directly affects the formation of
gaseous giant planets. Ground-based millimeter CO surveys are hampered
by freeze-out in the cold parts of the disk and
photodissociation in the upper layers (e.g., Aikawa et al.\ 2002),
whereas infrared CO lines probe only a small fraction of warm gas
(e.g., Najita et al.\ 2003). The lowest H$_2$ pure-rotational lines
provide a powerful alternative gas probe because it is the dominant
constituent of the gas, does not freeze-out onto grains, self-shields
effectively against photodissociation, and has optically thin
lines. The main disadvantage is that the pure rotational lines are
intrinsically weak and difficult to detect superposed on a strong
continuum at low spectral resolution. Also, the gas temperature needs
to be sufficiently high, $>$80 K.

Deep searches for the H$_2$ S(0) and S(1) lines at 28 and 17 $\mu$m
have been performed with the SWS toward a dozen young stars (Thi et
al.\ 2001a,b).  These sources were chosen to be isolated objects to
minimize contamination from surrounding cloud material. Detections of
lines have been claimed in several sources, including potentially some
older Vega-type objects, but confirmation with higher spatial and
spectral resolution data is needed (see Richter et al.\ 2002).  Since
the mid-infrared H$_2$ lines are sensitive to a fraction of $M_{\rm
Jup}$ of warm gas, they are well worth pursuing with future
instrumentation. Other gas tracers such as the [C II] 157.7 $\mu$m line
have been considered as well (Kamp et al.\ 2003).

\bigskip
\medskip
\centerline{\bf 11. CONCLUDING REMARKS}
\medskip
ISO has clearly demonstrated the power of full mid- and far-infrared
spectra to probe a wide variety of processes associated with star- and
planet formation, providing insight that cannot be obtained from
imaging data alone.  Using ISO, the inventory of ices has been
completed, a census of the PAHs features has been obtained, a new
diagnostic of warm gas --H$_2$-- has been opened up, key gas-phase
molecules (H$_2$O, CO$_2$) have been widely observed, and a wealth of
crystalline silicate features has been revealed. More importantly, ISO
has shown systematic variations in the features during the
star-formation process, providing a spectral classification scheme of
the earliest embedded stages which are difficult to probe otherwise
and stimulating new models. Related variations in the composition of
the material when it is cycled from one phase to another are also
found: freeze-out and evaporation of ices, ionization and
re-hydrogenation of PAHs, and amorphization, crystallization and
coagulation of silicates. At all stages, specific lines and line
ratios provide powerful diagnostics of temperatures and densities, as
well as the relative importance of shock- and photon heating of the
gas.  Since many of these same features and processes are also
ubiquitous in extragalactic sources out to the highest redshifts, a
good understanding of their diagnostic capabilities locally is
essential. The ISO archive will remain a valuable ---and in many ways
unique--- resource for future studies.

At the time of writing, the first results from the Spitzer Space
Telescope are just emerging. With its increased sensitivity up to two
orders of magnitude, Spitzer will extend mid- and far-infrared imaging
and spectroscopy into new regimes that ISO could not probe, such as
medium resolution spectroscopy of embedded low-mass protostars, of
disks around young stars down to the brown-dwarf regime, of debris
disks, and of H$_2$ and other lines in the most diffuse PDRs and
weakest shocks.  The ISO spectra summarized here form a guideline to
the interpretation of the Spitzer data.  Spitzer's spectroscopic
capabilities are limited, however, both in spectral resolution and
wavelength coverage. A combination of several future missions, in
particular the Stratospheric Observatory for Infrared Astronomy
(SOFIA), ASTRO-F, the Herschel Space Observatory, and the James Webb
Space Telescope (JWST), will be needed to obtain spectra over the full
2.4--200 $\mu$m range covered by ISO.  The highest spectral and
spatial resolution data will be provided at mid-infrared wavelengths
by large optical ground-based telescopes, and at submillimeter
wavelengths by the Atacama Large Millimeter Array (ALMA) and other
millimeter arrays.  ISO has paved the way for the exciting science to
be performed with these new facilities.

\bigskip
\medskip
\vfill \eject
\centerline{\bf Acknowledgments}
\medskip
The spectra presented here are a testimony to the ingenuity of the
instrument builders and their teams, both hardware and software.  I am
grateful to many colleagues for communicating results and figures
prior to publication and for countless stimulating discussions, making
ISO such a fun project. The writing of this review was supported by a
Spinoza grant from the Netherlands Organization for Scientific
Research (NWO) and by the Moore's Scholar Program of the California
Institute of Technology.

\bigskip
\medskip

%\vfill \eject

\centerline{\bf REFERENCES}
\bigskip
\parskip 5 pt

Aannestad PA, Emery RJ. 2001. {\it Astron. Astrophys.} 376:1040-53

Aannestad PA, Emery RJ. 2003. {\it Astron. Astrophys.} 406:155-64

Abergel A, Bernard JP, Boulanger F, Cesarsky C, Desert FX, et al.
1996. {\it Astron. Astrophys.} 315:L329-33

Abergel A, Teyssier D, Bernard JP, Boulanger F, Coulais A, et al.
2003. {\it Astron. Astrophys.} 410:577-85

Abergel A, Bernard JP, Boulanger F, Cesarsky D, Falgarone E, et al. 2002.
{\it Astron. Astrophys.} 389:239-51

Aikawa Y., van Zadelhoff GJ, van Dishoeck EF, Herbst E. 2002. {\it Astron.
Astrophys.} 386:622-32

Alexander RD, Casali MM, Andr\'e P, Persi P, Eiroa C. 2003.
{\it Astron. Astrophys.} 401:613-24

Allamandola LJ, Hudgins DM, Sandford SA. 1999. {\it Ap. J. Lett.}
511:L115-19

Allamandola LJ, Tielens AGGM, Barker JR. 1985. {\it Ap. J.} 290:L25-8

Andr\'e P, Ward-Thompson D, Barsony M. 2000. In {\it Protostars \& Planets
IV}, ed. V Mannings, AP Boss, SS Russell, pp. 59-96. Tucson: Univ. Arizona

Bacmann A, Andr\'e P, Puget JL, Abergel A, Bontemps S, Ward-Thomson D.
2000. {\it Astron. Astrophys.} 361:555-80

Bakes ELO, Tielens AGGM. 1994. {\it Ap. J.} 427:822-38

Bakes ELO, Tielens AGGM, Bauschlicher CW 2001. {\it Ap. J.} 556:501-14

Baluteau JP, Cox P, Cernicharo J, P\'equinot D, Caux E, et al. 1997.
{\it Astron. Astrophys.} 322:L33-6

Benedettini M, Pezzuto S, Spinoglio L, Saraceno P, Di Giorgio
AM. 2004.  In {\it Recent Research Development in Molecular and
Cellular Astronomy \& Astrophysics}, ed. xxx, in press (astro-ph/0310086)

Benedettini M, Viti S, Giannini T, Nisini B, Goldsmith PF, Saraceno P.
2002. {\it Astron. Astrophys.} 395:657-62

Bergin EA, Neufeld DA, Melnick GJ. 1998. {\it Ap. J.} 499:777-792

Bergin EA, Kaufman MJ, Melnick GJ, Snell RL, Howe JE. 2003.
{\it Ap. J.} 582:830-45

Bertoldi F. 1997. See Heras et al. 1997. pp. 67-72

Bertoldi F, Timmermann R, Rosenthal D, Drapatz S, Wright CM. 1999.
{\it Astron. Astrophys.} 346:267-77

Black JH, van Dishoeck EF. 1987. {\it Ap. J.} 322:412-49

Bockel\'ee-Morvan D, Gautier D, Hersant F, Hur\'e JM, Robert F. 2002.
{\it Astron. Astrophys.} 384-1107

Bockel\'ee-Morvan D, Lis DC, Wink JE, Despois D, Crovisier J, et al.\ 2000.
{\it Astron. Astrophys.} 353:1101-14

Boogert ACA, Ehrenfreund P. 2004. In {\it The Astrophysics of Dust},
ed. in press, San Francisco: Astron. Soc. Pac.

Boogert ACA, Ehrenfreund P, Gerakines P, Tielens AGGM, Whittet DCB, et al.
2000a. {\it Astron. Astrophys.}  353:349-62

Boogert ACA, Helmich FP, van Dishoeck EF, Schutte WA, Tielens AGGM, Whittet
DCB. 1998. {\it Astron. Astrophys.} 336: 352-58

Boogert ACA, Hogerheijde MR, Ceccarelli C, Tielens AGGM, van Dishoeck EF,
et al. 2002. {\it Ap. J.} 570:708-23

Boogert ACA, Tielens AGGM, Ceccarelli C, Boonman AMS, van Dishoeck EF, et al.
2000b. {\it Astron. Astrophys.} 360:683-98

Boonman AMS, van Dishoeck EF. 2003. {\it Astron. Astrophys.} 403:1003-10

Boonman AMS, Doty SD, van Dishoeck EF, Bergin EA, Melnick GJ, Wright CM,
Stark R. 2003a. {\it Astron. Astrophys.} 406:937-55

Boonman AMS, Stark R, van der Tak FFS, van Dishoeck EF, van der Wal PB, et al.
2001. {\it Ap. J. Lett.} 553:L63-6

Boonman AMS, van Dishoeck EF, Lahuis F, Doty SD. 2003b.
{\it Astron. Astrophys.} 399:1063-72

Boonman AMS, van Dishoeck EF, Lahuis F, Doty SD, Wright CM, Rosenthal D.
2003c. {\it Astron. Astrophys.} 399:1047-61

Boudin, N., Schutte WA, Greenberg JM. 1998. {\it Astron.
Astrophys.} 331:749-59

Boulanger F, Abergel A, Cesarsky D, Bernard JP, Miville-Desch\^enes
MA, Verstraete L, Reach, WT.  2000, in {\it ISO Beyond Point Sources:
Studies of Extended Infrared Emission}, eds. RJ Laureijs, K Leech, MF
Kessler, ESA-SP 455, p. 91

Boulanger F, Boissel P, Cesarsky D, Ryter C. 1998.
{\it Astron. Astrophys.} 339:194-200

Boulanger F, Falgarone E, Puget JL, Helou G. 1990. {\it Ap. J.} 364:136-45

Bouwman J, de Koter A, Dominik C, Waters LBFM. 2003.
{\it Astron. Astrophys.} 401:577-92

Bouwman J, de Koter A, van den Ancker ME, Waters LBFM. 2000.
{\it Astron. Astrophys.} 360:213-26

Bouwman J, Meeus G, de Koter A, Hony S, Dominik C, Waters LBFM. 2001.
{\it Astron. Astrophys.} 375:950-62

Bowey JE, Adamson AJ. 2002. {\it MNRAS} 334:94-106

Brooks KJ, Cox P, Schneider N, Storey JWV, Poglitsch A, Geis N, Bronfman L.
2004. {\it Astron. Astrophys.} in press

Cabrit S, Bontemps S, Lagage PO, Sauvage M, Boulanger F, et al. 1999.
See Cox \& Kessler 1999, pp. 449-52

Cabrit S, Lefloch B, Cernicharo J, Pineau des For\^ets G, Le Bourlot J, Flower
D. 2004. in {\it The Dense Interstellar Medium in Galaxies}, ed. S Pfalzner,
C Kramer, C Straubmeier, A Heithausen.
in press. Berlin: Springer

Calvet N, Magris GC, Patino A, d'Alessio P. 1992.
{\it Rev. Mex. Astron. Astrofis.} 24:27-42

Caux E, Ceccarelli C, Castets A, Vastel C, Liseau R et al. 1999.
{\it Astron. Astrophys.} 347:L1-4

Caux E, Ceccarelli C, Pagani L, Maret S, Castets A, Pardo JR. 2002.
{\it Astron. Astrophys.} 383:L9-13

Cazaux S, Tielens AGGM. 2002. {\it Ap. J. Lett.} 577:L29-32

Ceccarelli C, Baluteau JP, Walmsley M, Swinyard BM, Caux E, et
al. 2002a.  {\it Astron. Astrophys.} 383:603-13

Ceccarelli C, Boogert ACA, Tielens AGGM, Caux E, Hogerheijde MR, Parise B.
2002b. {\it Astron. Astrophys.} 395:863-71

Ceccarelli C, Castets A, Caux D, Loinard L, Molinari S, Tielens
AGGM. 2000.  {\it Astron. Astrophys.} 355:1129-37

Ceccarelli C, Caux E, Loinard L, Castets A, Tielens AGGM, et al. 1999.
{\it Astron. Astrophys.} 342: L21-4

Ceccarelli C, Caux E, Tielens AGGM, Kemper F, Waters LBFM, Phillips T.
2002c. {\it Astron. Astrophys.} 395:L29-33

Ceccarelli C, Caux E, White GJ, Molinari S, Furniss I, et al. 1998a.
{\it Astron. Astrophys.} 331:372-82

Ceccarelli C, Caux E, Wolfire M, Rudolph A, Nisini B, Saraceno P,
White GJ. 1998b. {\it Astron. Astrophys.} 331:L17-20

Cernicharo J, Goicoechea JR, Benilan Y. 2002. {\it Ap. J. Lett.}
580:L157-60

Cernicharo J, Goicoechea JR, Caux E. 2000a. {\it Ap. J. Lett.} 534:L199-02

Cernicharo J, Heras AM, Tielens AGGM, Pardo JR, Herpin F, Gu\'elin M, Waters
LBFM. 2001a. {\it Ap. J. Lett.} 546:L123-26

Cernicharo J, Heras AM, Pardo JR, Tielens AGGM, Gu\'elin M, et al. 
2001b. {\it Ap. J. Lett.} 546:L127-30

Cernicharo J, Noriega-Crespo A, Cesarsky D, Lefloch B, Gonz\'alez-Alfonso E, 
et al. 2000b. {\it Science} 288:649-52

Cernicharo J, Lim T, Cox P, Gonz\'alez-Alfonso E, Caux E, et al. 1997.
{\it Astron. Astrophys.} 323:L25-8

Cesarsky C, Abergel A, Agn\`ese P, Altieri B, Augu\`eres JL, et al.
1996. {\it Astron. Astrophys.} 315:L32-7

Cesarsky D, Cox P, Pineau des For\^ets G, van Dishoeck EF, Boulanger F,
Wright CM. 1999. {\it Astron. Astrophys.} 348:945-49

Cesarsky D, Jones AP, Lequeux J, Verstraete L. 2000a.
{\it Astron. Astrophys.} 358:708-16

Cesarsky D, Lequeux J, Ryter C, G\'erin M. 2000b. {\it Astron. Astrophys.}
354:L87-91

Chiang EI, Goldreich P. 1997. {\it Ap. J.} 490, 368-76

Chiang EI, Joung MK, Creech-Eakman MJ, Qi C, Kessler JE, Blake GA, van
Dishoeck EF. 2001. {\it Ap. J.} 547:1077-89

Chiar JE, Tielens AGGM, Whittet DCB, Schutte WA, Boogert ACA, et al.
2000. {\it Ap. J.} 537:749-62

Chi\`eze JP, Pineau des For\^ets G, Flower DR. 1998. {\it MNRAS} 295:672-682

Churchwell E. 2002. {\it Annu. Rev. Astron. Astrophys.} 40:27-62

Clegg PE, Ade PAR, Armand C, Baluteau JP, Barlow MJ, et al. 1996.
{\it Astron. Astrophys.} 315:L38-42

Cox P, Kessler MF. 1999. {\it The Universe as seen by ISO}
ESA SP-427. Noordwijk: ESTEC. 1090 pp.

Creech-Eakman MJ, Chiang EI, Joung RMK, Blake GA, van Dishoeck EF. 2002.
{\it Astron. Astrophys.} 385:546-62

Crovisier J, Leech K, Bockel\'ee-Morvan D, Brooke TY, Hanner MS, et al.
1997. {\it Science} 275:1904-7

Dartois E, Cox P, Roelfsema PR, Jones AP, Tielens AGGM, et al. 1998a.
{\it Astron. Astrophys.} 338:L21-24

Dartois E, Demyk K, d'Hendecourt L, Ehrenfreund P. 1999.
{\it Astron. Astrophys.} 351:1066-74

Dartois E, d'Hendecourt L. 2001. {\it Astron.
Astrophys.} 365:144-56

Dartois E, d'Hendecourt L, Boulanger F, Jourdain de Muizon M, Breitfellner M,
Puget JL, Habing HJ. 1998b. {\it Astron. Astrophys.} 331:651-60

Dartois E, Thi WF, Geballe TR, Deboffle D, d'Hendecourt L, van
Dishoeck E.  2003. {\it Astron.  Astrophys.} 399:1009-20

de Graauw Th, Haser LN, Beintema DA, Roelfsema PR, van Agthoven H, et
al. 1996a. {\it Astron. Astrophys.} 315:L49-54

de Graauw Th, Whittet DCB, Gerakines PA, Bauer O, Beintema DA, et al.
1996b. {\it Astron. Astrophys.}  315:L345-48

Demyk K, Dartois E, d'Hendecourt L, Jourdain de Muizon M, Heras AM,
Breitfellner M. 1998. {\it Astron. Astrophys.} 339:553-60

Demyk K, Jones AP, Dartois E, Cox P, d'Hendecourt L. 1999.
{\it Astron. Astrophys.} 349:267-75

d'Hendecourt L, Joblin C, Jones A. 1999, ed. {\it Solid Interstellar Matter:
The ISO Revolution}. Berlin: Springer. 315 pp.

Dominik C, Dullemond CP, Waters LBFM, Walch S. 2003.
{\it Astron. Astrophys.} 398:607-19

Draine BT. 1978. {\it Ap. J. Suppl.} 36:595-619

Draine BT. 2003. {\it Annu. Rev. Astron. Astrophys.} 41:241-89

Draine BT, Bertoldi F. 1996. {\it Ap. J.} 468:269-89

Draine BT, Bertoldi F. 1999. See Cox \& Kessler 1999, pp. 353-59

Draine BT, McKee CF. 1993. {\it Annu. Rev. Astron. Astrophys.} 31:373-432

Duley WW. 1996. {\it MNRAS} 279:591-4

Duley WW, Williams DA. 1981. {\it MNRAS} 196:269-74

Dullemond CP, Dominik C. 2004. {\it Astron. Astrophys.} in press

Dullemond CP, Dominik C, Natta A. 2001. {\it Ap. J.} 560:957-69

Ehrenfreund P, Boogert ACA, Gerakines PA, Jansen DJ, Schutte WA,
Tielens AGGM, van Dishoeck EF. 1996. {\it Astron. Astrophys.}
315:L341-44

Ehrenfreund P, Boogert ACA, Gerakines PA, Tielens AGGM, van Dishoeck EF.
1997a. {\it Astron. Astrophys.} 328:649-69

Ehrenfreund P, Dartois E, Demyk K, d'Hendecourt L. 1998.
{\it Astron. Astrophys.}  339:L17-20

Ehrenfreund P, d'Hendecourt L, Dartois E, Jourdain de Muizon M,
Breitfellner M, et al. 1997b. {\it Icarus} 130:1-15

Ehrenfreund P, Schutte WA. 2000. In {\it IAU Symp. 197, Astrochemistry},
ed. YC Minh, EF van Dishoeck, pp. 135-46. San Francisco: Astron. Soc. Pac.

Evans NJ, Lacy JH, Carr JS. 1991. {\it Ap. J.} 383:674-92

Feldt M, Henning Th, Lagage PO, Manske V, Schreyer K, Stecklum B. 1998.
{\it Astron. Astrophys.} 332:849-56

Feuchtgruber H, Helmich FP, van Dishoeck EF, Wright CM. 2000. {\it
Ap. J. Lett.}  535:L111-14 

Froebrich D, Smith MD, Eisl\"offel J. 2002. {\it Astron. Astrophys.}
385:239-56

Fuente A, Mart{\'\i}n-Pintado J, Rodr{\'\i}guez-Fern\'andez NJ,
Rodr{\'\i}guez-Franco A, de Vicente P, Kunze D. 1999. {\it Ap. J. Lett.}
518:L45-8

Fuente A, Mart{\'\i}n-Pintado J, Rodr{\'\i}guez-Fern\'andez NJ,
Cernicharo J, Gerin M. 2000. {\it Astron. Astrophys.} 354:1053-61

Genzel R, Cesarsky C. 2000. {\it Annu. Rev. Astron. Astrophys.} 38:761-814

Gerakines PA, Schutte WA, Greenberg JM, van Dishoeck EF. 1995. 
{\it Astron. Astrophys.} 296:810-18

Gerakines PA, Schutte WA, Ehrenfreund P. 1996. 
{\it Astron. Astrophys.} 312:289-305

Gerakines PA, Whittet DCB, Ehrenfreund P, Boogert ACA, Tielens AGGM, et al.
1999. {\it Ap. J.} 522:357-77

Giannini T, Lorenzetti D, Tommasi E, Nisini B, Benedettini M, et al.
1999. {\it Astron. Astrophys.} 346:617-25

Giannini T, Nisini B, Lorenzetti D. 2001. {\it Ap. J.} 555:40-57

Giannini T, Nisini B, Lorenzetti D, Di Giorgio AM, Spinoglio L, et al.
2000. {\it Astron. Astrophys.} 358:310-20

Gibb EL, Whittet DCB. 2002. {\it Ap. J. Lett.} 566:L113-16

Gibb EL, Whittet DCB, Boogert ACA, Tielens AGGM. 2004. {\it Ap. J.}
in press

Gillett FC, Forrest WJ. 1973. {\it Ap. J.} 179:483-91

Goicoechea JR, Cernicharo J. 2001a. {\it Ap. J. Lett.} 554:L213-6

Goicoechea JR, Cernicharo J. 2001b. 
In {\it The Promise of the Herschel Space Observatory}, ed. GL Pilbratt et al.
ESA-SP 460, pp. 413-6. Noordwijk:ESTEC

Goicoechea JR, Cernicharo J. 2002. {\it Ap. J. Lett.} 576:L77-81

Goldsmith PF, Melnick GJ, Bergin EA, Howe JE, Snell RL, et al.
2000. {\it Ap. J. Lett.} 539:L123-27 

Gonz\'alez-Alfonso E, Wright CM, Cernicharo J, Rosenthal D, Boonman AMS,
van Dishoeck EF. 2002. {\it Astron. Astrophys.} 386:1074-1102

Gonz\'alez-Alfonso E, Cernicharo J, van Dishoeck EF, Wright CM, Heras A. 1998.
{\it Ap. J. Lett.} 502:L169-72

Greenberg JM, Li A, Mendoza-Gomez CX, Schutte WA, Gerakines PA, de Groot M.
1995. {\it Ap. J. Lett.} 455:L177-80

Gry C, Boulanger F, Nehm\'e C, Pineau des For\^ets G, Habart E, Falgarone E.
2002. {\it Astron. Astrophys.} 391:675-80

G\"urtler J, Henning Th, K\"ompe C, Pfau W, Kr\"atschmer W, Lemke D.
1996. {\it Astron. Astrophys.} 315:L189-92

G\"urtler J, Schreyer K, Henning Th, Lemke D, Pfau W. 1999.
{\it Astron. Astrophys.} 346:205-10

G\"urtler J, Klaas U, Henning Th, \'Abrah\'am P, Lemke D, Schreyer K,
Lehmann K. 2002. {\it Astron. Astrophys.} 390:1075-87

Haas MR, Davidson JA, Erickson EF, eds. 1995. {\it Airborne
Astronomy Symposium on the Galactic Ecosystem: From Gas to Stars to Dust},
Vol. 73. San Francisco: Astron. Soc. Pac. xxx pp.

Habart E, Boulanger F, Verstraete L, Pineau des For\^ets G, Falgarone E,
Abergel A. 2003. {\it Astron. Astrophys.} 397:623-34

Habart E, Verstraete L, Boulanger F, Pineau des For\^ets G, Le Peintre F,
Bernard JP. 2001. {\it Astron. Astrophys.} 373:702-13

Habart E, Boulanger F, Verstraete L, Walmsley CM, Pineau des For\^ets G.
2004. {\it Astron. Astrophys.} submitted

Habing HJ. 1968. {\it Bull. Astron. Neth.} 19:421-32

Harwit M, Neufeld DA, Melnick GJ, Kaufman MJ. 1998. {\it Ap. J.
Lett.} 497:L105-8

Hasegawa T, Gatley I, Garden RP, Brand PWJL, Ohishi M, Hayashi M, Kaifu N.
1987. {\it Ap. J.} 318:L77-80

Helmich FP, van Dishoeck EF, Black JH, de Graauw T, Beintema DA, et al.
1996. {\it Astron. Astrophys.} 315:L173-6

Henning, Th. ed. 2003. {\it Astromineralogy}. Berlin: Springer. pp. 1-281

Heras AM, Leech K, Trams NR, Perry M, eds. 1997. {\it First ISO Workshop
on Analytical Spectroscopy}, ESA SP-419. Noordwijk: ESA-ESTEC. 319 pp.

Hollenbach D. 1997. In {\it Herbig-Haro Flows and the Birth of Stars},
IAU Symposium 182, ed. B Reipurth, C Bertout, pp. 182-98. Dordrecht:
Kluwer

Hollenbach D, Chernoff DF, McKee CF. 1989. In {\it Infrared Spectroscopy
in Astronomy}, ESA SP-290, pp. 245-58. Noordwijk: ESTEC

Hollenbach D, Salpeter EE. 1971. {\it Ap. J.} 163:155-64

Hollenbach DJ, Tielens AGGM. 1997. {\it Annu. Rev. Astron. Astrophys.}
35:179-216

Hollenbach DJ, Tielens AGGM. 1999. {\it Rev. Mod. Phys.} 71:173-230

Honda M., Kataza H, Okamoto YK, Miyata T, Yamashita T, et al. 2003.
{\it Ap. J. Lett.} 585:L59-62

Hony S, van Kerckhoven C., Peeters E, Tielens AGGM, Hudgins DM,
Allamandola LJ. 2001. {\it Astron. Astrophys.} 370:1030-43

Hudgins DM, Allamandola LJ. 1999. {\it Ap. J. Lett.} 516:L41-4

Hudgins DM, Sandford SA, Allamandola LJ, Tielens AGGM. 1993.
{\it Ap. J. Suppl.} 86:713-870

Jones AP, Frey V, Verstraete L, Cox P, Demyk K. 1999. See Cox et al. 1999,
pp. 679-82

Jura M. 1975. {\it Ap. J.} 197:575-80

Kahanp\"a\"a J, Mattila K, Lehtinen K, Leinert C, Lemke D. 2003.
{\it Astron. Astrophys.} 405:999-1012

Kamp I, van Zadelhoff GJ, van Dishoeck EF, Stark R. 2003. {\it
Astron. Astrophys.} 397:1129-41

Kaufman MJ, Neufeld DA. 1996. {\it Ap. J.} 456:611-30

Kaufman MJ, Wolfire MG, Hollenbach DJ, Luhman ML. 1999.
{\it Ap. J.} 527:795-813

Keane JV, Boonman AMS, Tielens AGGM, van Dishoeck EF. 2001a.
{\it Astron. Astrophys.} 376:L5-8

Keane JV, Tielens AGGM, Boogert ACA, Schutte WA, Whittet DCB. 2001b.
{\it Astron. Astrophys.} 376:254-70

Keller LP, Hony S, Bradley JP, Molster FJ, Waters LBFM, et al.
2002. {\it Nature} 417:148-50

Kemper F, J\"ager C, Waters LBFM, Henning Th, Molster FJ, Barlow MJ,
Lim T, de Koter A. 2002. {\it Nature} 415:295-7

Kemper F, Spaans M, Jansen DJ, Hogerheijde MR, van Dishoeck EF,
Tielens AGGM. 1999. {\it Ap. J.} 515:649-56

Kemper F, Vriend WJ, Tielens AGGM. 2004. {\it Ap. J.} submitted

Kessler MF, M\"uller TG, Leech K, Arviset C, Garc{\'\i}a-Lario P,
et al. 2003. {\it ISO - Mission \& Satellite Overview} ESA-SP 1262.
Noordwijk: ESA-ESTEC. 352 pp.

Kessler MF, Steinz JA, Anderegg ME, Clavel J, Drechsel G, et al. 1996.
{\it Astron. Astrophys.} 315:L27-31

Krause O, Lemke D, T\'oth LV, Klaas U, Haas M, Stickel M, Vavrek R.
2003. {\it Astron. Astrophys.} 398:1007-20

Lada CJ. 1999. In {\it The Origin of Stars and Planetary Systems},
eds. CJ Lada, ND Kylafis, pp. 143-92. Dordrecht: Kluwer

Lahuis F, van Dishoeck EF. 2000. {\it Astron. Astrophys.} 355:699-712

Larsson B, Liseau R, Men'shchikov AB. 2002. {\it Astron. Astrophys.}
386:1055-73

Le Bourlot J, Pineau des For\^ets G, Flower DR, Cabrit S. 2002.
{\it MNRAS} 332:985-93

Lefloch B, Cernicharo J, Rodr{\'\i}guez LF, Miville-Desch\^enes MA,
Cesarsky D, Heras AM. 2002. {\it Ap. J.} 581:335-56

Lefloch B, Cernicharo J, Cabrit S, Noriega-Crespo A, Moro-Mart{\'\i}n,
Cesarsky D. 2003. {\it Ap. J.} 590:L41-4

L\'eger A, Puget JL. 1984. {\it Astron. Astrophys.} 137:L5-8

Lemke D, Klaas U, Abollins J, {\'A}brah\'am P, Acosta-Pulido J, et al.
1996. {\it Astron. Astrophys.} 315:L64-70

Li A, Draine BT. 2002. {\it Ap. J.} 572:232-37

Li W, Evans NJ, Jaffe DT, van Dishoeck EF, Thi WF. 2002. {\it Ap. J.}
568:242-58

Lis DC, Keene J, Phillips TG, Schilke P, Werner MW, Zmuidzinas J. 2001.
{\it Ap. J.} 561:823-29

Liseau R, Ceccarelli C, Larsson B, Nisini B, White GJ, et al. 1996.
{\it Astron. Astrophys.} 315:L181-84

Liseau R, Giannini T, Nisini B, Saraceno P, Spinoglio L, et al. 1997.
In {\it Herbig-Haro Flows and the Birth of Stars}, IAU Symposium 182,
ed. B Reipurth, C Bertout, pp. 111-20. Dordrecht: Kluwer

Liseau R, White GJ, Larsson B, Sidher S, Olofsson G, et al. 1999.
{\it Astron. Astrophys.} 344:342-54

Lorenzetti D, Giannini T, Nisini B, Benedettini M, Creech-Eakman M, et al.
2000. {\it Astron. Astrophys.} 357:1035-44

Lorenzetti D, Giannini T, Nisini B, Benedettini M, Elia D, Campeggio L,
Strafella F. 2002. {\it Astron. Astrophys.} 395:637-45

Lorenzetti D, Tommasi E, Giannini T, Nisini B, Benedettini M, et al. 1999.
{\it Astron. Astrophys.} 346:604-16

Malfait K, Waelkens C, Bouwman J, de Koter A, Waters LBFM. 1999.
{\it Astron. Astrophys.} 345:181-6

Malfait K, Waelkens C, Waters LBFM, Vandenbussche B, Huygen E, de Graauw MS.
1998. {\it Astron. Astrophys.}  332:L25-8

Maret S, Ceccarelli C, Caux E, Tielens AGGM, Castets A. 2002. {\it
Astron. Astrophys.} 395: 573-85

Mattila K, Lemke D, Haikala LK, Laureijs RJ, L\'eger A, Lehtinen K,
Leinert C, Mezger PG. 1996. {\it Astron. Astrophys.} 315:L353-56

Meeus G, Waters LBFM, Bouwman J, van den Ancker ME, Waelkens C,
Malfait K. 2001. {\it Astron. Astrophys.} 365:476-90

Meeus G, Bouwman J, Dominik C, Waters LBFM, de Koter A. 2002.
{\it Astron. Astrophys.} 392:1039-46

Melnick G, Gull GE, Harwit M. 1979. {\it Ap. J. Lett.} 227:L29-33

Mennella V, Mu\~noz-Caro GM, Ruiterkamp R, Schutte WA, Greenberg JM,
Brucato JR, Colangeli L. 2001. {\it Astron. Astrophys.} 367:355-61

Miroshnichenko A, Ivezic Z, Vinkovic D, Elitzur M. 1999. {\it Ap. J.}
520:L115-18

Molinari S, Ceccarelli C, White GJ, Saraceno P, Nisini B, Giannini T, Caux E.
1999. {\it Ap. J.} 521:L71-4

Molinari S, Noriega-Crespo A. 2002. {\it Astron. J.} 123:2010-18

Molinari S, Noriega-Crespo A, Ceccarelli C, Nisini B, Giannini T, et al.
2000. {\it Ap. J.} 538:698-709

Molinari S, Noriega-Crespo A, Spinoglio L. 2001. {\it Ap. J.} 547:292-301

Molster FJ, Waters LBFM 2003. See Henning 2003, pp. 121-170

Molster FJ, Waters LBFM, Tielens AGGM, Koike C, Chihara H. 2002.
{\it Astron. Astrophys.} 382:241-55

Moneti A, Cernicharo J. 2000. See Salama et al. 2000, pp. 119-22

Moore MH, Hudson RL. 1998. {\it Icarus} 135:518-27

Moos HW, Sembach KR, Vidal-Madjar A, York DG, Friedman SD, et al. 2002.
{\it Ap. J. Suppl.} 140:3-17

Moro-Mart{\'\i}n A, Noriega-Crespo A, Molinari S, Testi L, Cernicharo
J, Sargent A. 2001. {\it Ap. J.} 155:146-59

Moutou C, Sellgren K, Verstraete L, L\'eger A. 1999a. {\it Astron. Astrophys.}
347:949-56

Moutou C, Verstraete L, Sellgren K, L\'eger A. 1999b. See Cox \& Kessler
1999, pp.\ 727-30

Moutou C, Verstraete L, L\'eger A, Sellgren K, Schmidt W. 2000.
{\it Astron. Astrophys.} 354:L17-20

Najita J, Carr JS, Mathieu RD. 2003. {\it Ap. J.} 589:931-52

Natta A, Meyer MR, Beckwith SVW. 2000. {\it Ap. J.} 534:838-45

Natta A, Prusti T, Neri R, Wooden D, Grinin VP, Mannings V. 2001.
{\it Astron. Astrophys.} 371:186-97

Neufeld DA, Melnick GJ, Harwit M. 1998. {\it Ap. J.} 506:L75-8

Neufeld DA, Zmuidzinas J, Schilke P, Phillips TG. 1997.
{\it. Ap. J.} 488:L141-44

Nisini B, Benedettini M, Giannini T, Clegg PE, Di Giorgio AM, et al.
1999b. {\it Astron. Astrophys.} 343:266-72

Nisini B, Benedettini M, Giannini T, Caux E, Di Giorgio AM, et al.
1999a. {\it Astron. Astrophys.} 350:529-40

Nisini B, Benedettini M, Giannini T, Codella C, Lorenzetti D, Di Giorgio AM,
Richer JS. 2000. {\it Astron. Astrophys.} 360:297-310

Nisini B, Giannini T, Lorenzetti D. 2002. {\it Ap. J.} 574:246-57

Nisini B, Lorenzetti D, Cohen M, Ceccarelli C, Giannini T, et al. 1996.
{\it Astron. Astrophys.} 315:L321-4

Noriega-Crespo A. 2002. {\it Rev. Mex. Astron. Astrofis.} 13:71-8

Nummelin A, Whittet DCB, Gibb EL, Gerakines PA, Chiar JE. 2001.  {\it
Ap. J.} 558:185-93

Okada Y, Onaka T, Shibai H, Doi Y. 2003. {\it Astron. Astrophys.} 412:199-212

Omont A, Gilmore GF, Alard C, Aracil B, August T, et al. 2003.
{\it Astron. Astrophys.} 403:975-92

Onaka T, Okada Y. 2003. {\it Ap. J.} 585:872-77

Parmar PS, Lacy JH, Achtermann JM. 1991. {\it Ap. J. Lett.} 372:L25-8

Peeters E, Allamandola LJ, Bauschlicher CW, Hudgins DM, Sandford SA, Tielens
AGGM. 2004a. {\it Ap. J. Lett.} in press

Peeters E, Allamandola LJ, Hudgins DM, Hony S, Tielens AGGM. 2004b. In {\it
The Astrophysics of Dust}, ed. AN Witt, GC Clayton, BT Draine, in
press, San Francisco: ASP

Peeters, E, Hony S, van Kerckhoven C, Tielens AGGM, Allamandola LJ, Hudgins
DM, Bauschlicher CW. 2002a. {\it Astron. Astrophys.} 390:1089-1113

Peeters, E, Mart{\'\i}n-Hern\'andez NL, Damour F, Cox P, Roelfsema PR, et al.
2002b. {\it Astron. Astrophys.} 381:571-605 

Pendleton YJ, Allamandola LJ. 2002. {\it Ap. J. Suppl.} 138:75-98

Pendleton YJ, Tielens AGGM, Tokunaga AT, Bernstein MP. 1999.
{\it Ap. J.} 513:294-304

Poglitsch A, Hermann F, Genzel R, Madden SC, Nikola T, et al. 1996.
{\it Ap. J. Lett.} 462:L43-6

Polehampton ET, Baluteau JP, Ceccarelli C, Swinyard BM, Caux E.
2002. {\it Astron. Astrophys.} 388:L44-7

Rho J, Reach W. 2003. {\it Rev. Mex. AA.} 15:263-6

Richter MJ, Jaffe DT, Blake GA, Lacy JH. 2002. {\it Ap. J. Lett.}
572:L161-4

Rietmeijer FJM, Hallenbeck SL, Nuth JA, Karner JM. 2002. {\it Icarus}
156:269-86

Rigopoulou D, Kunze D, Lutz D, Genzel R, Moorwoord AFM. 2002.
{\it Astron. Astrophys.} 389:374:86

Rosenthal D, Bertoldi F, Drapatz S. 2000. 
{\it Astron. Astrophys.} 356:705-23

Salama A, Kessler MF, Leech K, Schulz B, eds. 2000. {\it Second ISO Workshop
on Analytical Spectroscopy}, ESA-SP 456. Noordwijk: ESA-ESTEC. 389 pp.

Sandford SA, Bernstein MP, Allamandola LJ, Goorvitch D, Teixeira TCVS.
2001. {\it Ap. J.} 548:836-51

Saraceno P, Andr\'e P, Ceccarelli C, Griffin M, Molinari S. 1996.
{\it Astron. Astrophys.} 309:827-39

Saraceno P, Nisini B, Benedettini M, Di Giorgio AM, Giannini T, et al.
1999. See Cox \& Kessler 1999, pp. 575-78

Schneider N, Simon R, Kramer C, Kraemer K, Stutzki J, Mookerjea B. 2003.
{\it Astron. Astrophys.} 406:915-35

Schreyer K, Henning Th, van der Tak FFS, Boonman AMS, van Dishoeck EF. 2002.
{\it Astron. Astrophys.} 394:561-83

Schutte WA. 1999. See d'Hendecourt et al. 1999, pp. 183-98

Schutte WA, Khanna RK. 2003. {\it Astron. Astrophys.} 398:1049-62

Schutte WA, Boogert ACA, Tielens AGGM, Whittet DCB, Gerakines PA, et al.
1999. {\it Astron. Astrophys.} 343:966-76

Schutte WA, Tielens AGGM, Whittet DCB, Boogert ACA, Ehrenfreund P, et al.
1996. {\it Astron. Astrophys.}  315:L333-36

Schutte WA, van der Hucht KA, Whittet DCB, Boogert ACA, Tielens AGGM, et al.
1998. {\it Astron. Astrophys.} 337:261-74

Sellgren K. 1984. {\it Ap. J.} 277:623-33

Sempere MJ, Cernicharo J, Lefloch B, Gonz\'alez-Alfonso E, Leeks S. 2000.
{\it Ap. J.} 530:L123-27

Siebenmorgen R, Prusti T, Natta A, M\"uller TG. 2000.
{\it Astron. Astrophys.} 361:258-64

Smith MD, Froebrich D, Eisl\"offel J. 2003. {\it Ap. J.} 592:245-54

Smith RG, Sellgren K, Tokunaga AT. 1989.
{\it Ap. J.} 344:413-26

Snell RL, Howe JE, Ashby MLN, Bergin EA, Chin G, et al. 2000. {\it Ap. J.}
539:L101-4

Snow TP, Rachford BL, Tumlinson J, Shull JM, Welty DE, et al. 2000.
{\it Ap. J.} 538:L65-8

Spinoglio L, Giannini T, Nisini B, van den Ancker ME, Caux E, et al. 2000.
{\it Astron. Astrophys.} 353:1055-64

Stacey, GJ, Geis N, Genzel R, Lugten JB, Poglitsch A, Sternberg A, Townes CH.
1991. {\it Ap. J.} 373:423-44

Stantcheva T, Shematovich VI, Herbst E. 2002. 
{\it Astron. Astrophys.} 391:1069-80

Sternberg A, Dalgarno A. 1989. {\it Ap. J.} 338:197-233

Sternberg A, Neufeld DN. 1999. {\it Ap. J.} 516:371-80

Spitzer L, Cochran WD. 1973. {\it Ap. J. Lett.} 186:L23-8

Taban IM, Schutte WA, Pontoppidan KM, van Dishoeck EF. 2003.
{\it Astron. Astrophys.} 399:169-75

Teixeira TC, Devlin JP, Buch V, Emerson JP. 1999.  {\it Astron.
Astrophys.} 347:L19-22

Thi WF, Blake GA, van Dishoeck EF, van Zadelhoff GJ, Horn J, et al.
2001a. {\it Nature} 409:60-3

Thi WF, van Dishoeck EF, Black JH, Jansen DJ, Evans NJ, Jaffe DT. 1999.
See Cox \& Kessler 1999 pp.\ 767-70

Thi WF, van Dishoeck EF, Blake GA, van Zadelhoff GJ, Horn J, et al. 2001b,
{\it Ap. J.} 561:1074-94

Tielens AGGM, Hagen W. 1982. {\it Astron. Astrophys.} 114:245-60

Tielens AGGM, Hollenbach DJ 1985. {\it Ap. J.} 291:722-46

Tielens AGGM, Hony S, van Kerckhoven C, Peeters E. 2000.
In {\it IAU Symp. 197, Astrochemistry}, ed. YC Minh, EF van Dishoeck, pp.
349-62. San Francisco: Astron. Soc. Pac.

Tielens AGGM, Peeters E. in {\it The Dense Interstellar Medium in
Galaxies}, ed. S Pfalzner, C Kramer, C Straubmeier, A Heithausen, in
press. Berlin: Springer

Tielens AGGM, Tokunaga AT, Geballe TR, Baas F. 1991. {\it Ap. J.} 381:181-99

Timmermann R. 1998. {\it Ap. J.} 498:246-55

Timmermann R, Bertoldi F, Wright CM, Drapatz S, Draine BT, Haser L,
Sternberg A.  1996. {\it Astron. Astrophys.} 315:L281-4

Timmermann R, K\"oster B, Stutzki J. 1998. {\it Astron. Astrophys.} 336:L53-6

Valentijn EA, van der Werf PP. 1999. {\it Ap. J. Lett.} 522:L29-33

Valentijn EA, van der Werf PP, de Graauw Th, de Jong T. 1996,
{\it Astron. Astrophys.} 315:L145-8

van Broekhuizen FA, Keane JV, Schutte WA. 2004. {\it Astron. Astrophys.}
in press (astro-ph/0311617)

van den Ancker ME. 2000. PhD thesis, University of Amsterdam

van den Ancker ME, Bouwman J, Wesselius PR, Waters LBFM, Dougherty SM,
van Dishoeck EF. 2000a. {\it Astron. Astrophys.} 357:325-29

van den Ancker ME, Tielens AGGM, Wesselius PR. 2000b. {\it Astron. Astrophys.}
358:1035-48

van den Ancker ME, Wesselius PR, Tielens AGGM. 2000c. {\it Astron. Astrophys.}
355:194-210

van den Ancker ME, Wesselius PR, Tielens AGGM, van Dishoeck EF, Spinoglio L.
1999. {\it Astron. Astrophys.} 348:877-87

Vandenbussche B, Dominik C, Min M, van Boekel R, Waters LBFM, et al.
2004. {\it Astron. Astrophys.} submitted

Vandenbussche B, Ehrenfreund P, Boogert ACA, van Dishoeck EF, Schutte
WA, et al. 1999. {\it Astron. Astrophys.} 346:L57-60

van der Tak FFS, van Dishoeck EF, Caselli P. 2000a. {\it Ap. J.}
361:327-39

van der Tak FFS, van Dishoeck EF, Evans NJ, Blake GA. 2000b. 
{\it Ap. J.} 537:283-303

van Dishoeck EF. 1998. {\it Faraday Disc.} 109:31-46

van Dishoeck EF. 2003. In {\it Chemistry as a Diagnostic of Star Formation},
eds. CL Curry, M Fich, in press. Ottawa: NRC press (astro-ph/0301518)

van Dishoeck EF, Helmich FP. 1996. 
{\it Astron. Astrophys.} 315:L177-80

van Dishoeck EF, Helmich FP, de Graauw Th, Black JH, Boogert ACA, et al.
1996. {\it Astron. Astrophys.} 315:L349-52

van Dishoeck EF, Tielens, AGGM. 2001. In {\it The Century of Space Science},
ed. JAM Bleeker, J Geiss, MCE Huber, pp. 607-45. Dordrecht: Kluwer

van Dishoeck EF, Black JH, Boogert ACA, Boonman AMS, Ehrenfreund P, et
al. 1999. See Cox \& Kessler 1999,
pp. 437-48. 

van Dishoeck EF, Wright CM, Cernicharo J, Gonz\'alez-Alfonso E, Helmich FP,
de Graauw TH, Vandenbussche B. 1998. {\it Ap. J. Lett.} 502:L173-76

van Kerckhoven C, Hony S, Peeters E, Tielens AGGM, Allamandola LJ, et al.
2000. {\it Astron. Astrophys.} 357:1013-19

Vastel C, Caux, E, Ceccarelli C, Castets A, Gry C, Baluteau JP. 2000.
{\it Astron. Astrophys.} 357:994-1000

Vastel C., Spaans M, Ceccarelli C, Tielens AGGM, Caux E. 2001.
{\it Astron. Astrophys.} 376:1064-72

Vastel C, Polehampton ET, Baluteau JP, Swinyard BM, Caux E, Cox P. 2002.
{\it Ap. J.} 581:315-324

Verstraete L, Pech C, Moutou C, Sellgren K, Wright CM, et al. 2001.
{\it Astron. Astrophys.} 372:981-97

Verstraete L, Puget JL, Falgarone E, Drapatz S, Wright CM, Timmermann R.
1996. {\it Astron. Astrophys.} 315:L337-40

Waelkens C, Waters LBFM, de Graauw Th, Huygen E, Malfait K, et al. 1996.
{\it Astron. Astrophys.} 315:L245-48

Walmsley CM, Pineau des For\^ets G, Flower DR. 1999. {\it Astron. Astrophys.}
342:542-50

Ward-Thompson D, Andr\'e P, Kirk JM. 2002. {\it MNRAS} 329:257-76

Waters LBFM, Waelkens C. 1998. {\it Annu. Rev. Astron. Astrophys.} 36:233-66

Waters LBFM, Waelkens C, van der Hucht KA, Zaal PA, eds. 1998
{\it ISO's View on Stellar Evolution}. Dordrecht:Kluwer. 530 pp.

White GJ, Liseau R, Men'shchikov AB, Justtanont K, Nisini B, et al. 2000.
{\it Astron. Astrophys.} 364:741-62

Willner SP, Gillett FC, Herter TL, Jones B, Krassner J, et al. 1982.
{\it Ap. J.} 253:174-87

Wright CM. 2000. In {\it Astrochemistry: from Molecular Clouds to
Planetary Systems}, IAU Symposium 197, ed. YC Minh, EF van Dishoeck,
pp.\ 177-190. San Francisco: Astron. Soc. Pac.

Wright CM, Drapatz S, Timmermann R, van der Werf PP, Katterloher R, de Graauw
Th. 1996. {\it Astron. Astrophys.} 315:L301-4

Wright CM, Timmermann R, Drapatz S. 1997. See Heras et al. 1997.
pp.\ 311-2

Wright CM, van Dishoeck EF, Black JH, Feuchtgruber J, Cernicharo J,
Gonz\'alez-Alfonso E, de Graauw Th. 2000. {\it Astron. Astrophys.}
358:689-700

Wright CM, van Dishoeck EF, Cox P, Sidher SD, Kessler MF. 1999.
{\it Ap. J. Lett.} 515:L29-33

Yun J, Liseau R., eds. 1998. {\it Star Formation with the Infrared Space
Observatory}, Vol. 132. San Francisco: Astron. Soc. Pac. 448 pp.

\vfill \eject

\bye